\documentclass[useAMS,usenatbib]{mn2e}
 \usepackage{times}
\usepackage[utf8]{inputenc}
\usepackage{txfonts}
\usepackage{pdflscape}
\usepackage{multirow}
\usepackage{graphicx}
\usepackage{epstopdf}

\begin{document}
\date{Accepted -. Received -}
\title{A VLBI survey of compact Broad Absorption Lines (BAL) quasars with BALnicity Index BI=0 }

\author[M. Ceg\l{}owski et al.]{M. Ceg\l{}owski$^{1}$\thanks{E-mail:ceglowski@astri.uni.torun.pl (MC)},
M. Kunert-Bajraszewska$^{1}$ , C. Roskowi\'nski$^{1}$\\
$^{1}$ Toru\'n Centre for Astronomy, Faculty of Physics, Astronomy
and Informatics, NCU, Grudziacka 5, 87-100 Toru\'n, Poland}

\pagerange{\pageref{firstpage}--\pageref{lastpage}} \pubyear{2002}

\maketitle
\label{firstpage}

\begin{abstract}
We  present high-resolution observations, using both the European VLBI Network (EVN) at 1.7\,GHz, and the Very Long Baseline Array (VLBA) at 5 and 8.4\,GHz
to image radio structures of 14 compact sources classified as broad absorption line (BAL) quasars based on the absorption index (AI). 
All source but one  were resolved, with the majority showing core-jet morphology typical for radio-loud quasars.
We discuss in details the most interesting cases. The high radio luminosities and small linear sizes of the observed objects indicate they are strong young AGNs.
 Nevertheless, the distribution of the radio-loudness parameter, log\,$\rm R_{I}$, of a larger sample of AI quasars shows that the objects observed 
by us constitute the most luminous, small subgroup of AI population. Additionally we report that for the radio-loudness parameter, the distribution
of AI quasars and those selected by using the traditional balnicity index (BI) – BI quasars differ significantly.
Strong absorption is connected with the lower log\,$\rm R_{I}$, and thus probably with larger viewing angles.
Since, the AI quasars have on average larger log\,$\rm R_{I}$, the orientation can cause that we see them less absorbed. 
However, we suggest that the orientation is not the only parameter that affects the  detected absorption. 
The fact that the strong absorption is associated with the weak radio emission is equally important and worth exploring.

\end{abstract}

\section{Introduction}
\label{intro}

Broad absorption line quasars have been observed for over two decades. They 
spectra have blue-shifted absorption troughs. This features are usually
linked with the outflow of highly ionized plasma. The velocities of outflow can reach up to 0.3 c \citep{hewett03}. Noteworthy is that, in general absorption, 
can have various origin \citep{richard03}, 
from structures closely bonded by gravity to central engine, followed by 
host galaxy environment and matter which lies in the line of sight between
observer and AGN. Therefore, it is extremely difficult to exclude one forme of absorption
from another and draw unambiguous threshold. To tackle this issue, \citet{weymann91}
proposed the balnicity index (BI), which accounts only for troughs 2000 km/s wide
and blue-shifted more than 3000 km/s, to quantify the true broad absorption. 
However, BI is too restrictive since it excludes large number of
quasars with significant absorption lines troughs or the so-called mini-BALs with BI=0. 
Hence, \citet{hall02} suggested a more relaxed definition, the absorption index (AI), which includes troughs no smaller than 
1000 km/s. The fraction of broad absorption line quasars (BALQSOs) among the whole quasar population varies
from 15\,\% to 26\,\% depending on the definition used \citep{hewett03, trump06, dai08, maddox08, shankar2008, knigge2008, gibson2009}.

Recently, the radio, optical and X-ray studies of traditional BI quasars and absorption line (AI) quasars showed that they might be 
two independent classes of objects \citep{knigge2008, strebl10}.
However, this do not necessarily means that the classical BALs and much weaker and narrower absorption lines must
be produced in different line-forming regions. Theoretical model of \citet{elvis00} unifies the broad and narrow
absorption features, suggesting they are formed in the same disc wind but the orientation of the source with respect to the 
observer changes their appearance or can even prevent us from detection of these lines. 
Discovery of the existence of radio-loud BALQSOs gave us another opportunity to study the BAL phenomenon. 
The radio emission is an additional tool to understand their orientation and age by the VLBI imaging 
(detection of radio jets and their direction plus size determination), through the radio-loudness parameter distribution and variability study.

\begin{table*}
\caption[]{Radio-loud AI quasars observed with VLBA and EVN.}
\label{basic}
{\small
\begin{tabular}{cccccccccccc}
\hline

RA(J2000) & Dec(J2000) &
\multicolumn{1}{c}{\it z}&
\multicolumn{1}{c}{AI}&
\multicolumn{1}{c}{${\rm S_{1.4\,GHz}}$}&
\multicolumn{1}{c}{log${\rm L_{1.4\,GHz}}$}&
\multicolumn{1}{c}{${\rm S_{4.85\,GHz}}$}&
\multicolumn{1}{c}{${\rm S_{8.4\,GHz}}$}&
 \multicolumn{1}{c}{ $\alpha^{1.4 GHz}_{4.85 GHz}$}&
 \multicolumn{1}{c}{ $\alpha^{4.85 GHz}_{8.4 GHz}$}&
\multicolumn{1}{c}{log $\rm R_{I}$}
\\

h~m~s & $\degr$~$\arcmin$~$\arcsec$ & 
\multicolumn{1}{c}{}&
\multicolumn{1}{c}{($\rm km~s^{-1}$)} &
\multicolumn{1}{c}{(mJy)}&
\multicolumn{1}{c}{($\rm W~Hz^{-1}$)} & 
\multicolumn{1}{c}{(mJy)}&
\multicolumn{1}{c}{(mJy)}&
&
&
\\

(1)& (2)& (3) &(4)&   
\multicolumn{1}{c}{(5)}&
\multicolumn{1}{c}{(6)}& 
\multicolumn{1}{c}{(7)}&
\multicolumn{1}{c}{(8)}&  
\multicolumn{1}{c}{(9)}&
(10)&(11)\\

\hline
02	17	28.614	&	00	52	26.88	&	2.46	&	791	&       217	&	28.01	&	104	&	--	&      0.59	&--		&3.44\\
07	56	28.260	&	37	14	55.80   &	2.51	&	841	&	233	&	28.08	&	236	&	136	&	0.04	&0.99		&3.46\\
08 00 16.065 & 40 29 55.990 & 2.02& 1114 & 199 & 27.77& 68 & -- &0.86 & --&2.93 \\
08	15	34.184	&	33	05	29.28	&	2.43	&	747	&	342	&	28.19	&	112	&	--	&	0.90	&--		&3.20\\
09	28	24.133	&	44	46	04.68	&	1.90	&	293	&      162	&	27.60	&	256	&	229	&      -0.37	&0.20		&2.64\\
10	05	15.961	&	48	05	33.21	&	2.37	&	783	&      209	&	27.95	&	70	&	--	&       0.88  	&--		&2.47\\
10	13	29.931	&	49	18	41.11	&	2.20	&	361	&	269	&	27.98	&	167	&	106	&       0.38	&0.83		&3.09\\
10	18	27.837	&	05	30	29.90	&	1.94	&	441	&	296	&	27.89	&654 ( 330$^{*}$ )&	300	&-0.64 ( -0.10 )&1.42 ( 0.17 )	&3.13\\
10	42	57.598	&	07	48	50.60	&	2.67	&	1011	&	381	&	28.33	&	167	&	--	&	0.66	&--		&3.20\\
10	57	26.608	&	03	24	48.47	&	2.83	&	440	&	157	&	28.01	&	70	&	--	&	0.65	&--		&3.03\\
11	03	44.536	&	02	32	09.74	&	2.51	&	460	&      165	&	27.91	&	107	&	60	&       0.35	&	1.05	&2.67\\
11	59	44.832	&	01	12	06.87	&	2.00	&	2887	&      268	&	27.88	&	125	&	147	&	0.61	&	-0.30	&2.40\\
12	23	43.165	&	50	37	53.49	&	3.49	&	413	&      228	&	28.40	&	110	&	117	&       0.59	&	-0.11	&2.42\\
14	05	07.795	&	40	56	58.06	&	1.99	&	780	&      214	&	27.77	&	278	&	197	&      -0.21	&	0.63	&3.14\\
14	32	43.322	&	41	03	28.04	&	1.97	&	343	&	261	&	27.85	&	102	&	--	&	0.76	&	--	&2.60\\
15 28 21.684 & 53 10 30.68 & 2.82& 1701 & 232 & 28.09& 76 & --&0.90 & --& 3.32\\

\hline
\end{tabular}
\begin{minipage}{175mm} 
{Description of the columns:
(1) \& (2) source coordinates (J2000) extracted from FIRST, (3) redshift as measured from the SDSS, (4) absorption index taken from \citet{trump06}, 
(5) total flux density at 1.4\,GHz extracted from FIRST (VLA measurement), (6) log of the radio luminosity at 1.4\,GHz, (7) total flux density at 4.85\,GHz taken
from  \citet{becker91} or \citet{grif95} if marked as $^{*}$ (single dish measurements),(8) total flux density at 8.4\,GHz taken
from  \citet{heal07} (VLA measurement) ,(9) spectral index between
1.4 and 4.85\,GHz calculated using flux densities in columns (5) and (7), (10) spectral index between
4.85 and 8.4\,GHz calculated using flux densities in columns (7) and (8), (11) radio-loudness, the radio-to-optical (i-band) ratio of the quasar core \citep{kimball2011}, which were calculated
from {\it z}, ${\rm S_{1.4\,GHz}}$, $\rm M_{i}$ taken from \citet{trump06}, and the assumption of a radio core spectral index of 0 and
an optical spectral index of -0.5}
\end{minipage}
}
\end{table*}

It has been suggested \citep{becker00} that, because of their small sizes,
most of the radio-loud BALQSOs belong to the class of compact radio
sources, namely compact steep spectrum (CSS) objects and gigahertz peaked
spectrum (GPS) objects. 
Currently, there are only few surveys focused on radio imaging of compact radio-loud BALQSOs 
using global interferometric arrays as well as local-interferometry technique
\citep{jiang03,kun07,liu08,monte08,doi09,doi13,kun10a,gawr10,bruni12,bruni13}. About half of them have still
unresolved radio structures even in the high resolution observations. If resolved, most of the compact
BALQSOs have a core-jet morphology and some, have more complex structures, indicating the dense 
environment or restarted activity \citep{kun07, kun10a, bruni13, taka13}. All of them are potentially smaller
than their host galaxies.
The analysis of the spectral shape,
variability and polarization properties of some of them shows that
they are indeed similar to CSS and GPS objects \citep{monte08, liu08, kun10a}.
Moreover, they are not oriented along a
particular line of sight, although they are more often observed farther from the jet axis compared to
normal quasars \citep{dipompeo11,bruni12}.

On the other hand, since the GPS and CSS sources are considered to be young radio
objects, progenitors of large-scale radio-loud AGNs, the evolution scenario has been proposed for 
BALQSOs \citep{becker00, gregg06}. The BAL phase appear together with the birth of a quasar and is 
systematically destroyed by the radio jets during quasar lifetime \citep{urrutia09}. 
It is still uncertain which scenario, maybe mixture of both, is most suitable. 

This paper presents our VLBI observations and analysis of a sample of BALQSOs 
selected from the \citet{trump06} catalogue. The authors classify the Sloan
Digital Sky Survey Third Data Release (SDSS/DR3) sources
as BALQSOs based on the more liberal absorption index, although the values
of both, the AI and BI indices, are calculated for all quasars. Therefore all sources from the catalogue in a 
natural way falls into two groups: 1) objects with AI$>$0 and BI=0, the mini-BALs (hereafter AI quasars) and 2) 
objects with AI$>$0 and BI$>$0, the so called 'true BALQSOs' (hereafter BI quasars). The study
of the sources from the second group will be reported in a separate paper. Here we focus on the AI quasars.

\section{Sample selection and radio observations}
\label{obs}

\begin{figure}
\centering      
\includegraphics[width=\columnwidth]{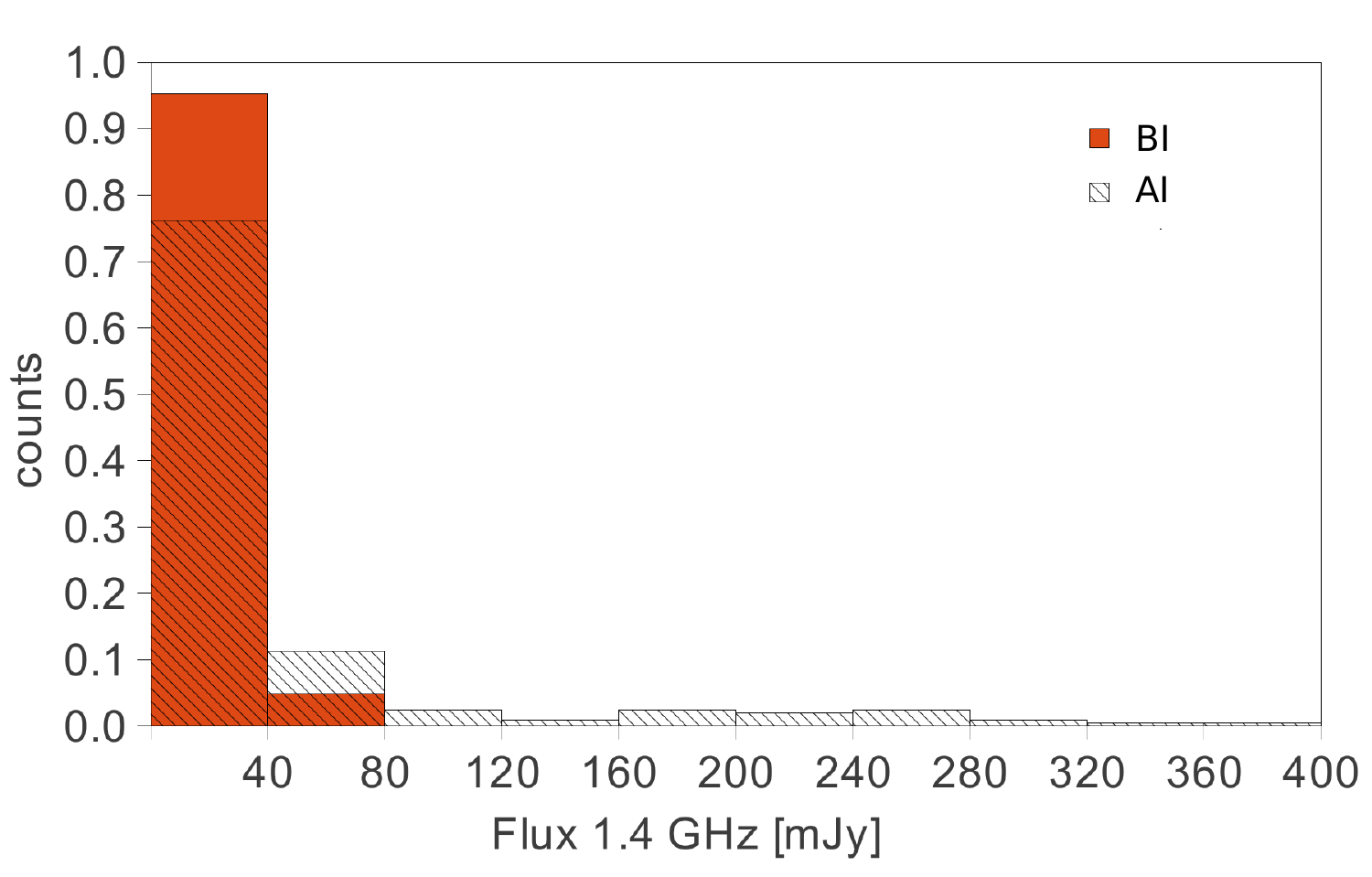}
\caption{1.4\,GHz flux distribution for radio-loud BALQSO selected as specified in section \ref{obs}.
Histogram represents compact BALQSOs - both AI (204 objects; B=0) and BI (105 objects; BI$>$0) sources - found by us during the selection process.
}
\label{counts1}
\end{figure}

We  carefully prepared our sample based on final release of the FIRST survey \citep{white97}
and {\it A catalogue of Broad Absorption Line Quasars} from the {\it Sloan
Digital Sky Survey Third Data Release} made by \citet{trump06}\footnote{
Our observations were made in 2008, therefore,
we could use only data available from \citet{trump06} as no additional catalogue existed back then.
}.
The selection process required several stages and resulted in pinpointing genuine unresolved radio objects. 
In the first step we have matched the optical 
positions of BALQSOs from the \citet{trump06} catalogue to FIRST coordinates
in a radius of 10 arcseconds. 
Primary we excluded sources with radio emission less than 2 mJy and/or side lobe probability more than 0.1, as the likelihood of false detection was the highest in this case.
At this stage our sample counted 350 sources.
Afterwards we have excluded from the sample objects with additional radio counterparts within 60 arcseconds of SDSS position as those possessing
large scale structures which constituted $<\,12\%$ of the above number. This approach allowed us to avoid ambiguity in identification of the
radio core, which is important in the statistical studies.
We recall here that \citet{trump06} quantify the sources as BALQSOs
using absorption index (AI), so their final catalogue contains also sources with the traditional balnicity index BI=0. As has been
already discussed in Sec.~\ref{intro} the AI (AI$>$0 and BI=0) and BI (AI$>$0 and BI$>$0) quasars differ significantly in many properties and
thus we treat them also as separate groups. 
Finally our sample of compact BALQSOs - which we define as the 'parent sample' - consists of 309 sources, out of which 105 are the BI quasars and 204 are the AI objects (Fig.~\ref{counts1}). 

In the next step we selected comparably numbered samples of AI and BI sources with the largest 1.4\,GHz flux densities for further
high resolution VLBI observations. This resulted in 16 AI quasars with flux densities $\rm S_{1.4\,GHz}>150\,mJy$ and 15 BI 
objects with flux densities $\rm S_{1.4\,GHz}>20\,mJy$. Note that statistically, 
compact BI objects are fainter than AI objects and all BI quasars from our sample have flux densities $\rm S_{1.4\,GHz}<80\,mJy$ (Fig.~\ref{counts1}). 
In this paper we analyse the observations and properties of AI quasars (Table~\ref{basic}). 

We observed the AI quasars with EVN at 1.7\,GHz and with VLBA at 5 and 8.4\,GHz. However, during the sample preparations
we noticed that five of our objects had  already been observed with VLBA at 5\,GHz as part of the VLBA Imaging and
Polarimetry Survey (VIPS; \citet{helm07}). Therefore we continued observations with the VLBA 
in the same fashion in order to acquire radio maps with comparable dynamical range for the rest of our sources. Each target source was observed using EVN or VLBA for approximately two hours on each frequency in phase-referencing mode.  
We have observed all 16 
objects at 8.4\,GHz and 11 sources at 5\,GHz with VLBA. Ten sources were observed at 1.7\,GHz with EVN. 
We failed to proper image 2 objects (0800+4029, 1528+5310) due to detection problems at 1.7\,GHz with EVN and at 5\,GHz with VLBA.

L-band observations were carried out with EVN on 27-28th of October in 2008. 
Antennas in Jodrell Bank, Westerbork, Effelsberg, Onsala, Medicina, Torun, Shanghai, Urumqi, Noto and Robledo took part in our experiment. The data were correlated at the Joint Institute for VLBI
in Europe (JIVE) correlator in Dwingeloo (The Netherlands).

C and X-band observations were performed by VLBA array in two runs,
first on 27-30th of August and second on 6-7th of September in 2008. 
The correlation was performed with the VLBA correlator at the National Radio Astronomy Observatory (NRAO)
in Socorro (US). The data were then processed with the Distributed FX (DiFX) software correlator
\citep{dell07}.

Data reduction including calibration and fringe-fitting was performed using Astronomical Image Processing System - AIPS
package\footnote{http://www.aips.nrao.edu}. Imaging and self-calibration part
was performed using Difmap software \citep{shep93}. Source components were fitted with
circular Gaussian model on the final, self-calibrated visibility data using the MODELFIT programme.
The final images of the radio-loud BALQSOs are presented in Fig.~\ref{images1} and Fig.~\ref{images2} and the modelfit parameters
are listed in Table~\ref{observations}.

Throughout the paper, we assume a cosmology with
${\rm H_0}$=70${\rm\,km\,s^{-1}\,Mpc^{-1}}$, $\Omega_{M}$=0.3, $\Omega_{\Lambda}$=0.7. The adopted convention for the spectral index
definition is $S\propto\nu^{-\alpha}$.

\subsection{Possible biases in the sample}

This paper in detail presents analysis of BALQSOs with BI=0, called by us AI sources. 
However, to capture a wider perspective of the blushifted BAL phenomenon we complement this information of findings for BI sources - objects with BI$>$0.
Notice that we have focused mainly on the objects which where unresolved in FIRST survey and do not have 
any additional components within 1 arcminute. Therefore, the statistical 
analysis of the BALQSOs is devoid of error connected with pinpointing the radio core
in sources which posses more than one counterpart on VLA.
Definitely most of the BALQSOs are single, unresolved objects on the VLA resolution and we found only 41 sources out of 350 (both AI and BI sources) that
could be classified as large-scale ones. Even
after the rejecting of large-scale BALQSOs from our sample we still draw the statistical conclusions based on the significant fraction of BALQSO population.

\begin{figure}
\centering
\includegraphics[width=\columnwidth]{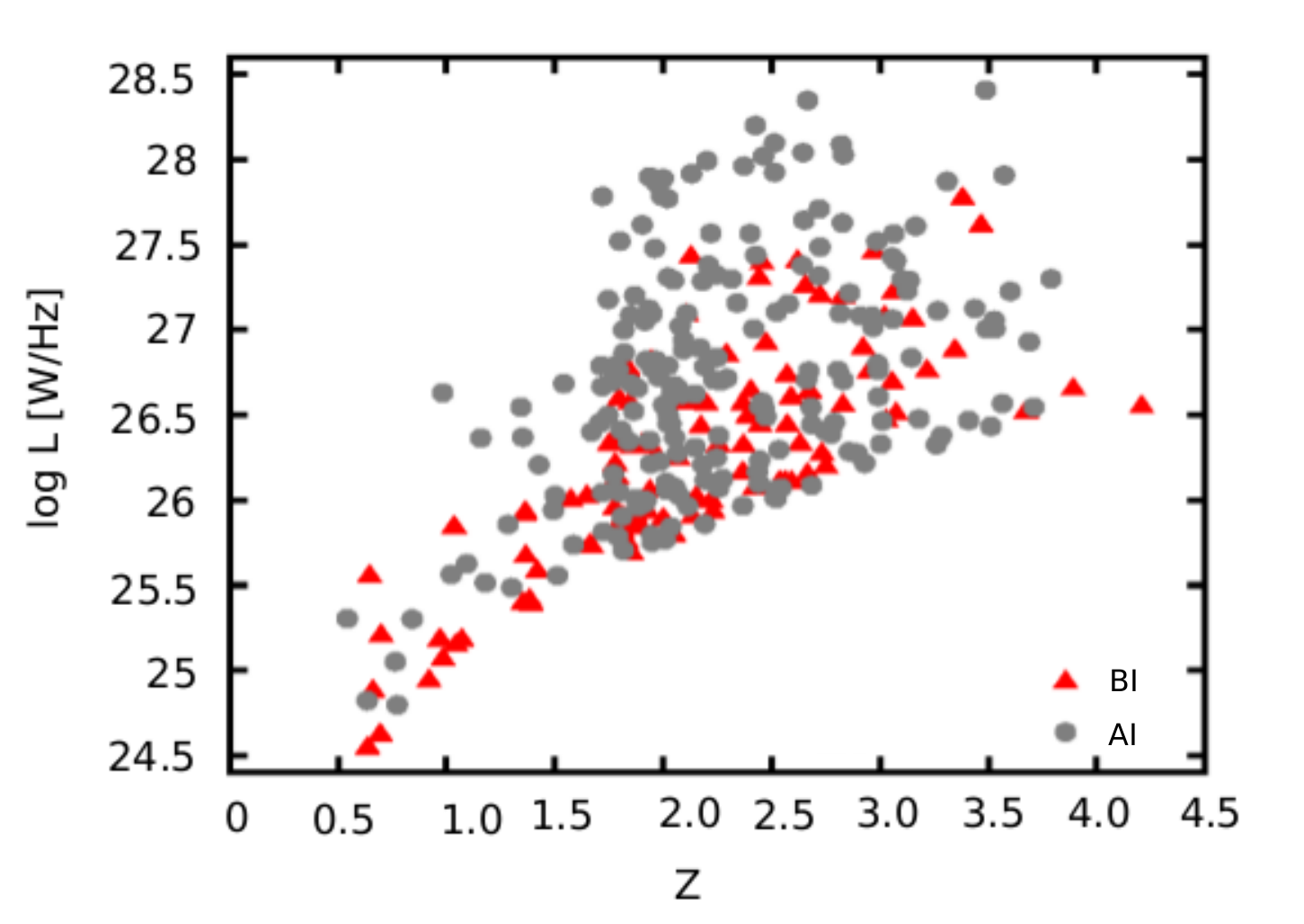}
\caption{1.4\,GHz luminosity-redshift diagram for compact AI (circles) and BI quasars (triangles)
selected as specified in section \ref{obs}.}
\label{mal}
\end{figure}

We have also found that the number of extended objects decreases with increasing the value of the flux density in both sub-samples and amounts 
to 7\% and 17\% for BI and AI quasars, respectively.
Note that possible bias can account for the above percentage difference. Radio luminosity of objects from BI sample strongly
correlates with redshift (Fig.~\ref{mal}), with Pearson correlation coefficient r=0.77. The correlation in AI sample is weaker, r=0.46. This could be the result
of flux density limit superimposed on our samples and generally present in radio surveys. We might simply not by detecting weak extended sources. 

Finally, we report one more bias that can be present in our sample and is connected with the process of identification of BALQSOs.
\citet{ganguly07} has performed their own classification of BALQSOs from SDSS DR3 survey and reported that the catalogue of \citet{trump06} 
suffers a 15\% rate of false positives for BI quasars.
However, the method used in \citet{ganguly07} to classify sources as BALQSOs is subjective (by visual inspection)
and therefore it is difficult to estimate its accuracy. 
Additionaly, our analysis is limited to a small sample of radio-loud BI quasars and we argue that the effect of this bias is not significant.

\section{Notes on individual sources}
\label{not}

In this section we briefly describe morphology of BALQSOs based on radio maps obtained from our VLBI observations. 
Moreover, we summarize information from literature considering previous
interferometric observations as well as single dish flux measurements (Table~\ref{basic}). During our observations major axis of beam on L, C and X-band 
were approximately 25, 3, 1.5 mas respectively. 3$\sigma$ level 
was usually 1.1, 0.9, and 0.5 mJy beam$^{-1}$ for 1.7, 5, 8.4 GHz respectively.
The classification of the radio components was feasible due to multi-frequency observations, which provide us with spectral indices.
We assigned different letters to describe components depending on the source structure.
In the first place we used C to indicate component consisting of radio core, or E for eastern, S for southern, W for western and N for northern component. When 
counterparts were further resolved on higher frequency, number was added, e.g C $\rightarrow$ C1, C2 $\rightarrow$ C1.1, C1.2 and C2.1 C2.2.

We classified our sources in 3 categories: Single, Core-jet and Other.
Single means the source is a point-like object unresolved even at the highest observed frequency. 
Core-jet is a source with a bright central component and a one-sided jet. The final category Other consists of 
objects with complex morphologies which are impossible to reconcile within Core-jet or Single division.
Additionally, 4 of our sources were observed by \citet{taka13} in similar fashion hence, dynamic range of radio maps should be comparable.
Moreover, all of our sources have been observed by the Optically ConnecTed Array for VLBI Exploration project (OCTAVE) \cite{doi09}.
However, this project was significantly limited in resolution. Therefore, most sources in OCTAVE survey
remains unresolved.

{\bf 0217-0052}. Source on both 5- and 8.4-GHz maps has a dual structure. Component C with a spectral index $\alpha^{5}_{8.4} $ = 0.32 is 
probably a core. Western component is likely to be a radio-jet. Our 5\,GHz image account only for a $\sim 35\%$ of the total flux density reported
in the single dish observations, what implies the existence of more extended emission resolved out during the observations.
\citet{guai} classified this object as a steep spectrum radio quasar (SSRQ).
We suggest it has a core-jest morphology.

{\bf 0756+3714}. This source was observed with the VLBA at 5\,GHz as part of the VIPS project \citep{helm07}. Our 1.7-GHz EVN observations did not 
resolve the object. The 5-GHz VIPS image and our 8.4-GHz observations show
this source has three components, 
with C2 ($\alpha^{5}_{8.4} $ = 0.68) being the brightest. Previous 5-GHz VLBA observations of this source made by \citet{bruni13}
suggest double morphology, which they interpret as two flat spectrum hot spots in the mini-lobes of a young radio source. Spectral energy
distribution (SED) of 0756+3714 modelled between 408\,MHz and 8.4\,GHz 
peaks at 2.5\,GHz, which suggest a young age. 
On the other hand \citet{zhou06} reported significant flux density changes between NVSS and FIRST 1.4\,GHz observations in this object and classified  
0756+3714 as a candidate for variable source and polar BAL quasar.  
Note that there is also significant discrepancy ($\sim 35\%$) in flux density at 5\,GHz between the VIPS observations and that 
reported by \citet{bruni13} over a time interval a few years, which can be interpreted in the favour of the variability scenario. 

Although it is not straightforward, we suggest core-jet morphology of this source, where the radio core is still hidden in the C2 component.
The brightness temperature calculated for C2 is not high, but close to the equipartition value proposed by \citet{read94} and in the frame of above 
discussion should be treated as a lower limit (Table~\ref{observations}).

{\bf 0815+3305}. The map at 1.7\,GHz shows a structure resolved into two main components, which are separated by $\sim$ 6\,kpc.
This source is the largest one in our whole sample of AI quasars. 
Both, the northern and southern parts consists of 3 components and are elongated in the direction which suggest there
might be an extended bridge of emission between them.
At 5 and 8.4-GHz only the more compact southern part is
detected with components C1.2 or C1.3 being probably a radio core.
Indeed, the EVN and VLBA observations of 0815+3305 can account only for $\sim 43\%$ and $\sim 20\%$ of the total flux at 1.7 and 5\,GHz
respectively indicating the existence of additional structure on angular scales not sampled by the EVN/VLBA.
The overall spectrum of 0815+3305 is steep and does not present a peak in the gigahertz range. The classification of this source is 
difficult and thus we marked it as 'Other' in Table~\ref{observations}. 

{\bf 0928+4446}. This source has been observed with the VLBA at 5\,GHz as part of the VIPS project \citep{helm07} and we present this image in Fig.~\ref{images1}. 
Our 1.7- and 8.4-GHz observations are consistent with those observations showing flat spectrum central component C being a radio core and steep
spectrum one-sided jet E. The same results were obtained by VLBA observations made by \citet{taka13}. 
We report however, $\sim 50\%$ and $\sim 23\%$ increase of the 1.7- and 8.4-GHz flux density, respectively 
in our observations compared to single dish measurements (Table~\ref{basic}). 
This could be connected with the nature of 0928+4446. This source has been classified as flat spectrum radio quasar (FSRQ) by \citet{hewe10} 
and indicated as a good candidate for variable, blazar-type object by \citet{taka13}.
We classified the morphology of 0928+4446 as a core-jet.

{\bf 1005+4805}. A point-like object on the 1.7-GHz image has been resolved into a more complex structure in 
the 5- and 8.4-GHz maps. The brightest component C1 is probably a radio core and the steep spectrum components
C2 and S are parts of the one-sided radio jet. The overall spectrum of 1005+4805 is steep and does not present a peak in the gigahertz range.
The large fraction of the total flux density, $\sim 85\%$ at 1.7\,GHz and $\sim 49\%$ at 5\,GHz, 
is lost in our observations implying the existence of extended emission not sampled
by EVN/VLBA spacings. Structure and spectral indices of this object suggest core-jet morphology.

{\bf 1013+4918}. The radio structure of 1013+4918 visible in our 1.7- and 8.4-GHz observations is consistent with that obtained in the 
VIPS project \citep{helm07} and we present all three images in Fig.~\ref{images1}.
The inverted-spectrum C1 component is a radio core and three northern steep spectrum components (N1, N2, N3) are parts of the one-sided radio jet.
The overall spectrum of 1013+4918 is steep. Its morphology can be classified as core-jet.

{\bf 1018+0530}. This source was observed only at 5 and 8.4 GHz with VLBA (Fig.~\ref{images2}). 
The same results have been obtained in VLBA observations made by \citet{taka13}.
The brightest component C1 is a radio core and steep spectrum C2 is probably a radio jet. The angular size of 1018+0530 indicates it is one of the
most compact objects in our sample of AI sources.
1018+0530 has been classified as flat spectrum radio quasar (FSRQ) by \citet{hewe10}.
The large flux variations observed in this quasar \citep{Gorshkov2008} make 
it a good candidate for variable, blazar-type object. It has also a FERMI-LAT detection \citep{nolan12}.
We have classified this source as a core-jet.

\setcounter{figure}{0}
\begin{figure*}
\centering

      \includegraphics[width=0.32\textwidth, height=0.23\textheight]{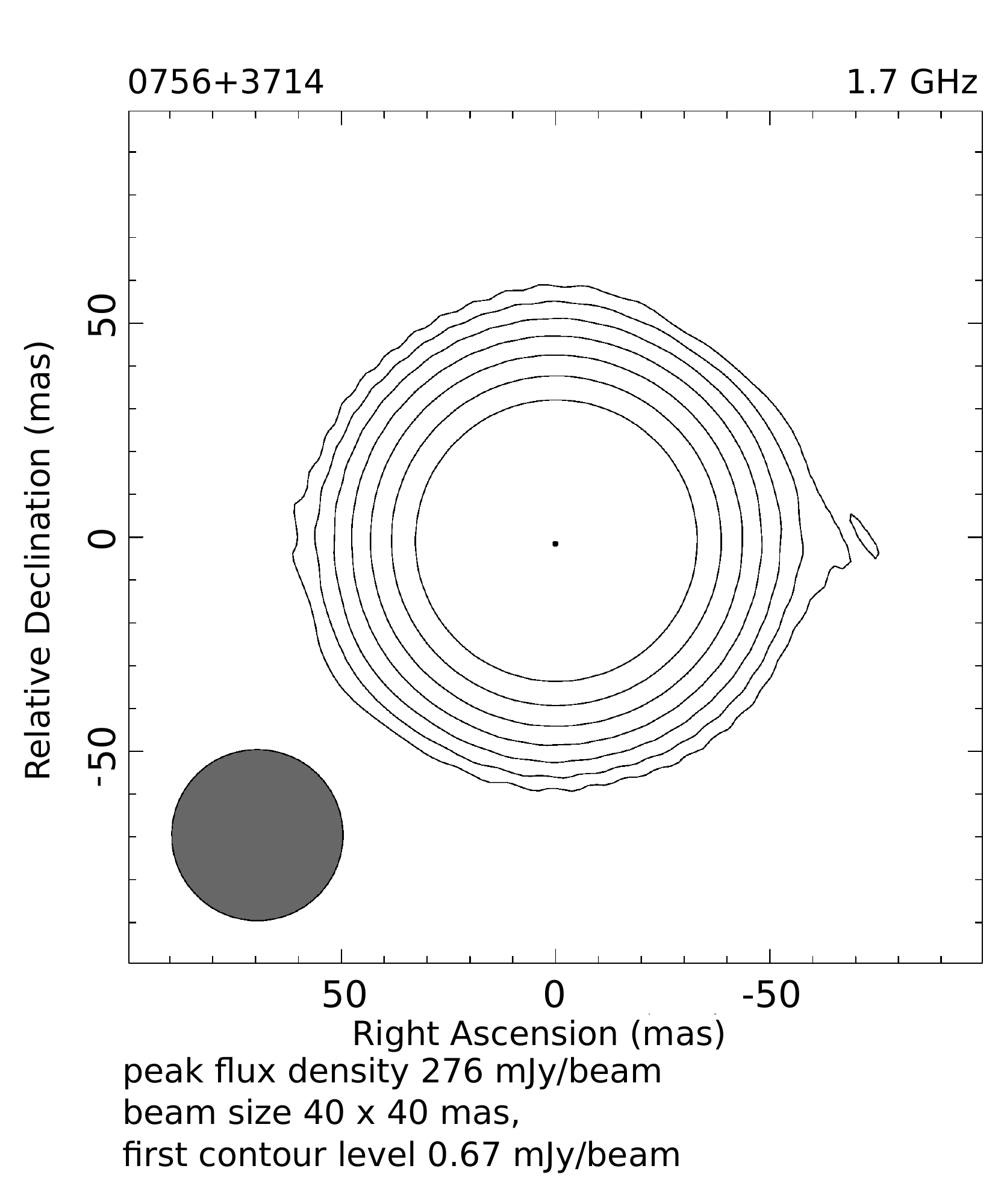}
       \includegraphics[width=0.32\textwidth, height=0.23\textheight]{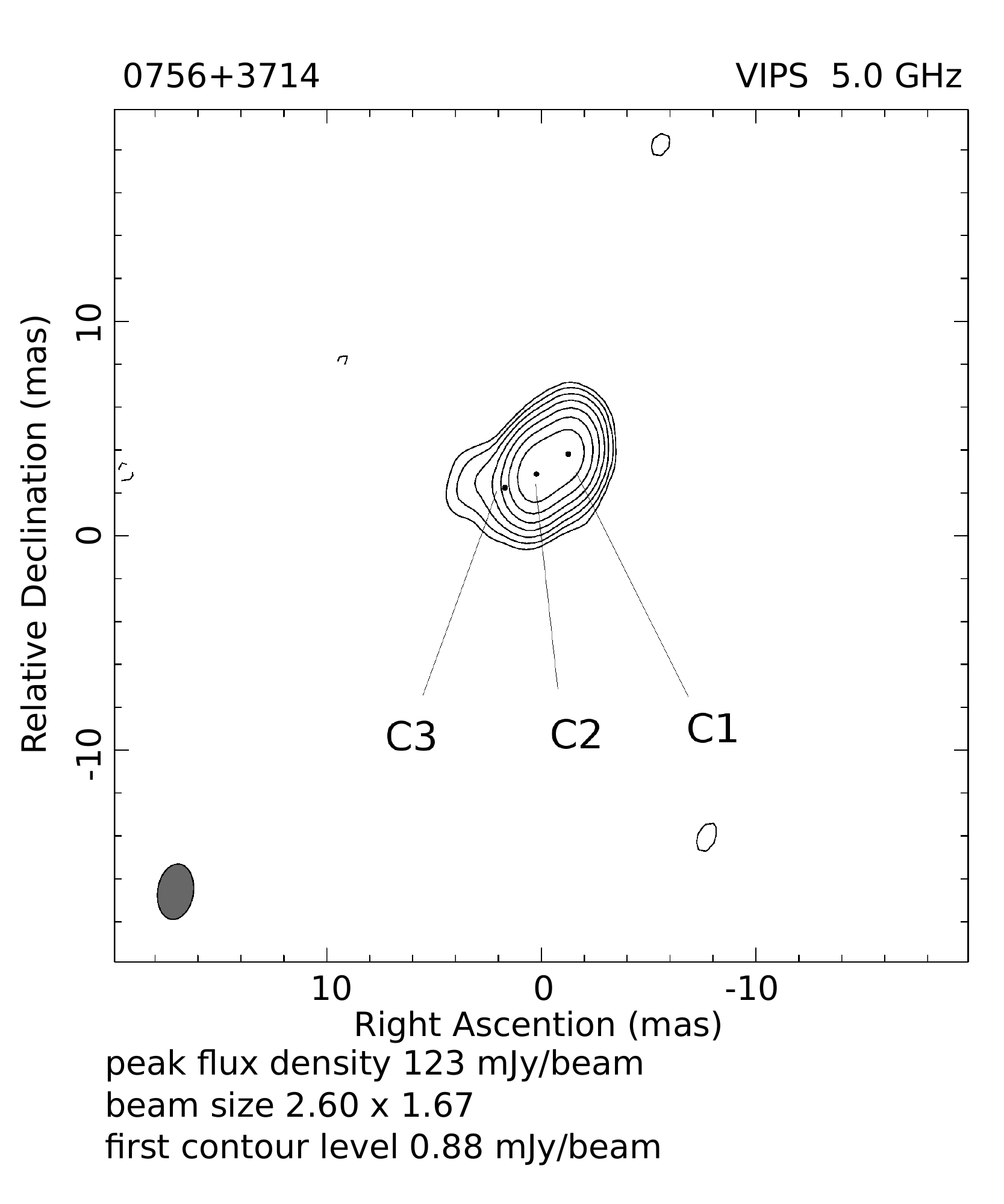}
       \includegraphics[width=0.32\textwidth, height=0.23\textheight]{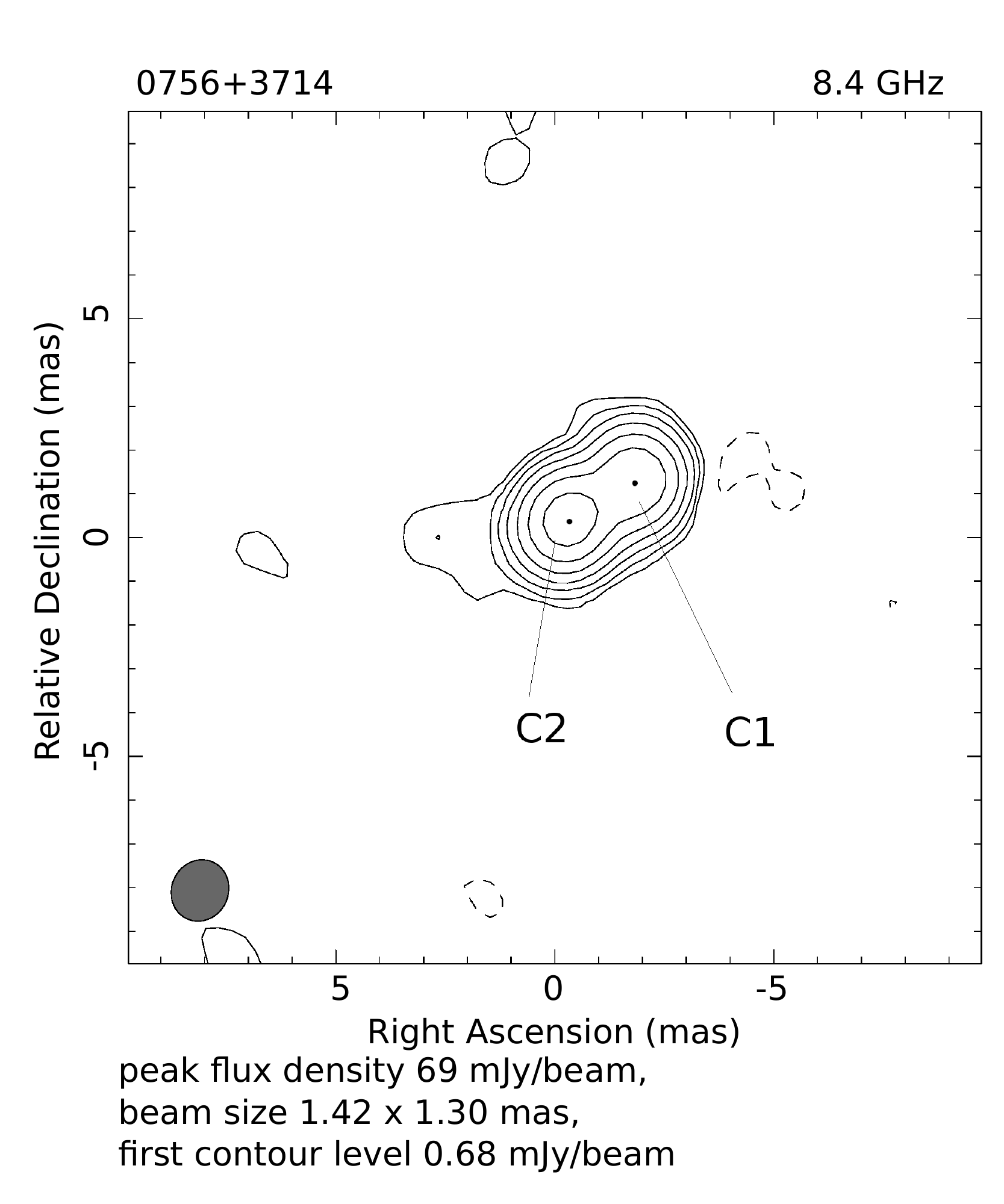}

     \includegraphics[width=0.32\textwidth, height=0.23\textheight]{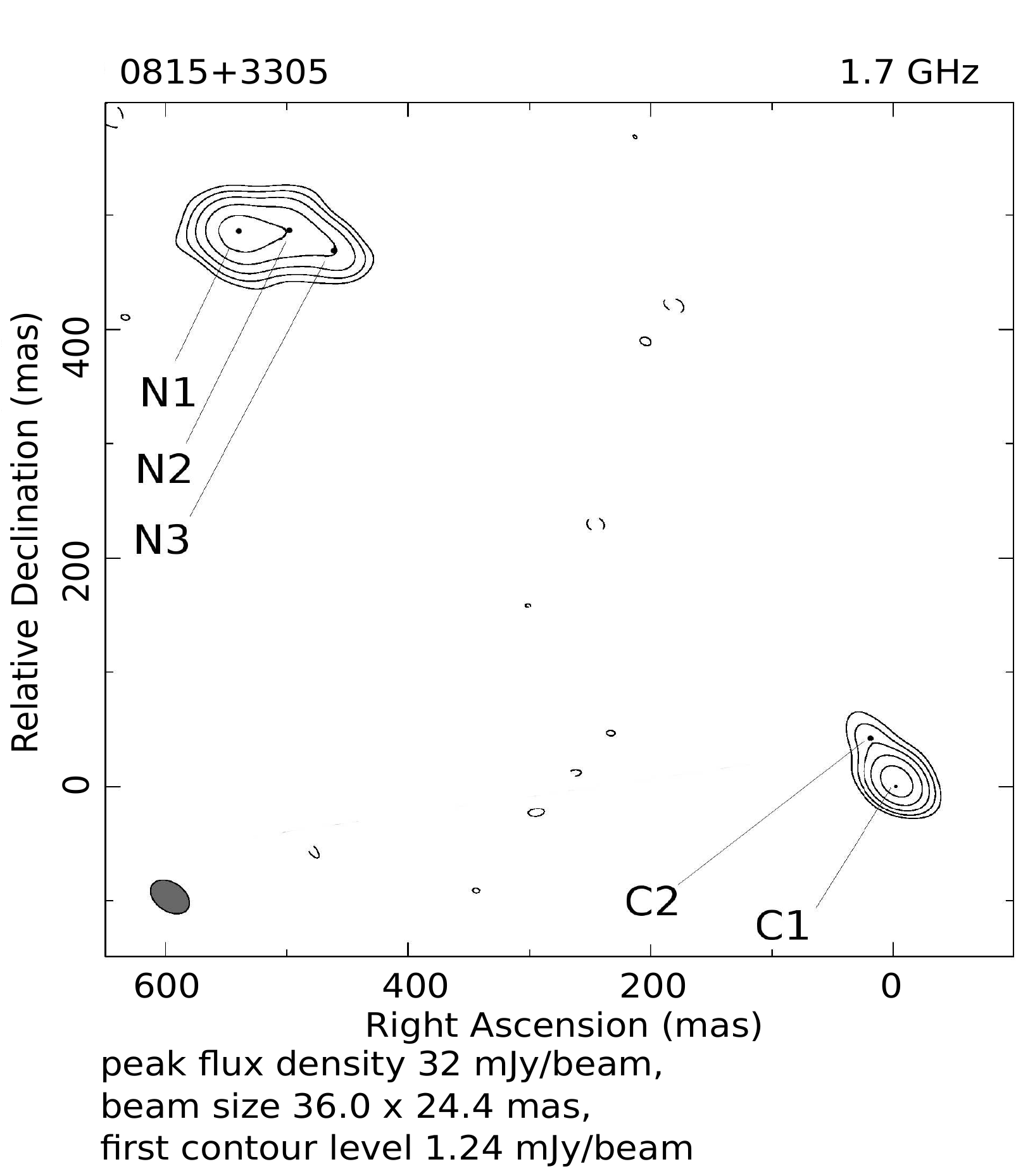}
     \includegraphics[width=0.32\textwidth, height=0.23\textheight]{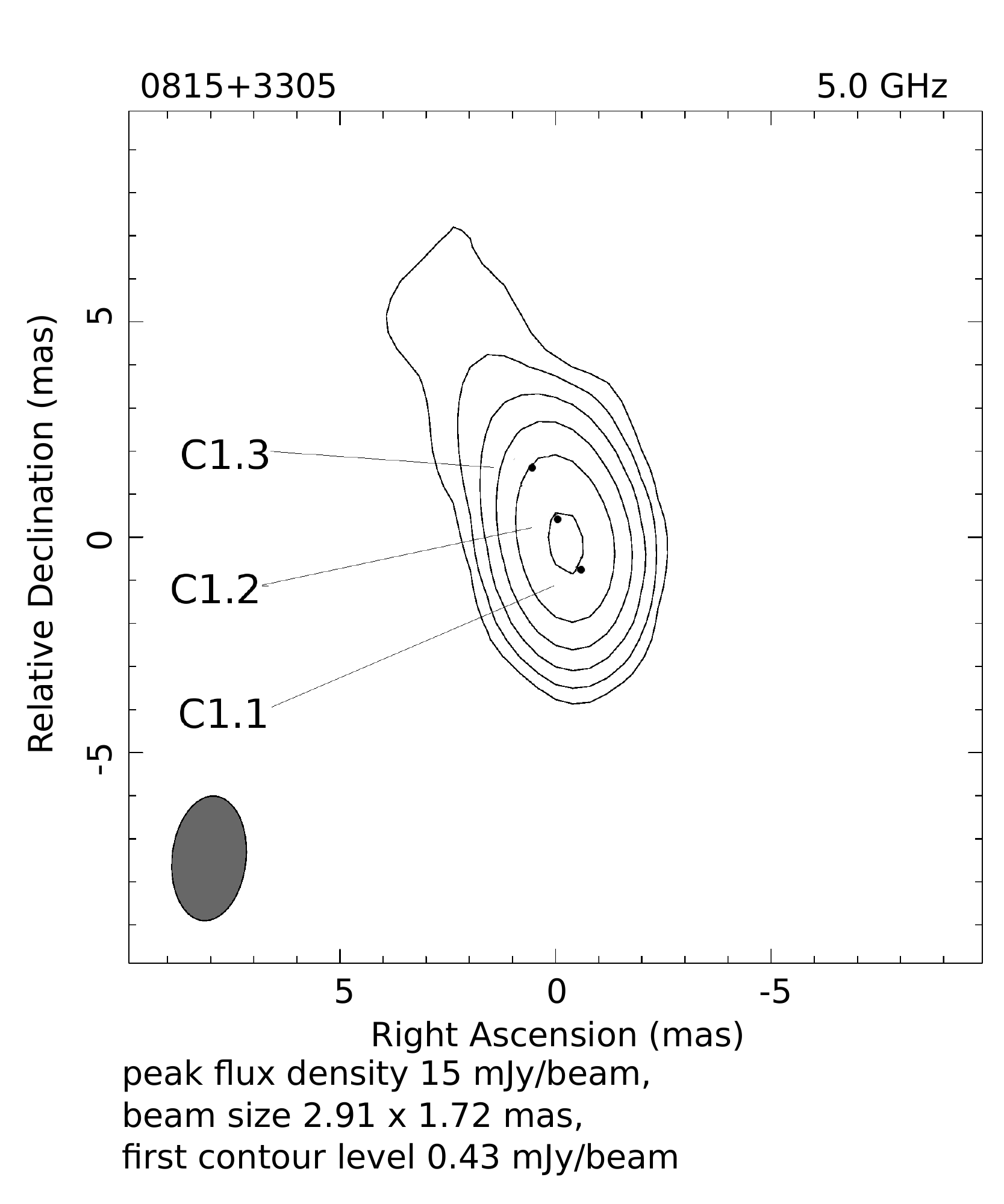}
     \includegraphics[width=0.32\textwidth, height=0.23\textheight]{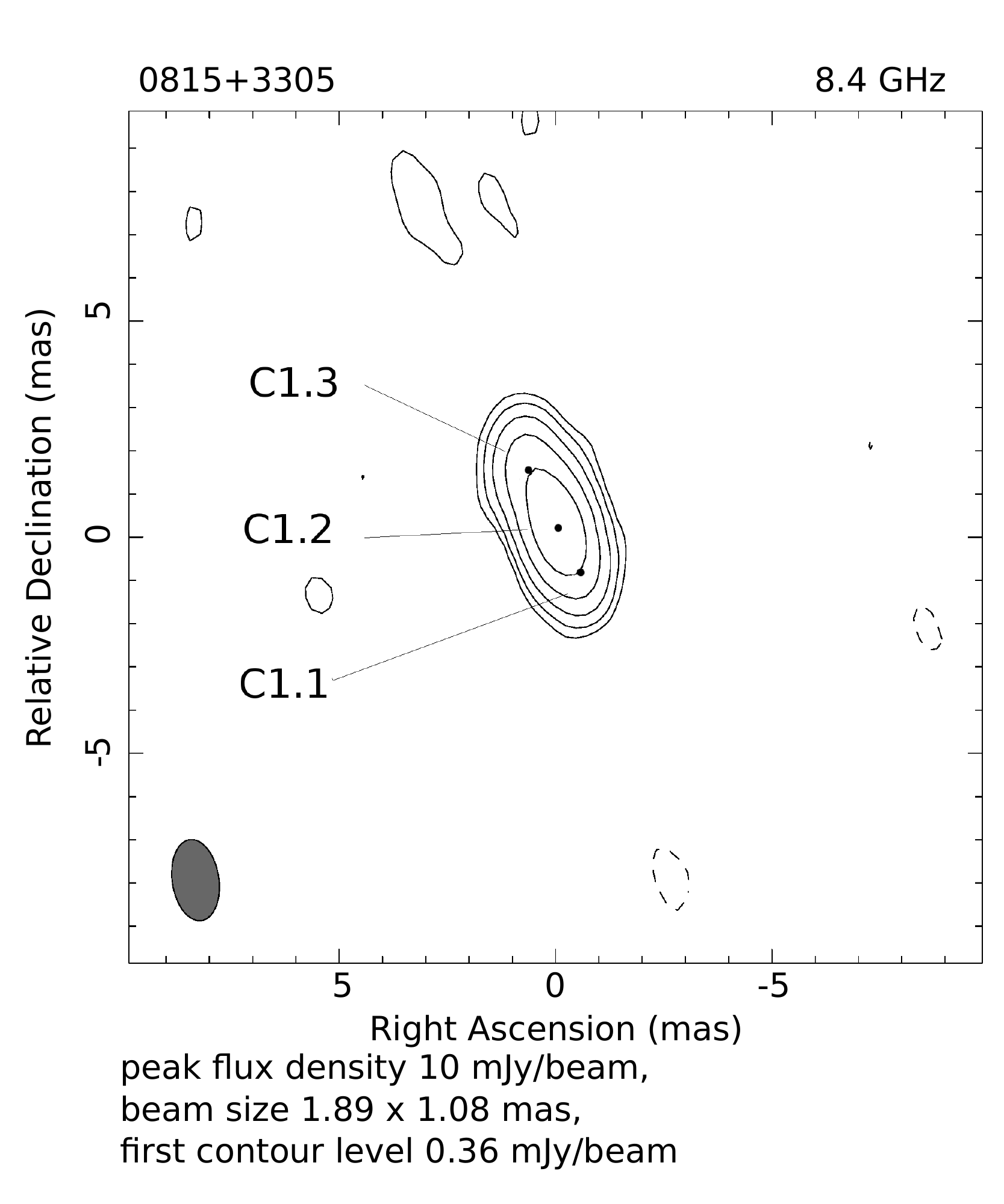}

    \includegraphics[width=0.32\textwidth, height=0.23\textheight]{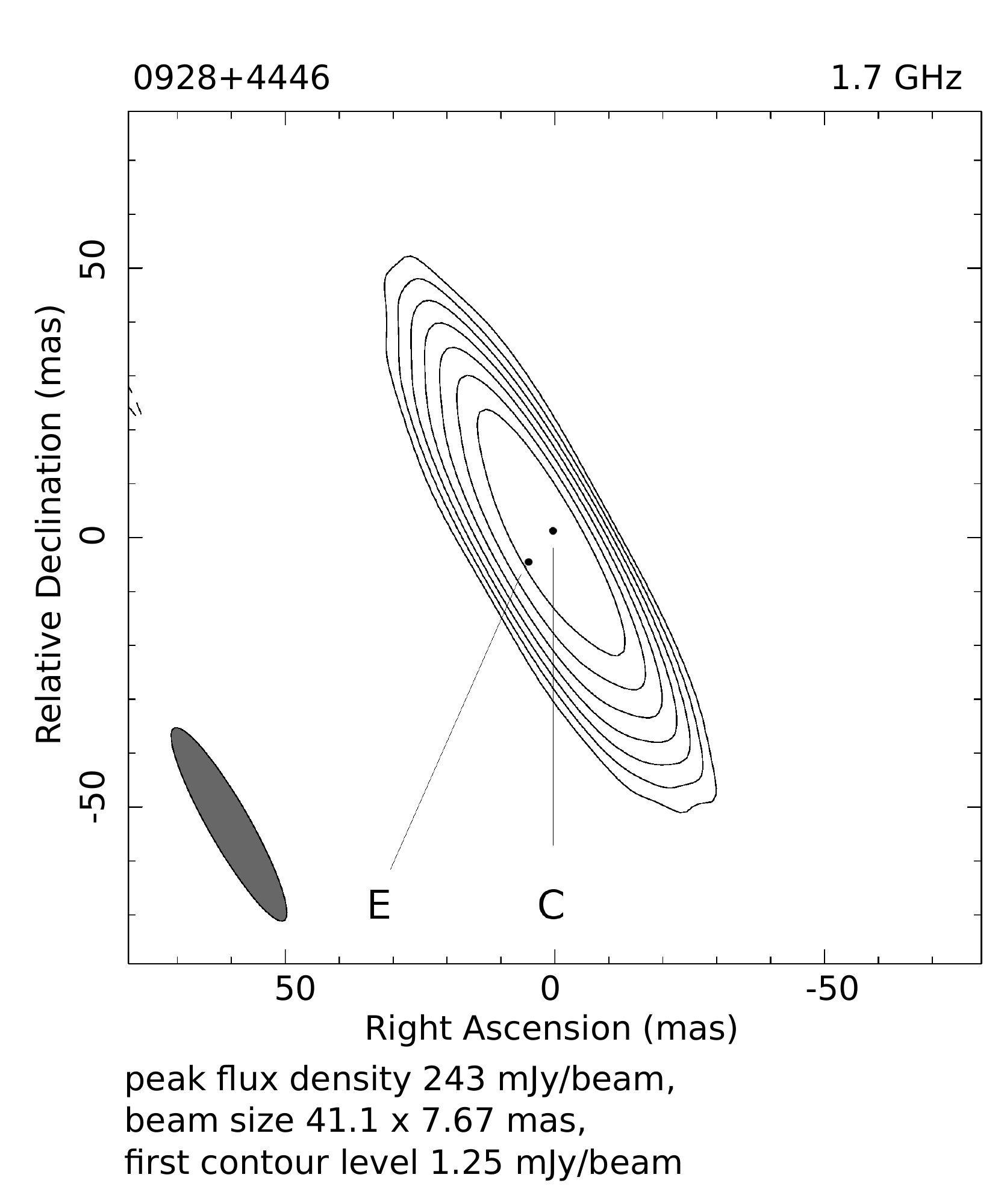}
     \includegraphics[width=0.32\textwidth, height=0.23\textheight]{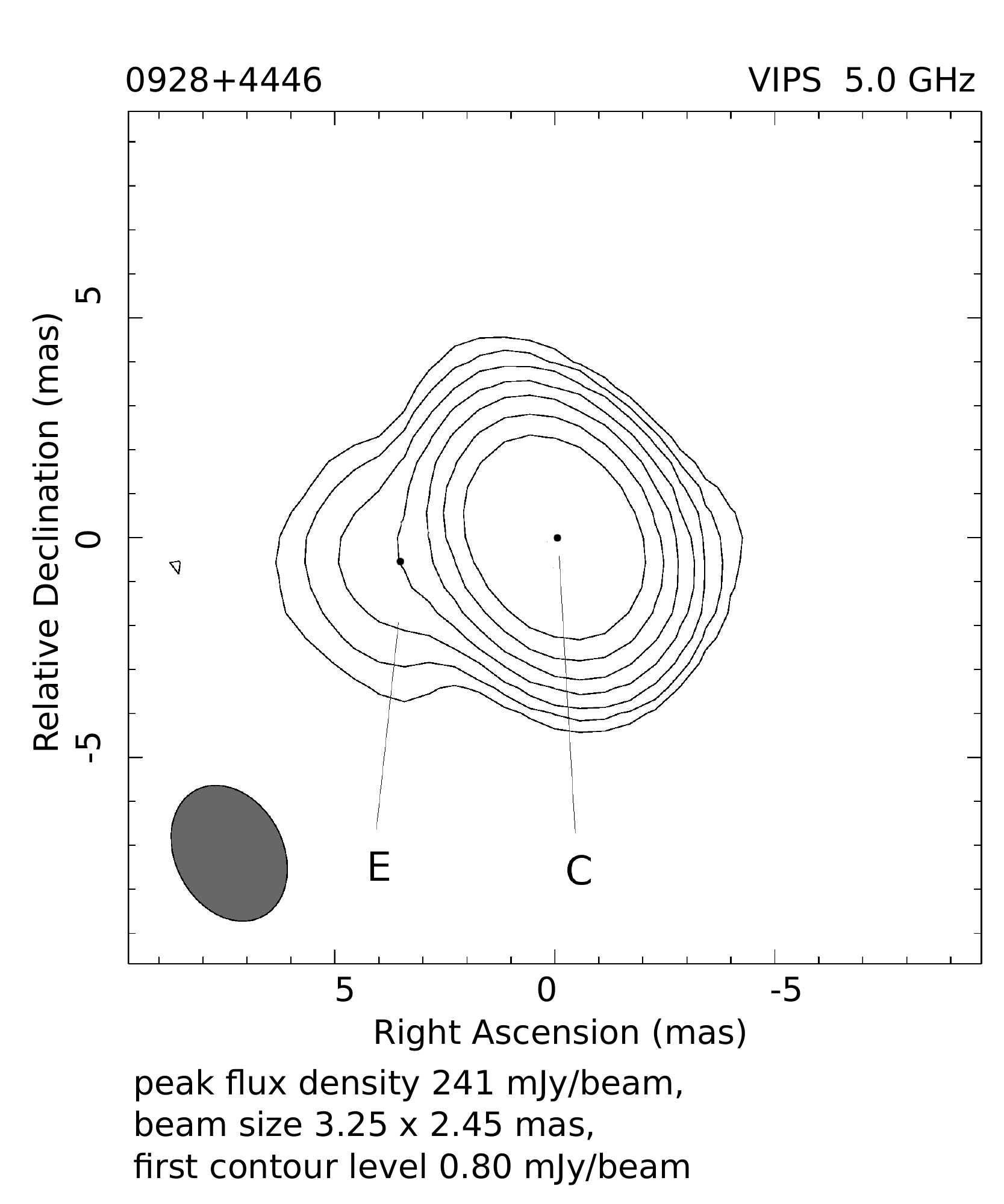}
     \includegraphics[width=0.32\textwidth, height=0.23\textheight]{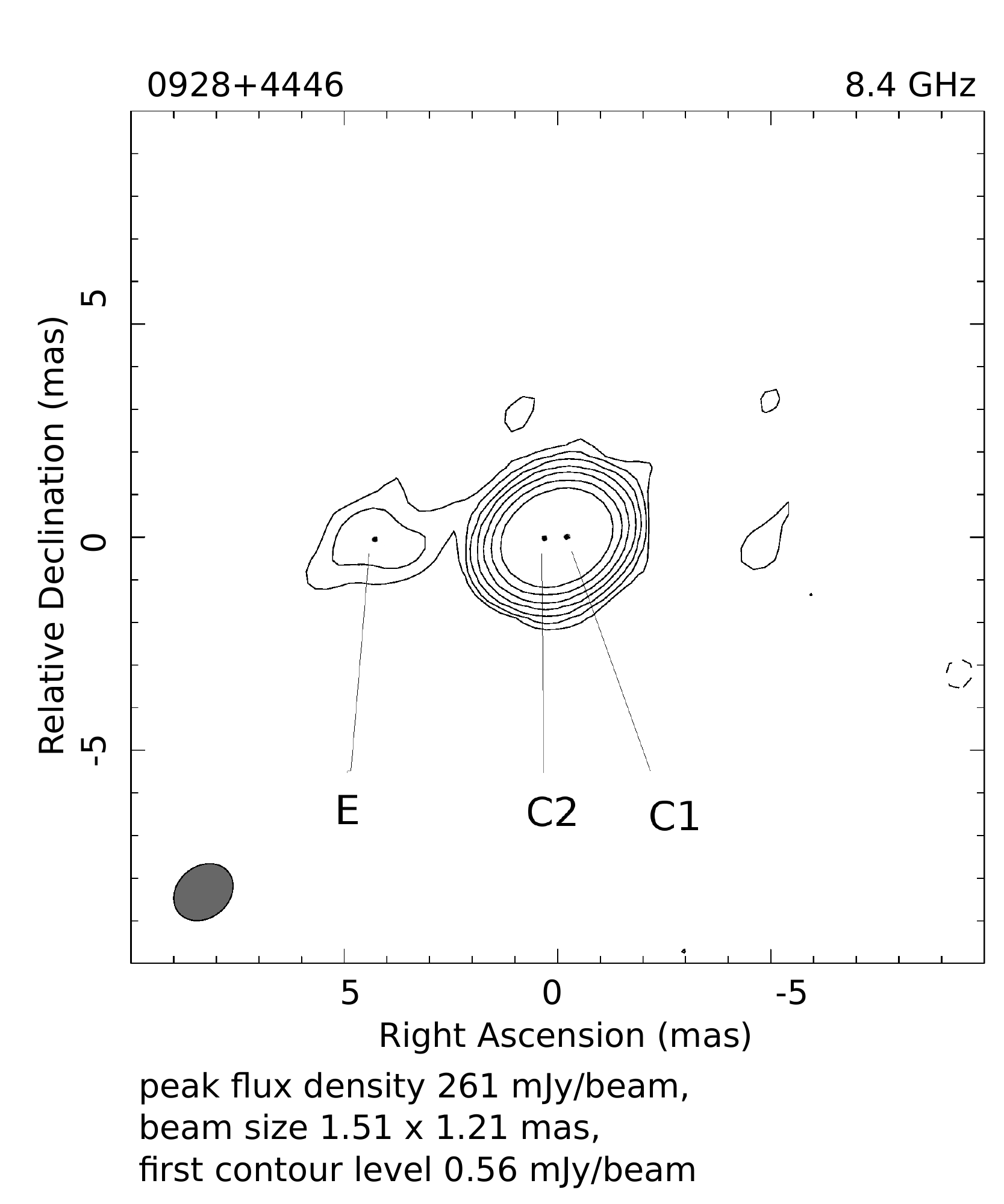}

       \includegraphics[width=0.32\textwidth, height=0.23\textheight]{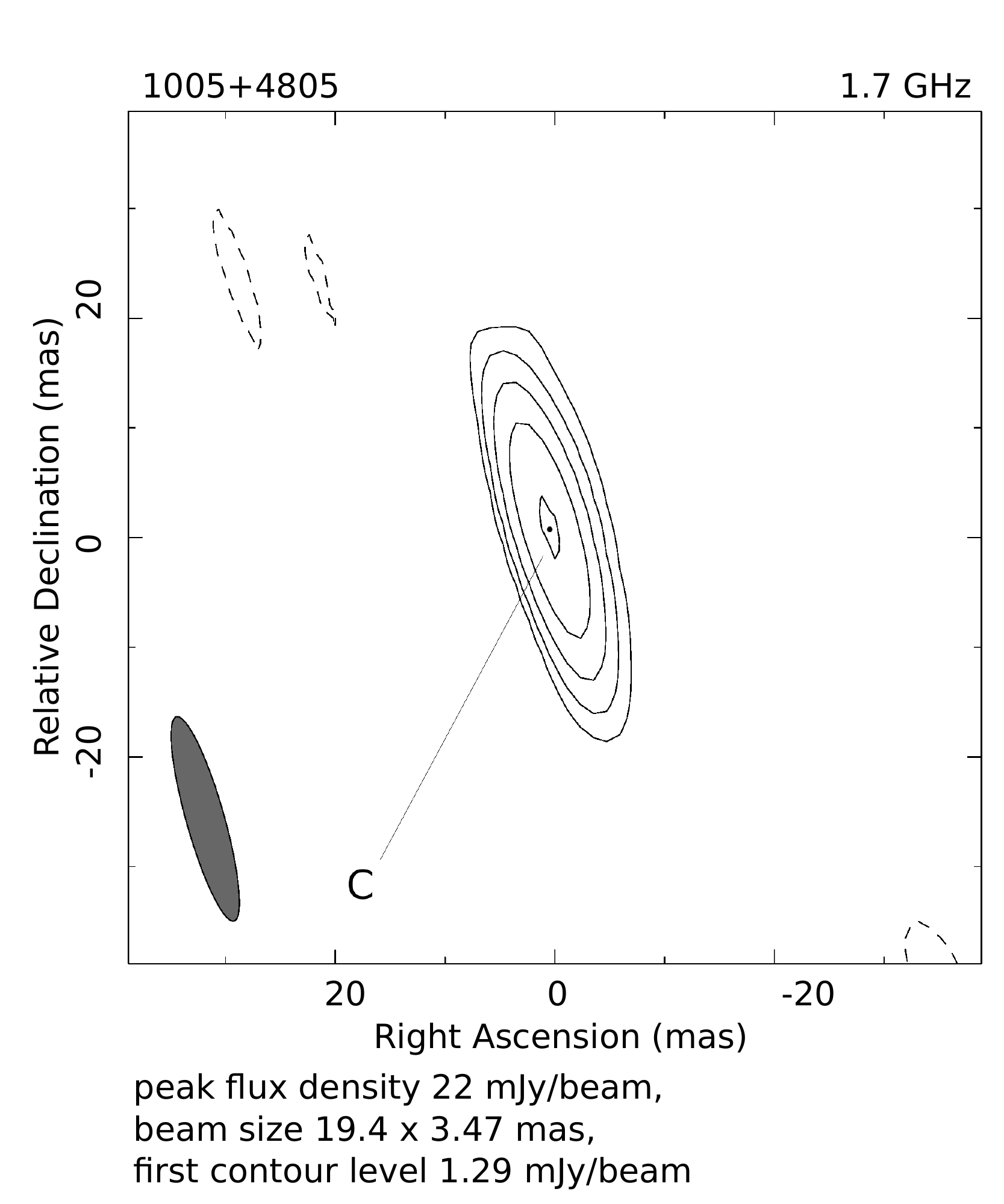}
      \includegraphics[width=0.32\textwidth, height=0.23\textheight]{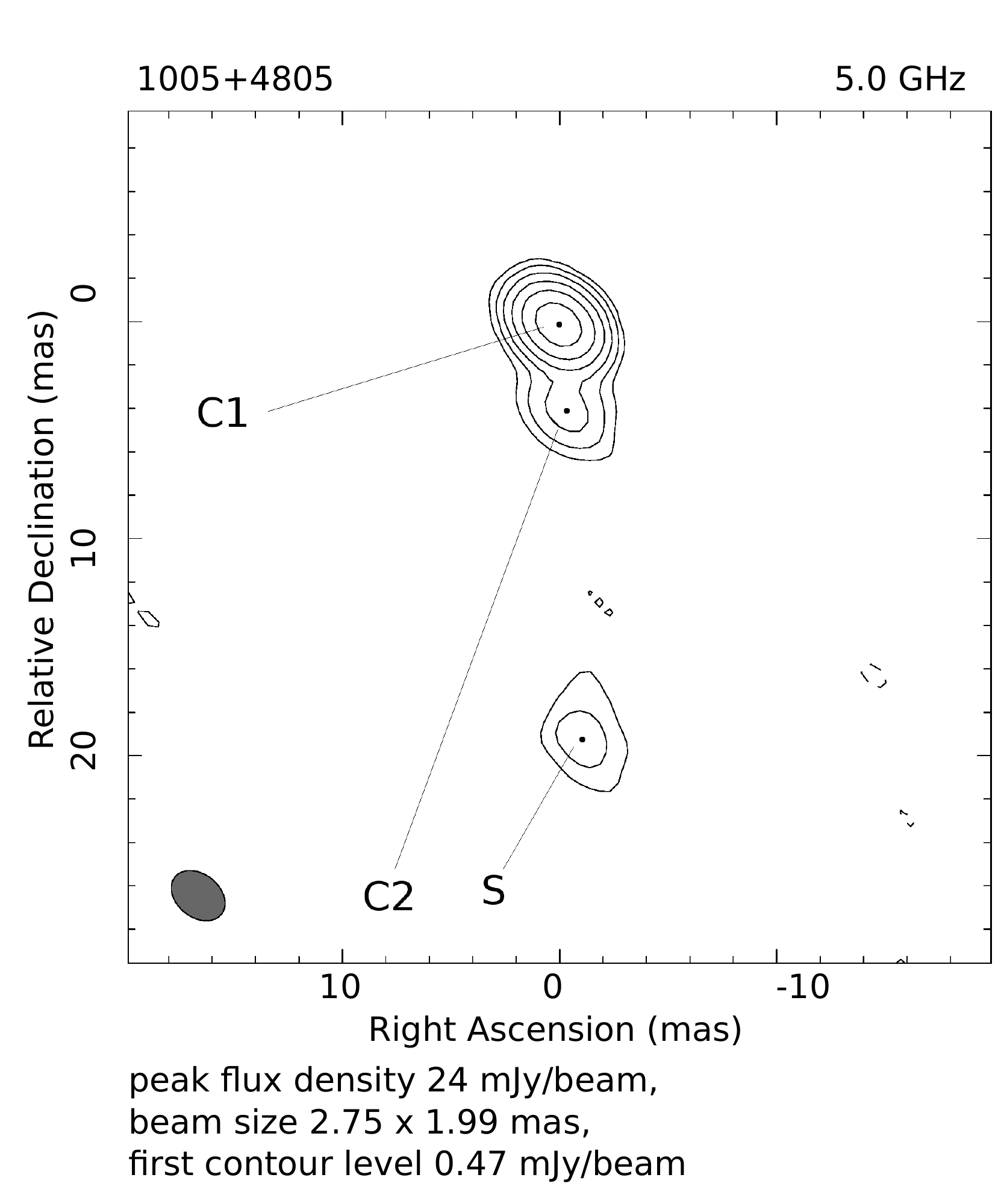}
      \includegraphics[width=0.32\textwidth, height=0.23\textheight]{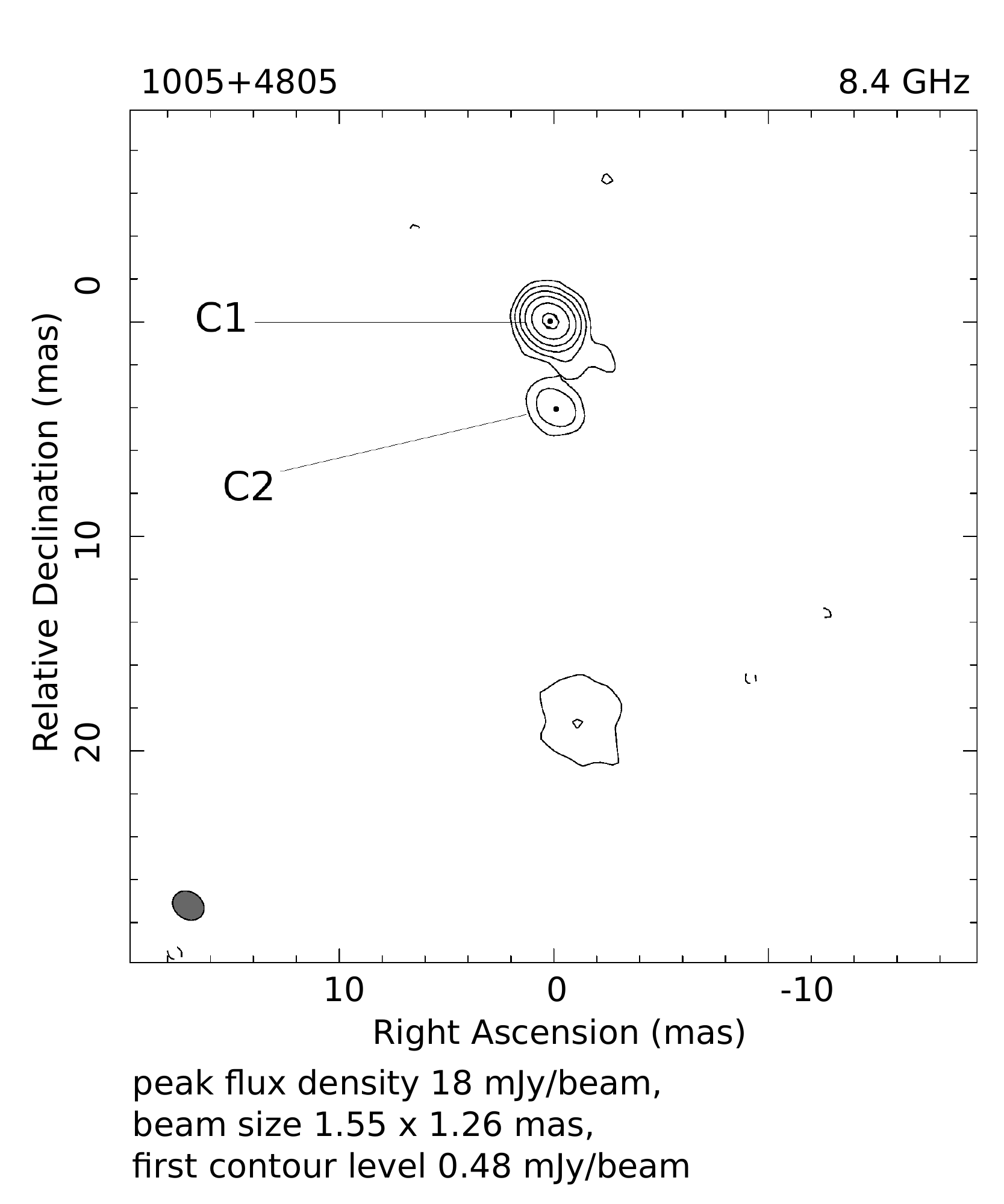}

   \caption{Sources with observations at all three frequencies: EVN 1.7\,GHz and VLBA 5 and 8.4\,GHz. 
   Contours increase by a factor 2, and the first contour level corresponds to $\approx 3\sigma$.}
\label{images1}
\end{figure*}

\setcounter{figure}{0}
\begin{figure*}
 
     \includegraphics[width=0.32\textwidth, height=0.23\textheight]{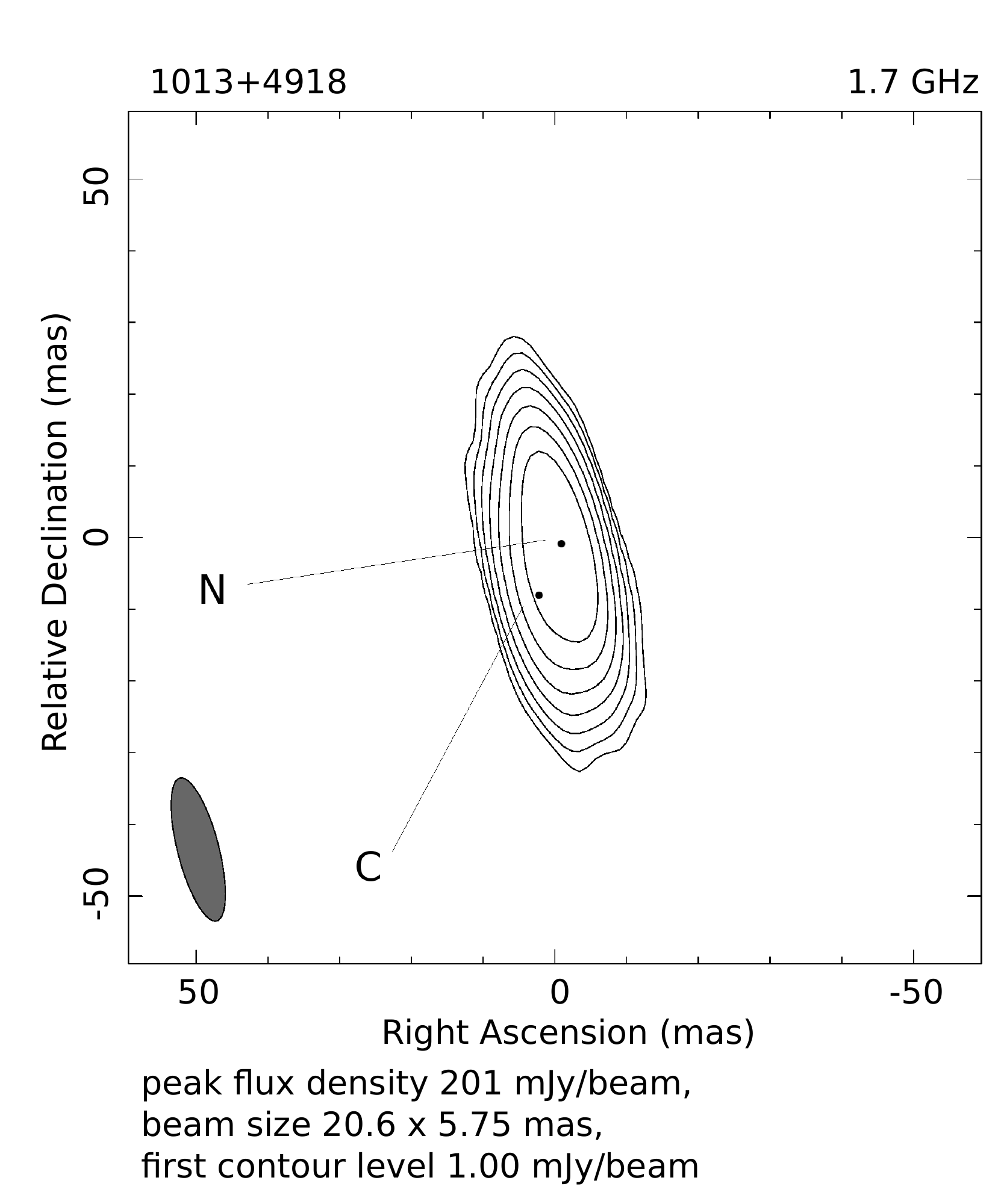}
     \includegraphics[width=0.32\textwidth, height=0.23\textheight]{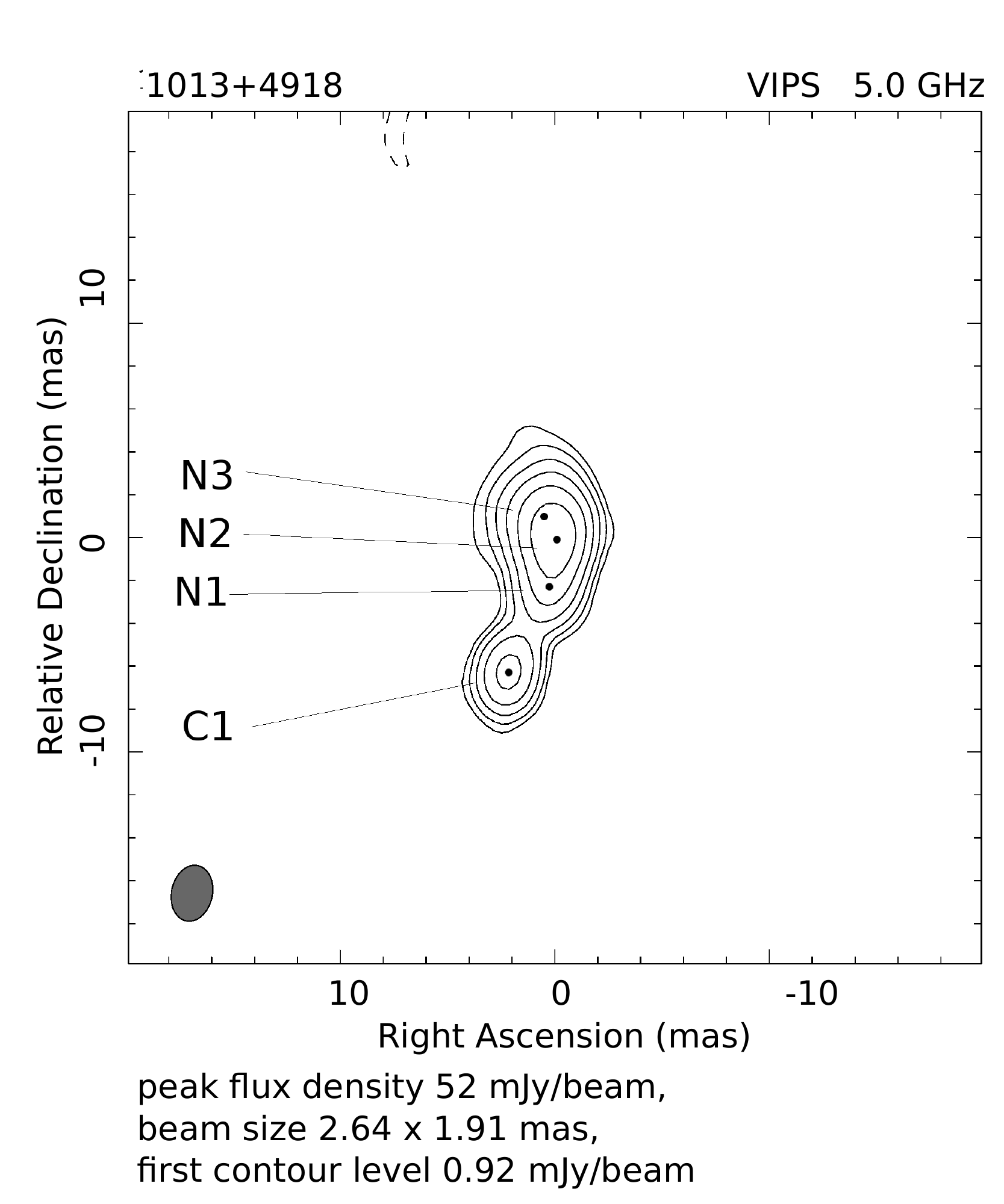}
   \includegraphics[width=0.32\textwidth, height=0.23\textheight]{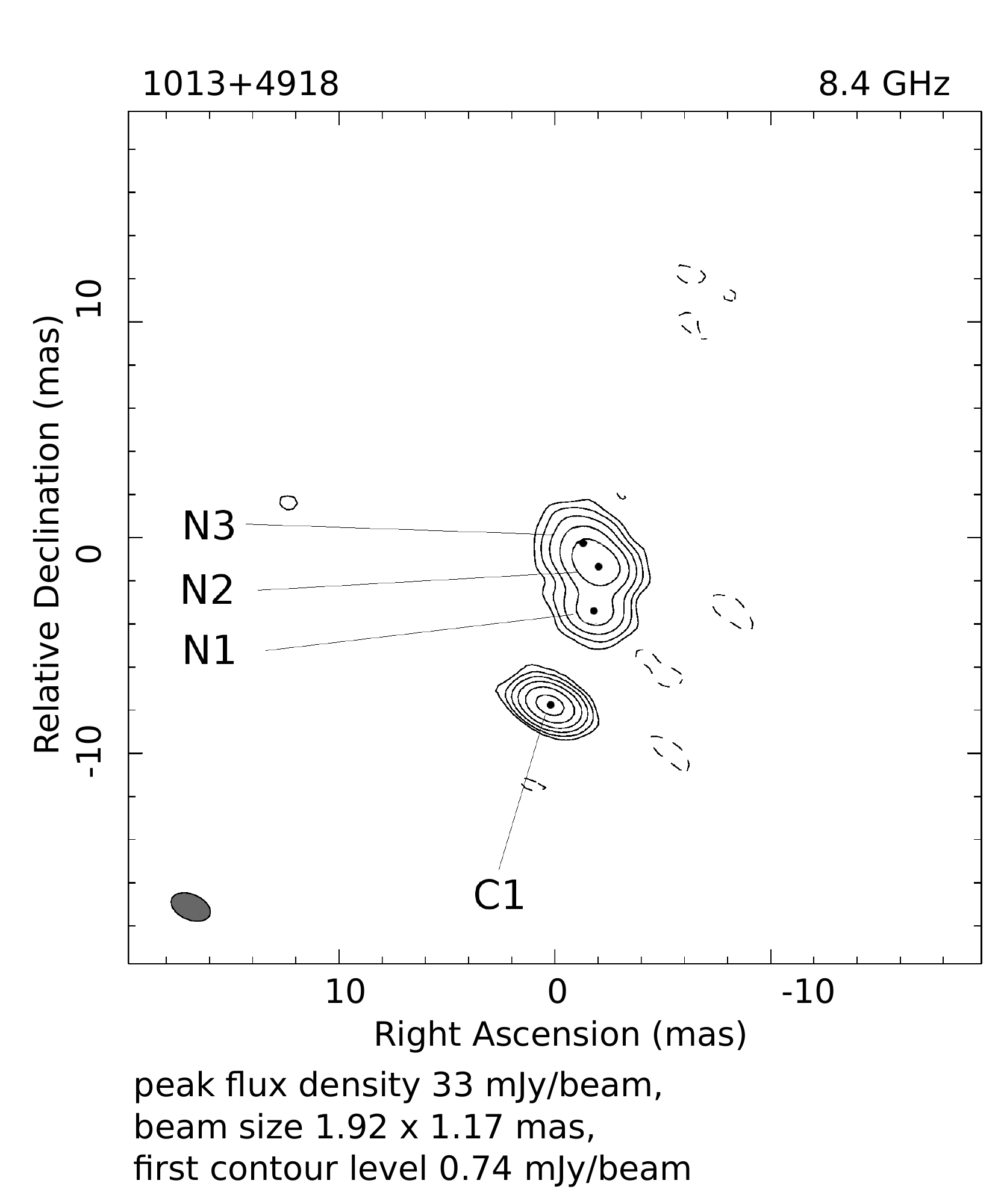}

    \includegraphics[width=0.32\textwidth, height=0.23\textheight]{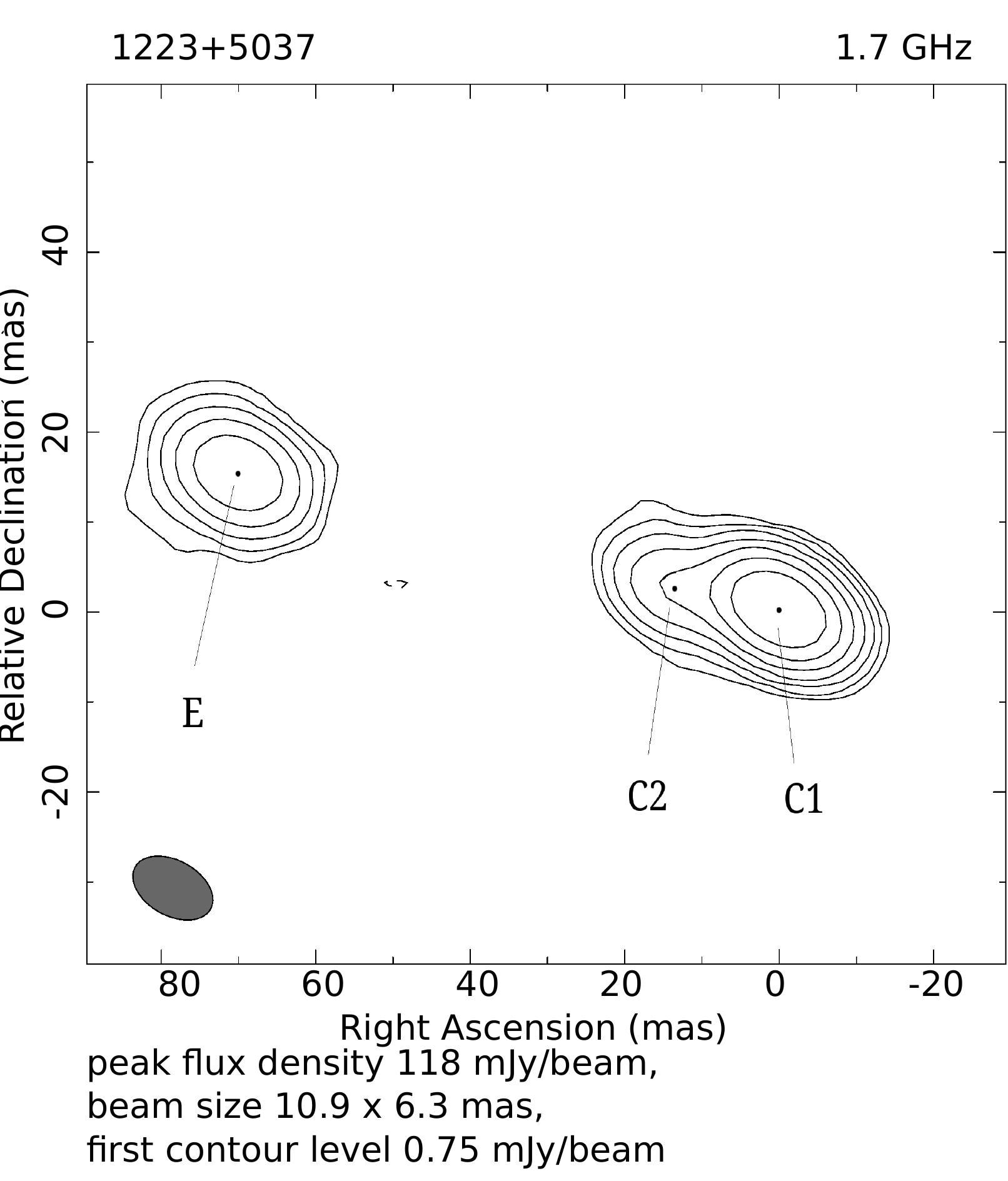}
    \includegraphics[width=0.32\textwidth, height=0.23\textheight]{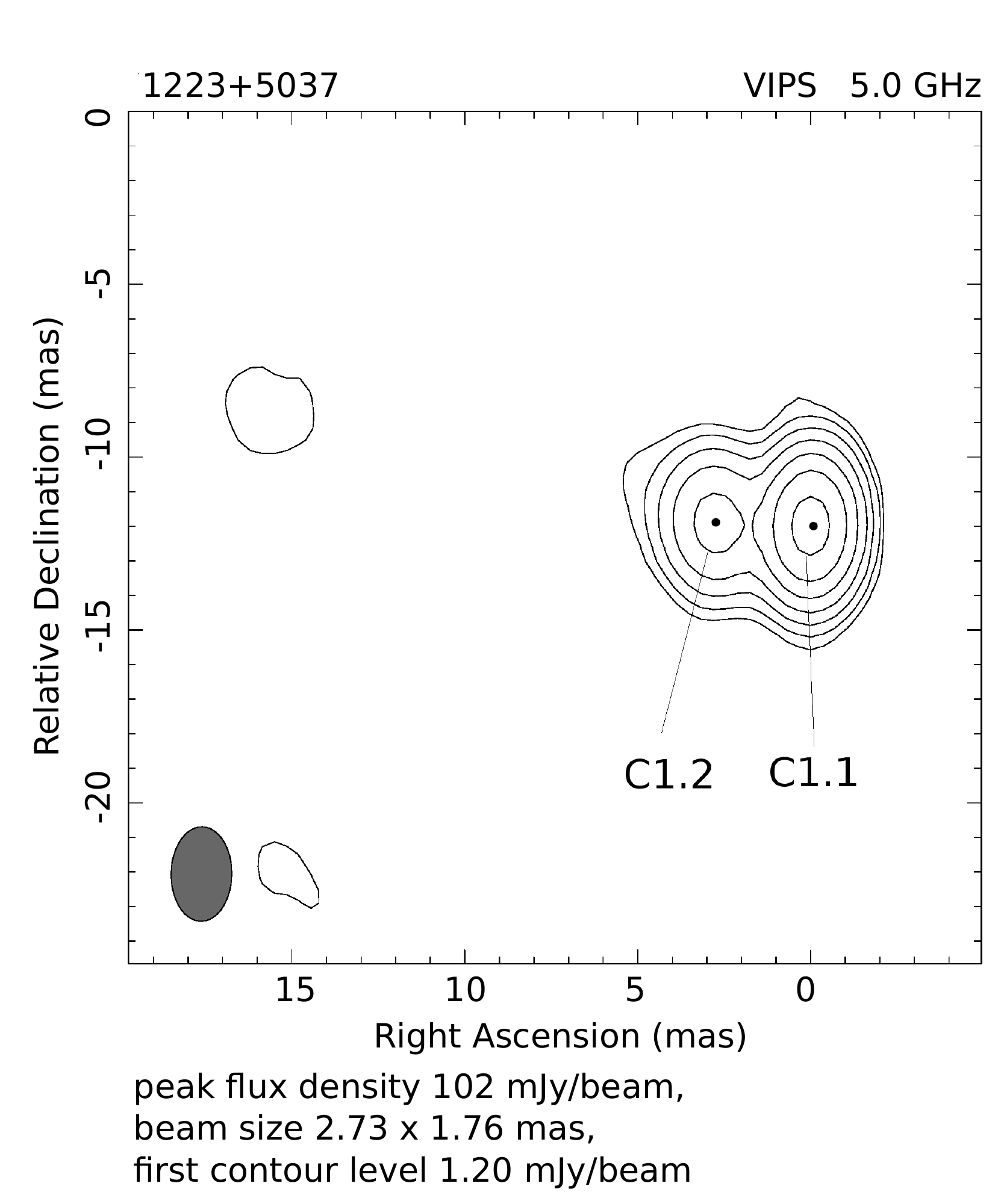}
      \includegraphics[width=0.32\textwidth, height=0.23\textheight]{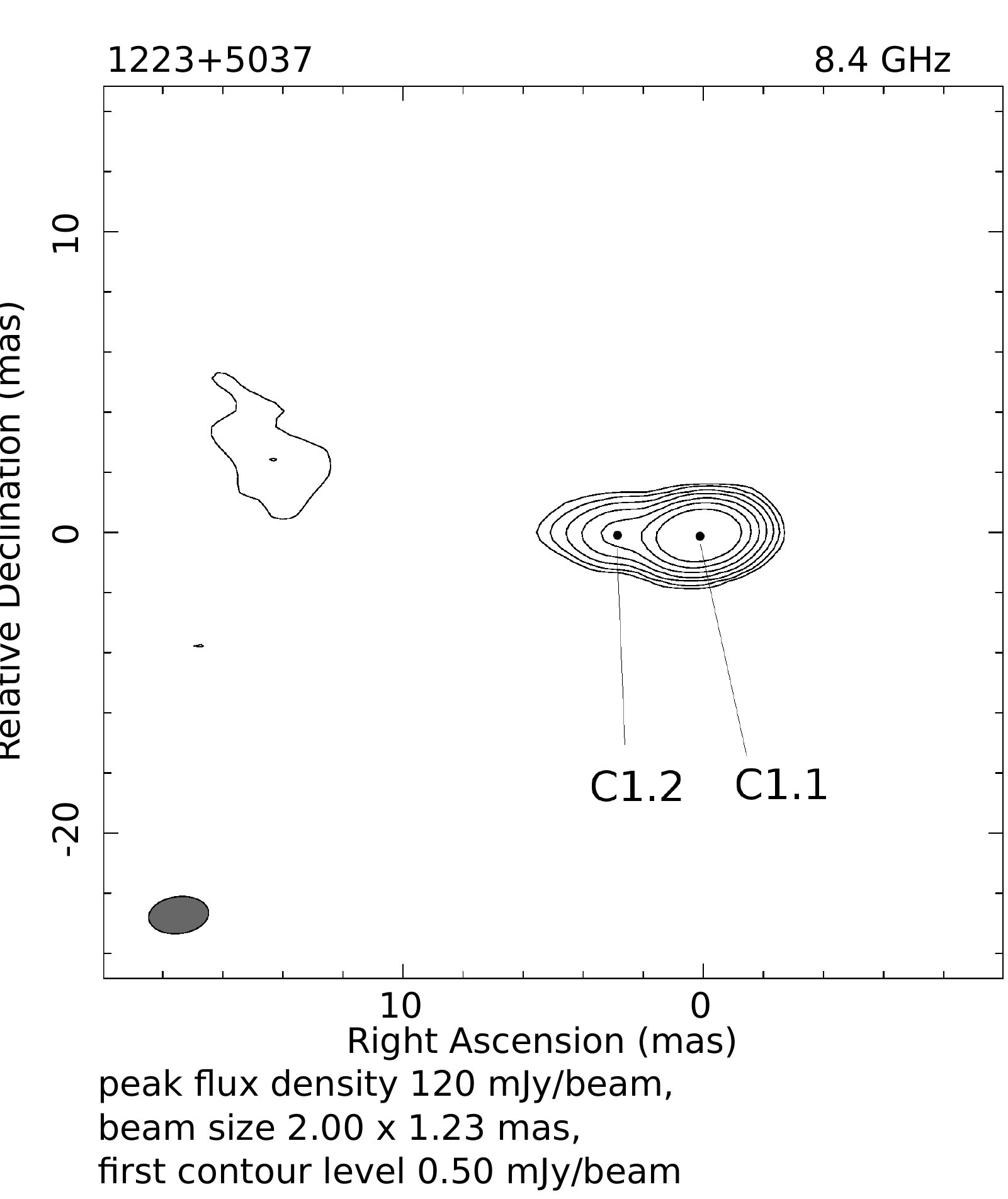}
     
    \includegraphics[width=0.32\textwidth, height=0.23\textheight]{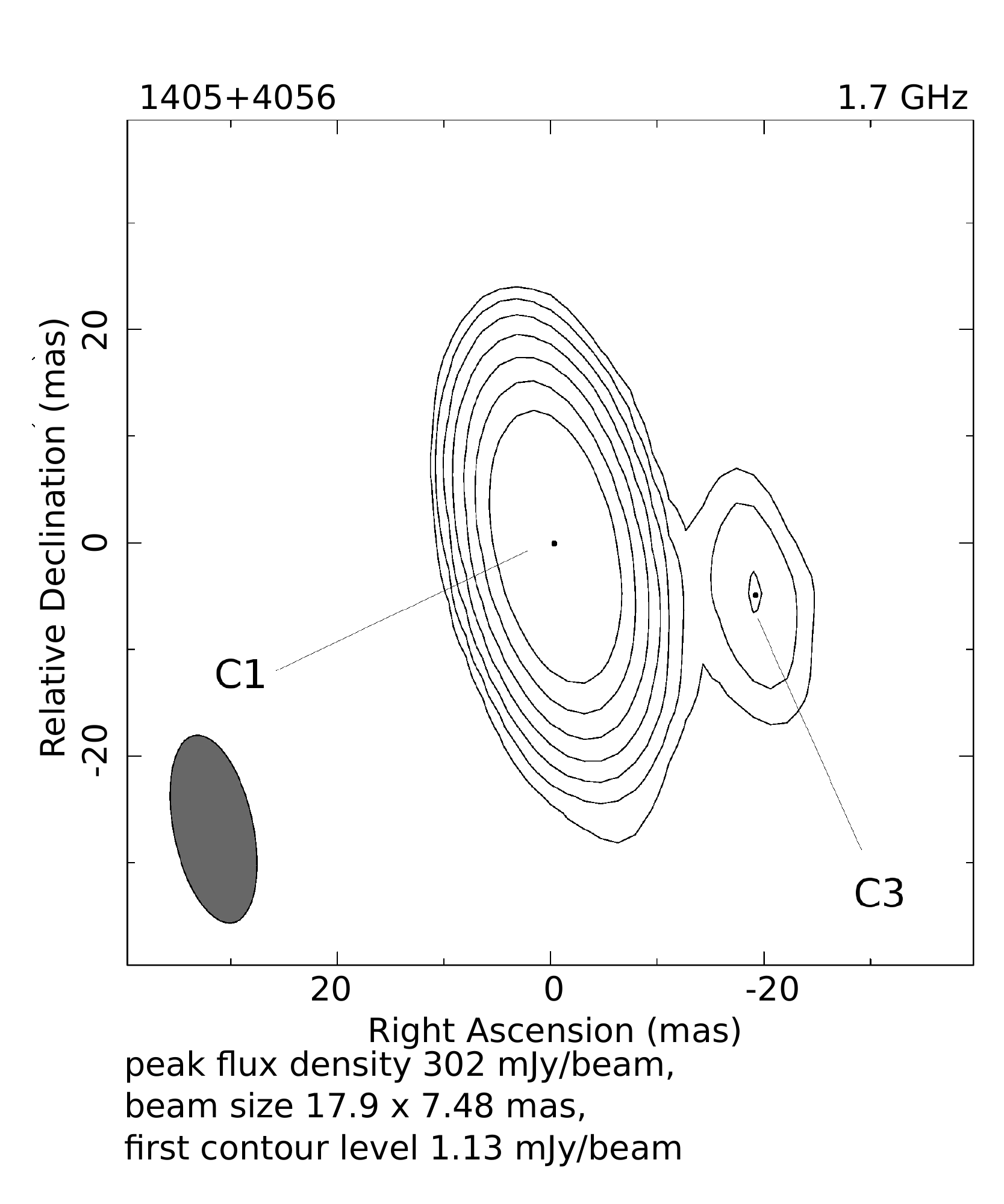}
    \includegraphics[width=0.32\textwidth, height=0.23\textheight]{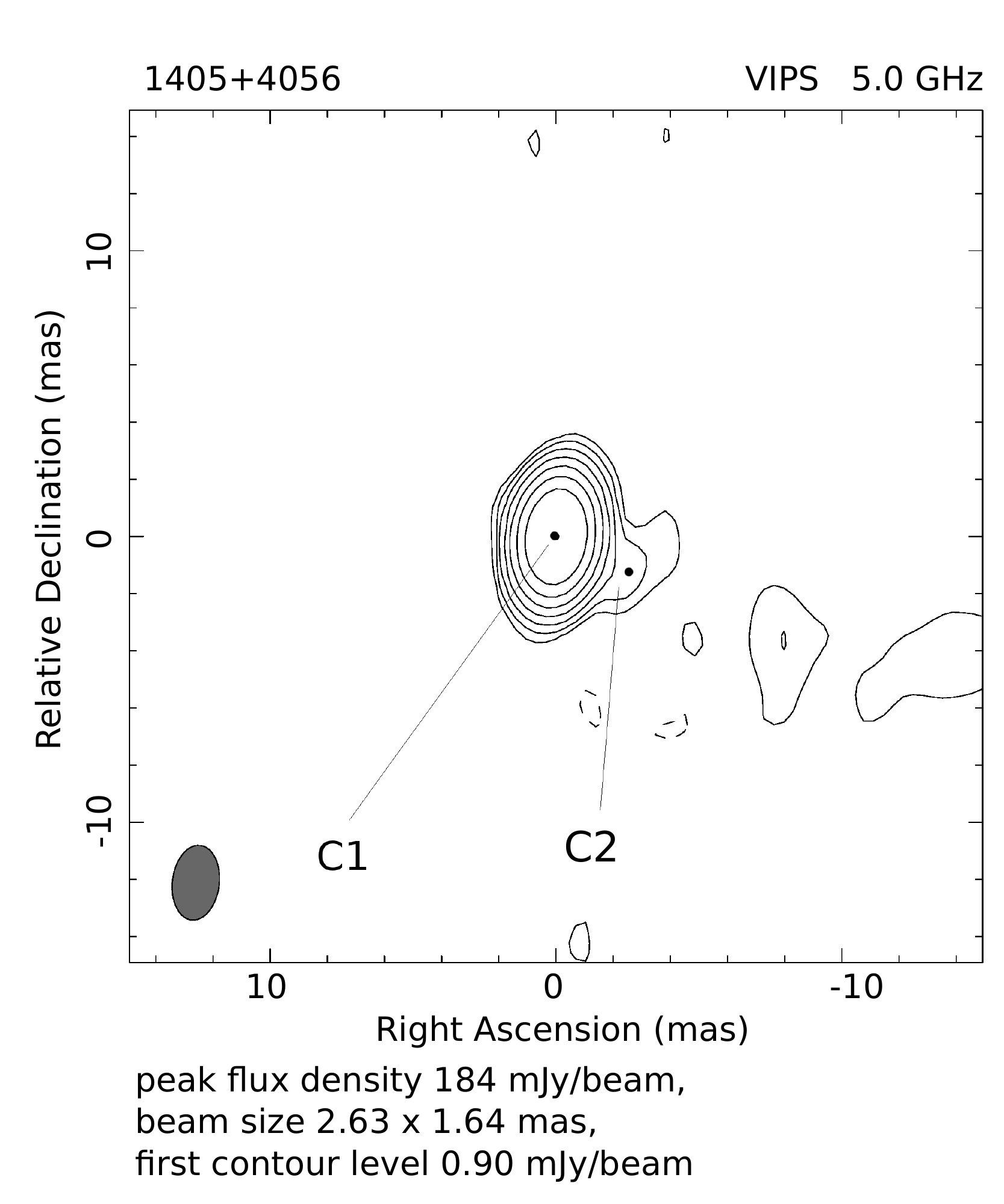}
      \includegraphics[width=0.32\textwidth, height=0.23\textheight]{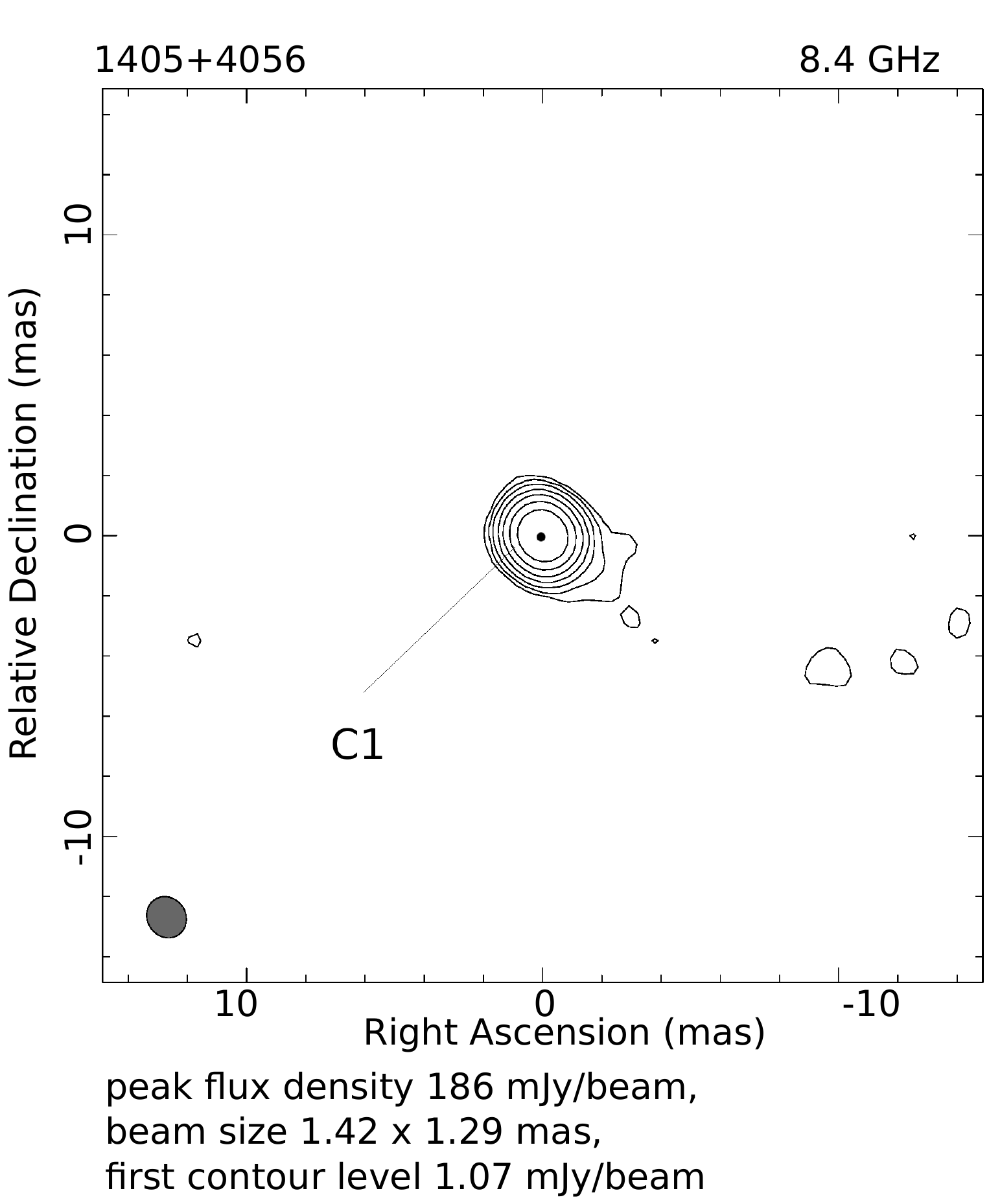}

   \includegraphics[width=0.32\textwidth, height=0.23\textheight]{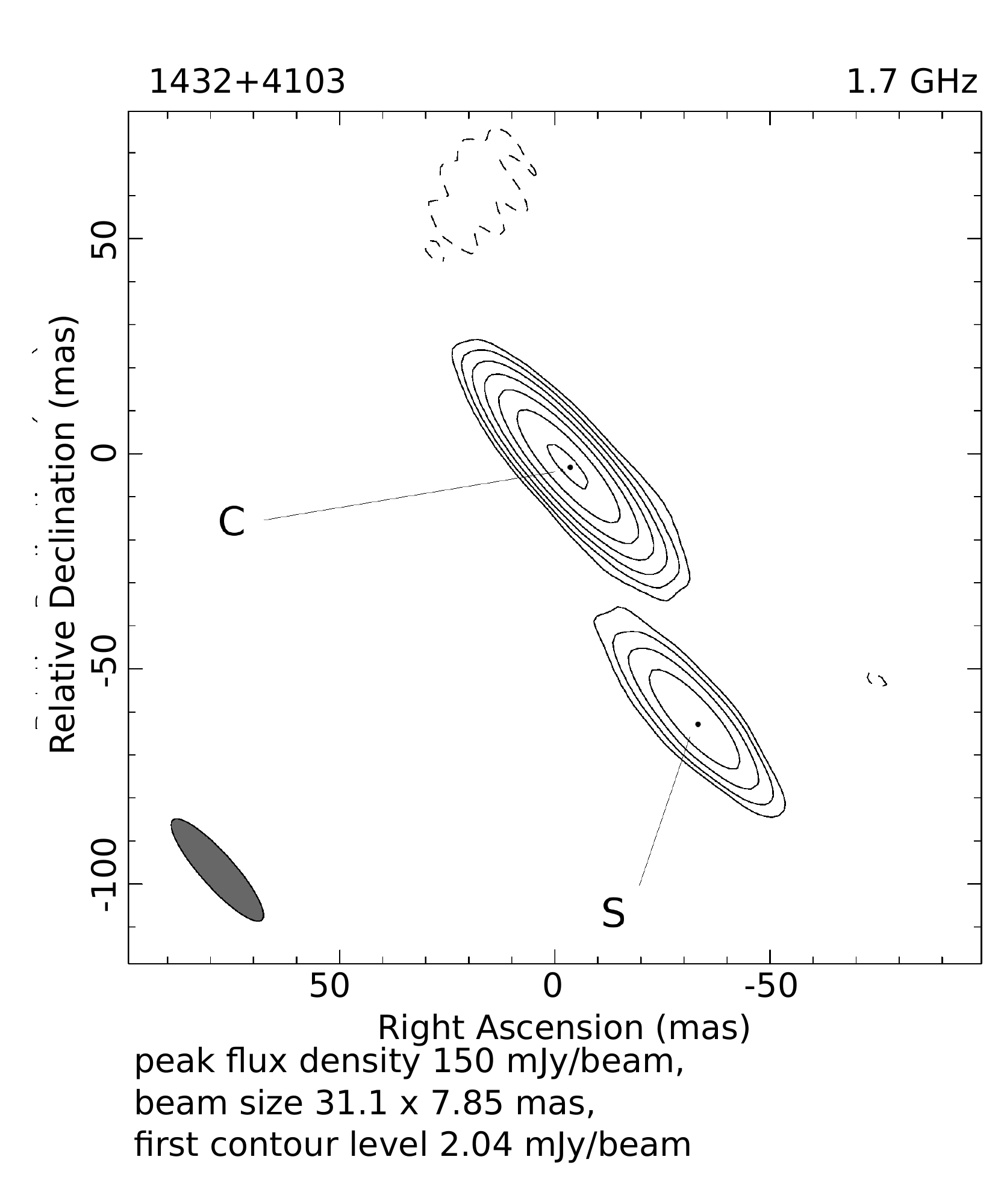}
   \includegraphics[width=0.32\textwidth, height=0.23\textheight]{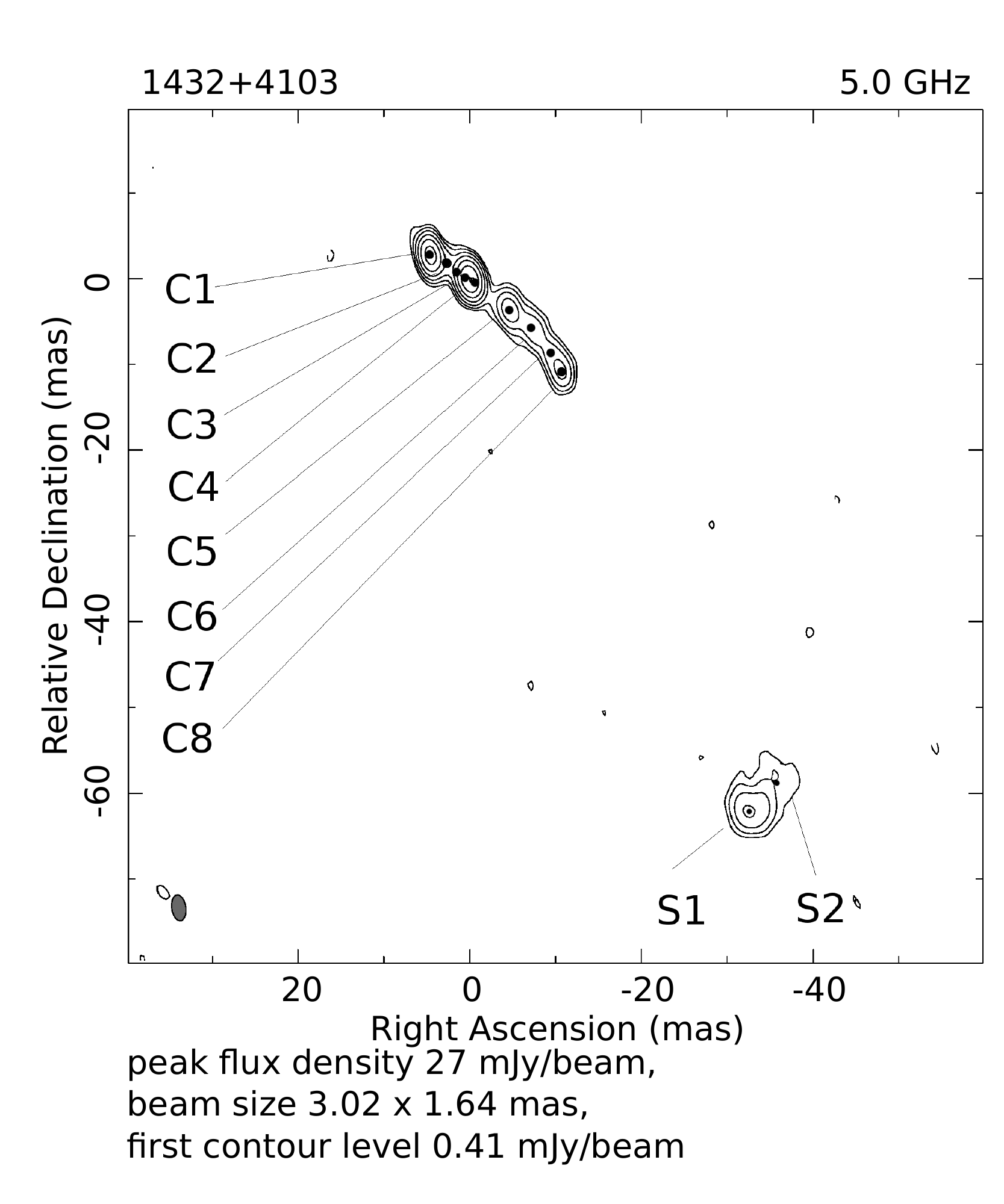}
    \includegraphics[width=0.32\textwidth, height=0.23\textheight]{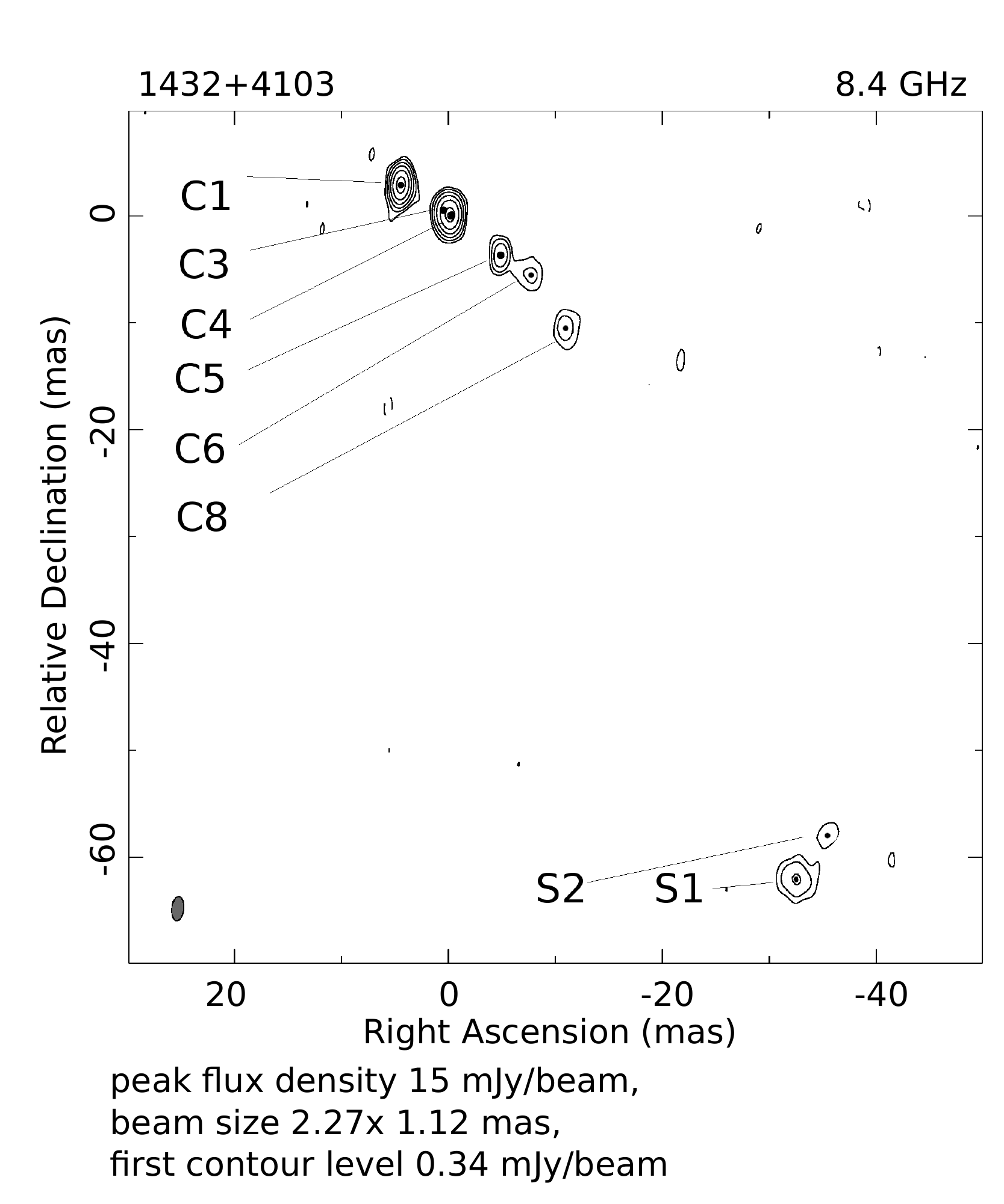}
   \caption{Sources with observations at all three frequencies (cont.).}
\end{figure*}

\begin{figure*}
\centering

    \includegraphics[width=0.32\textwidth, height=0.23\textheight]{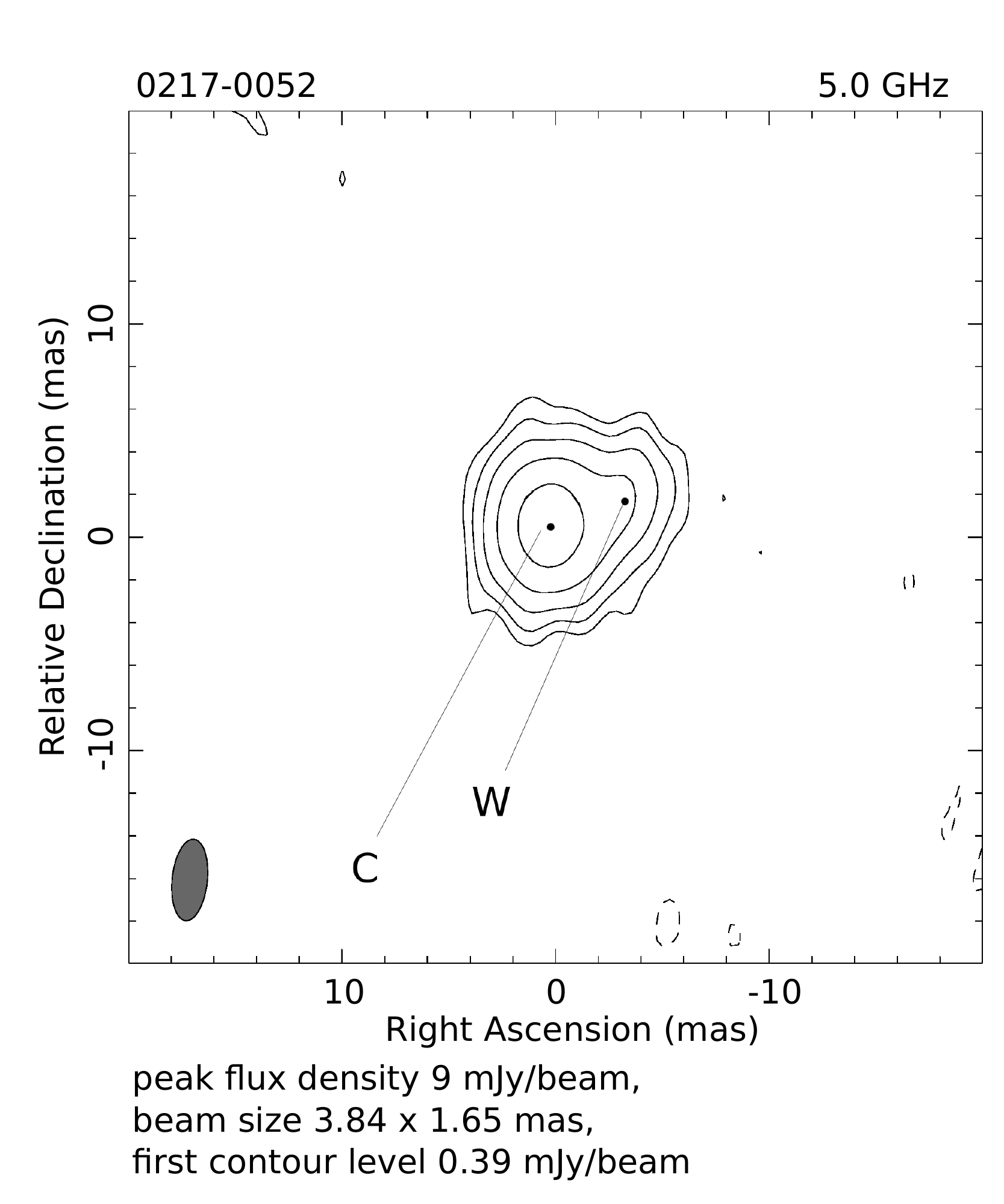}
    \includegraphics[width=0.32\textwidth, height=0.23\textheight]{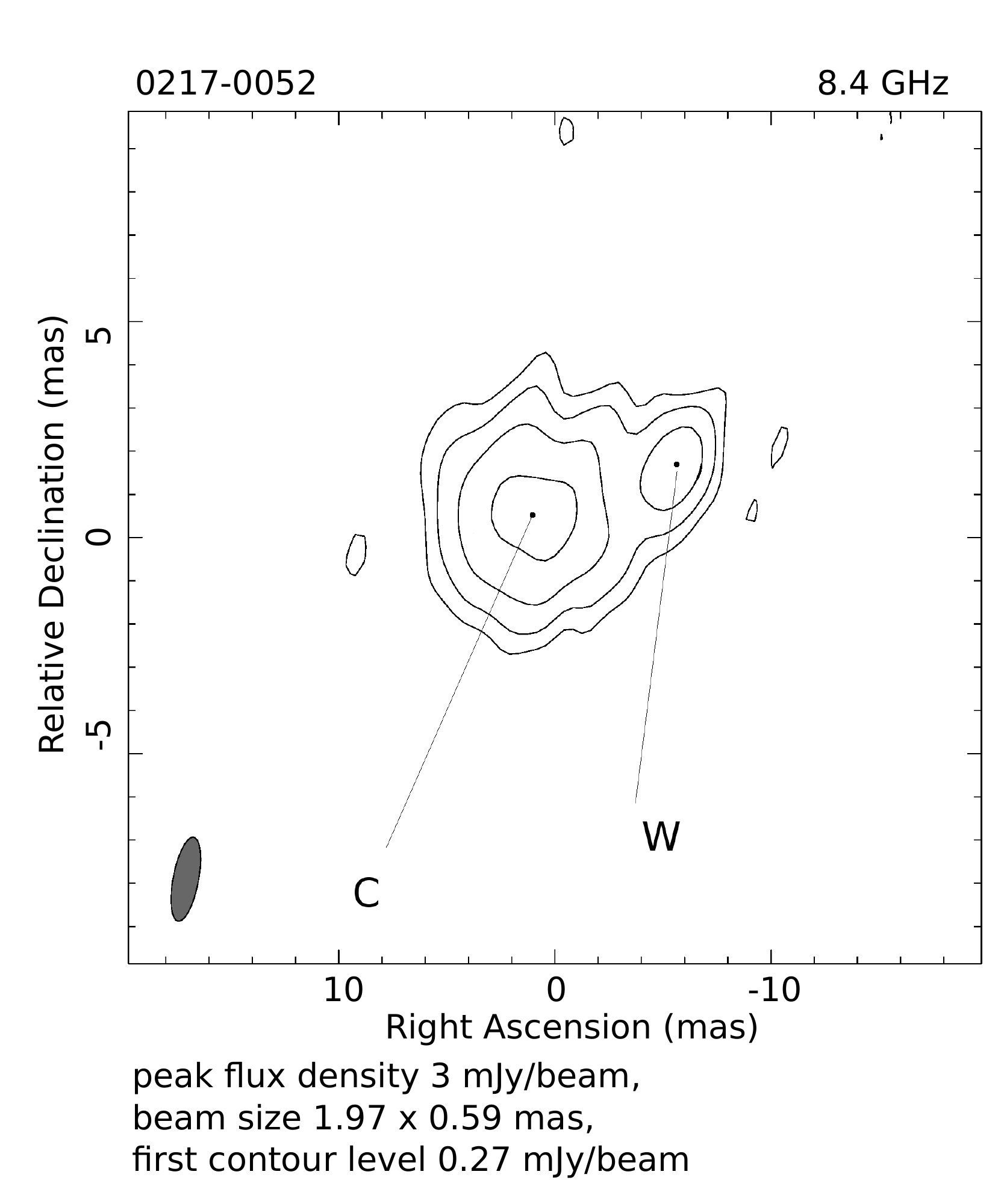}  
    \includegraphics[width=0.32\textwidth, height=0.23\textheight]{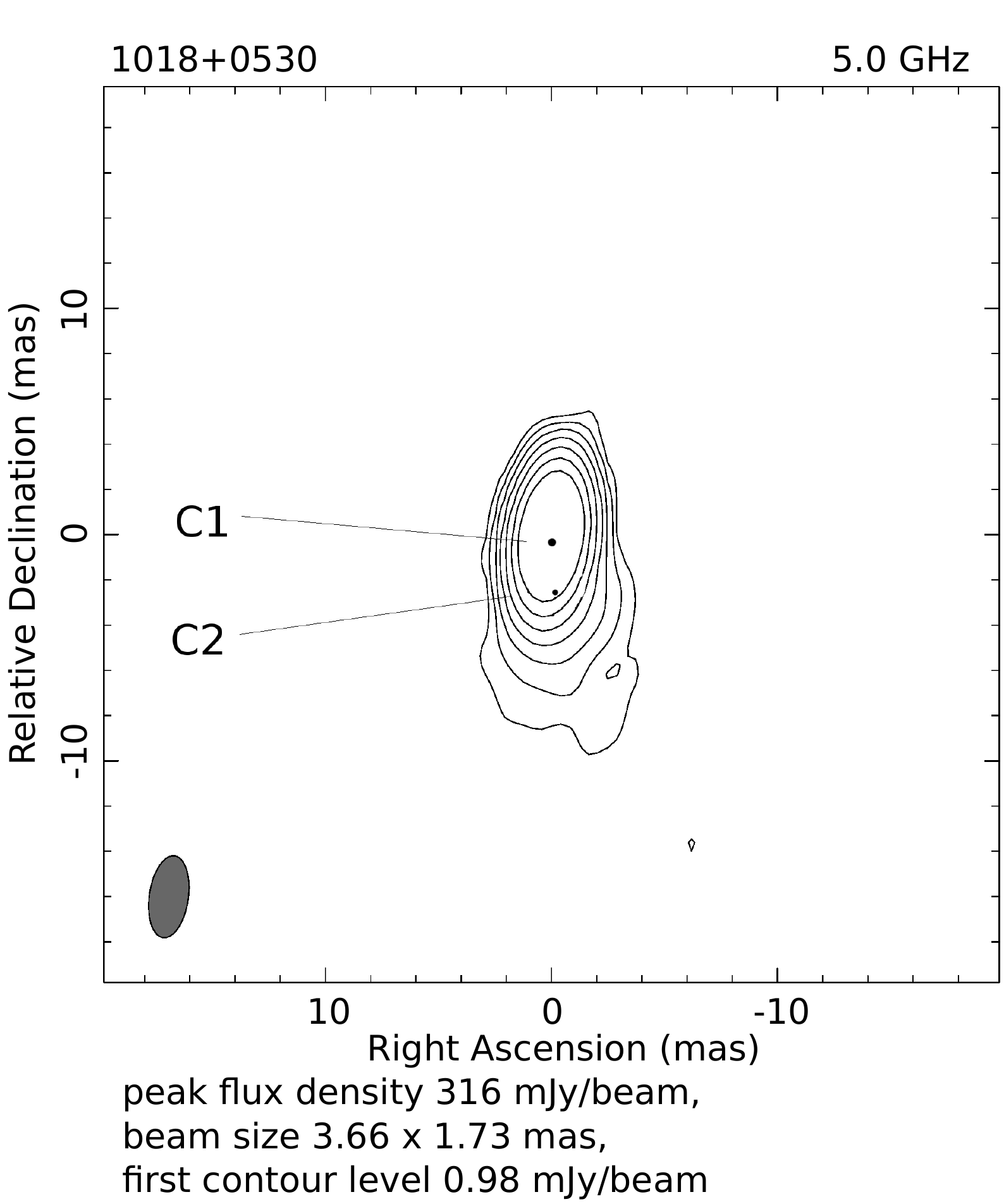}
     
    \includegraphics[width=0.32\textwidth, height=0.23\textheight]{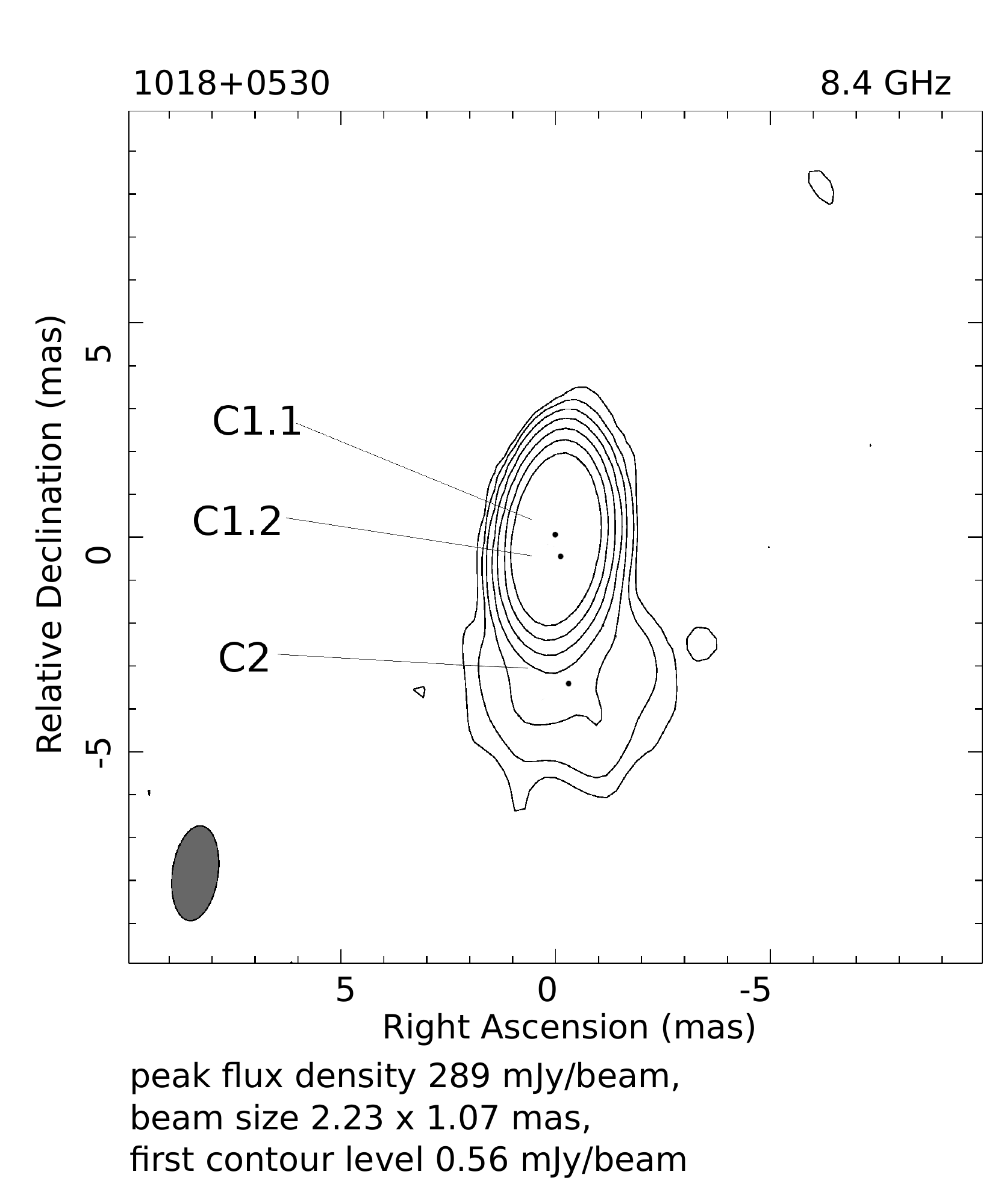}
    \includegraphics[width=0.32\textwidth, height=0.23\textheight]{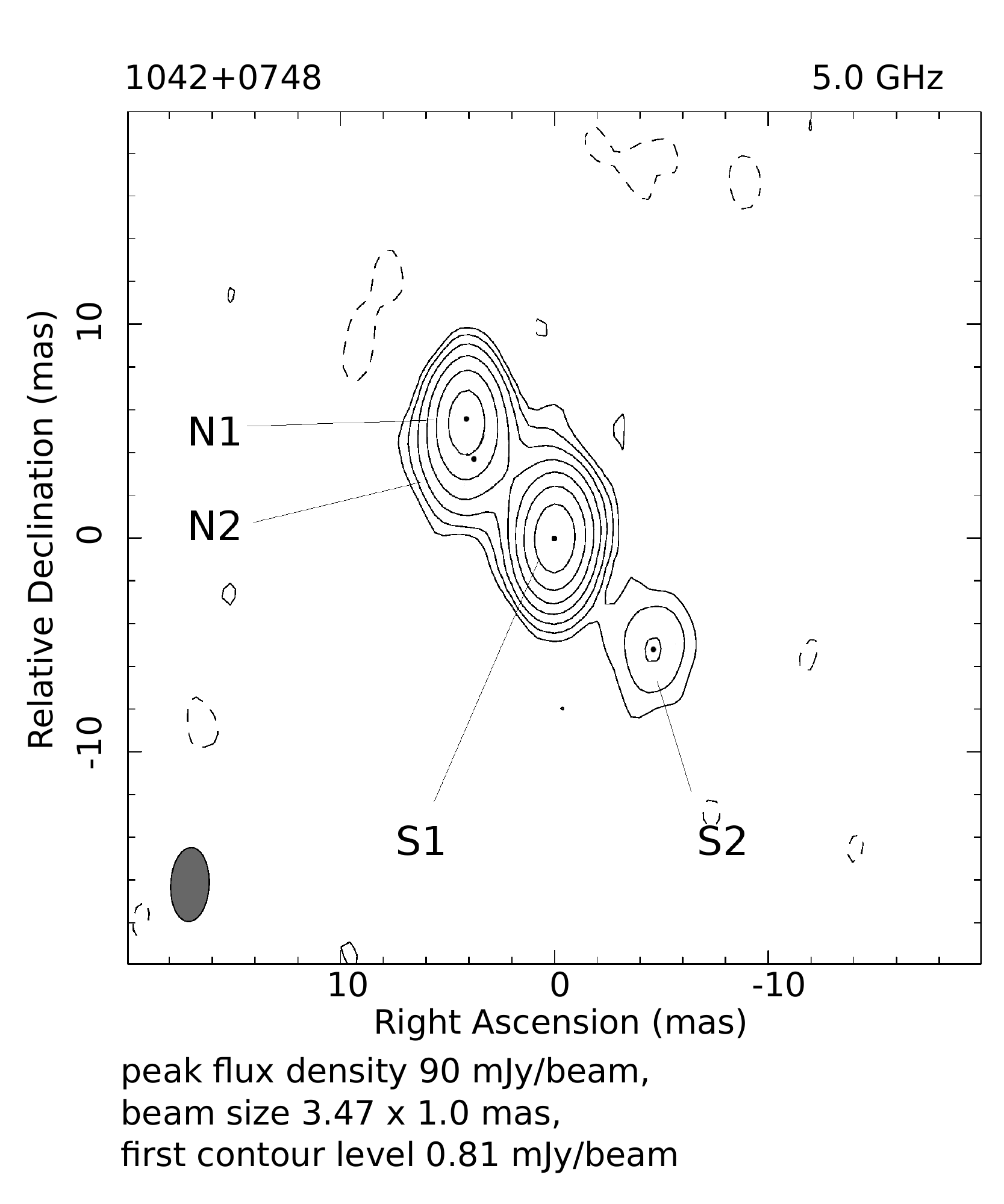}
    \includegraphics[width=0.32\textwidth, height=0.23\textheight]{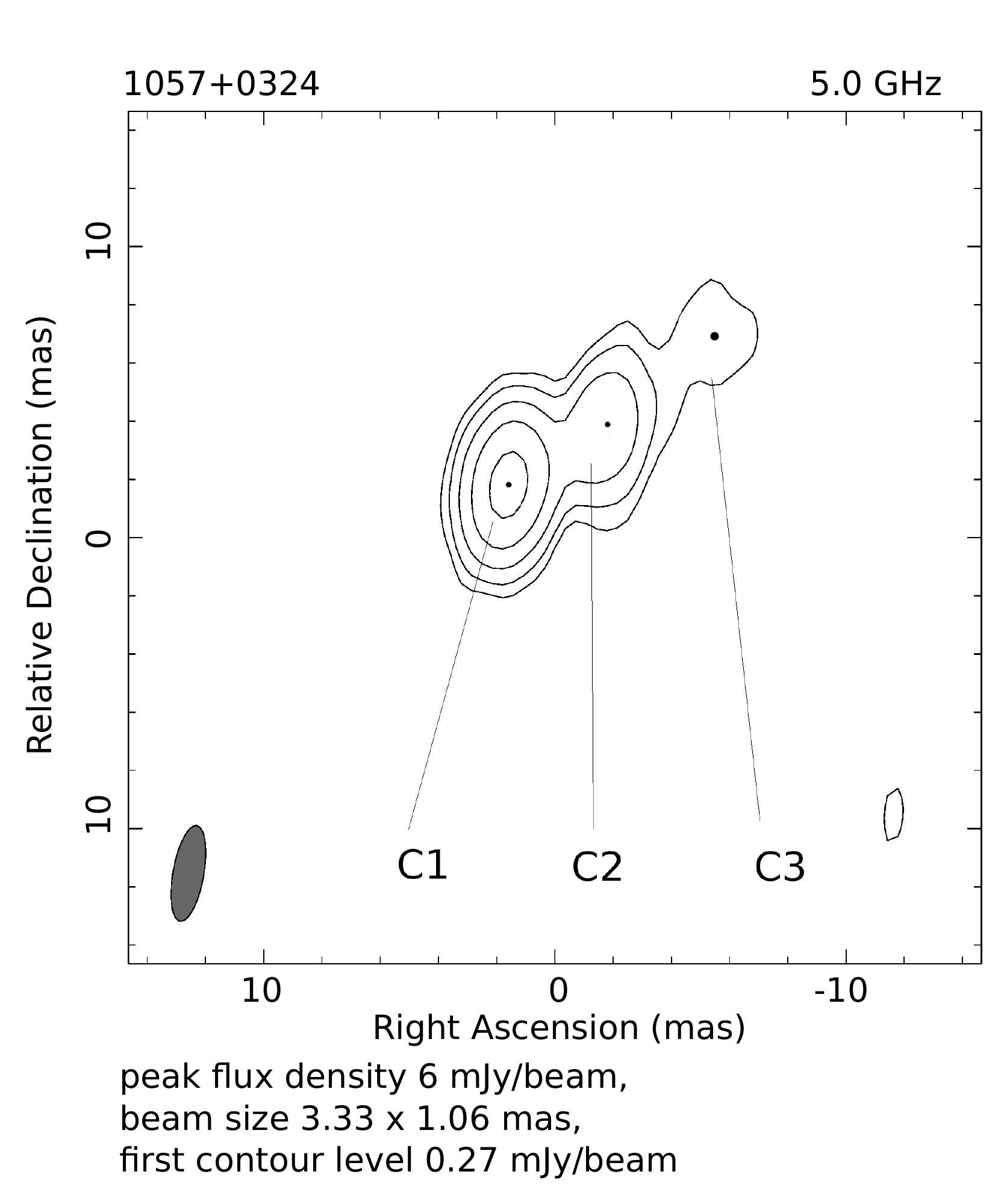}
   
    \includegraphics[width=0.32\textwidth, height=0.23\textheight]{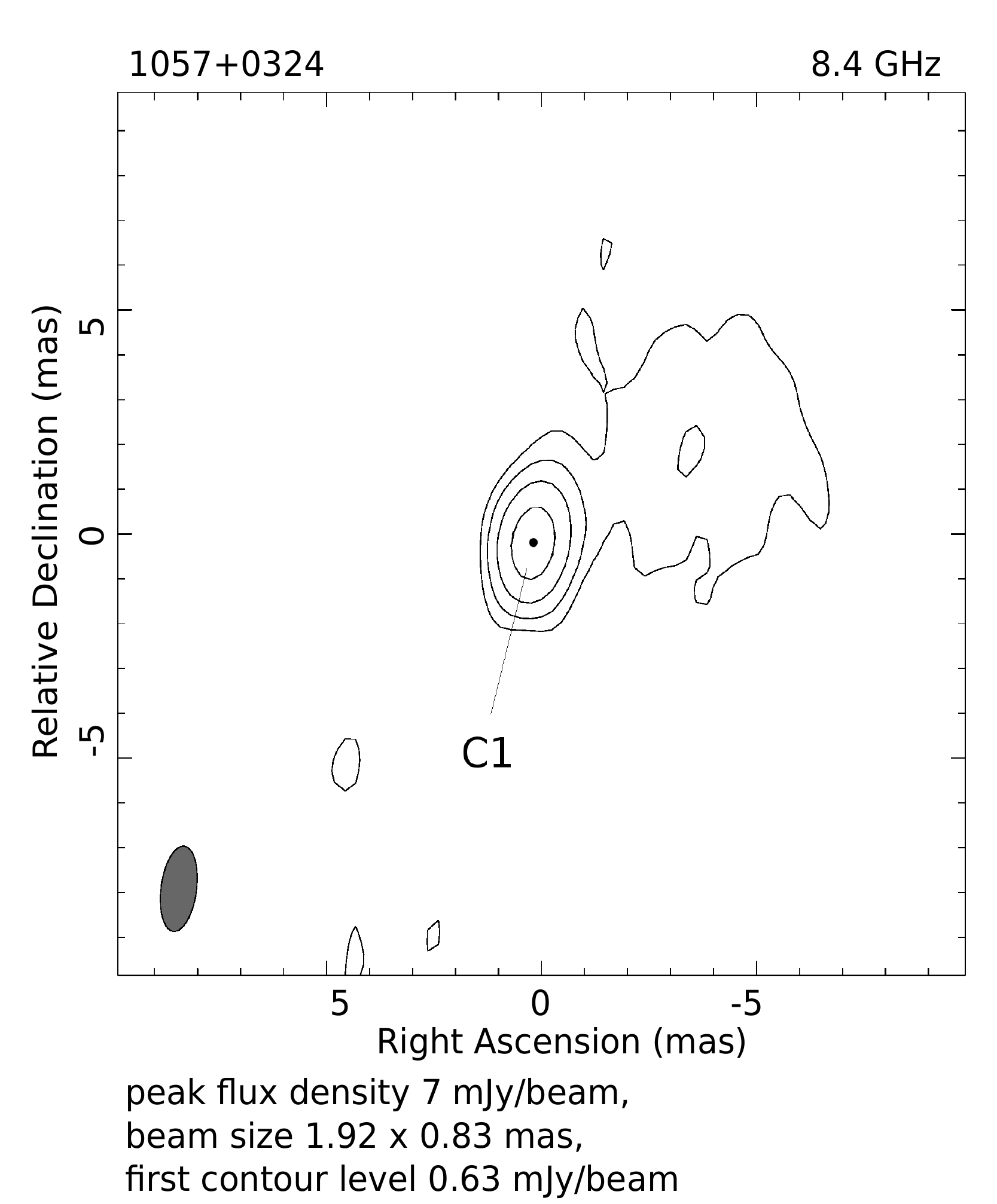}
    \includegraphics[width=0.32\textwidth, height=0.23\textheight]{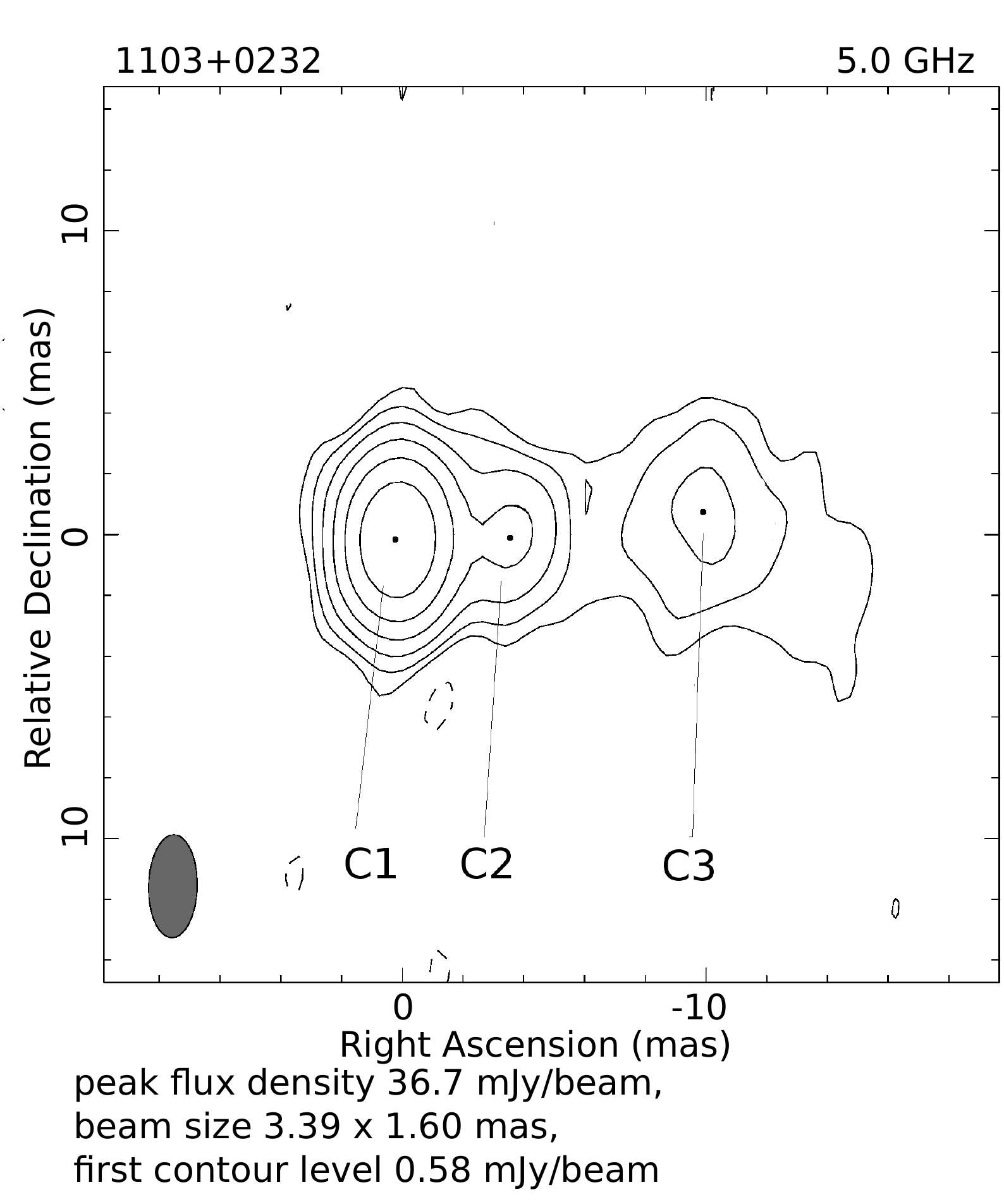}
    \includegraphics[width=0.32\textwidth, height=0.23\textheight]{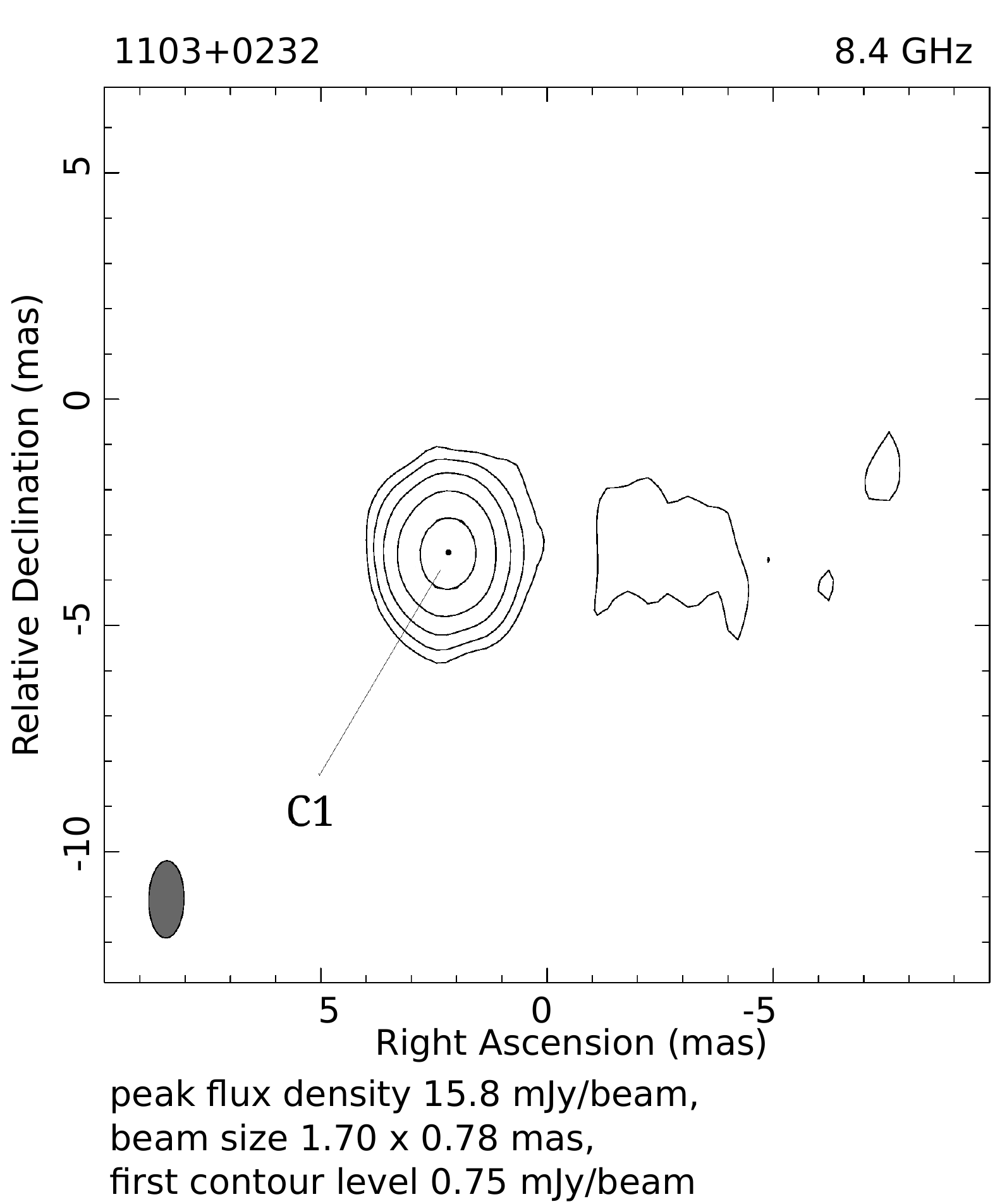}
   
    \includegraphics[width=0.32\textwidth, height=0.23\textheight]{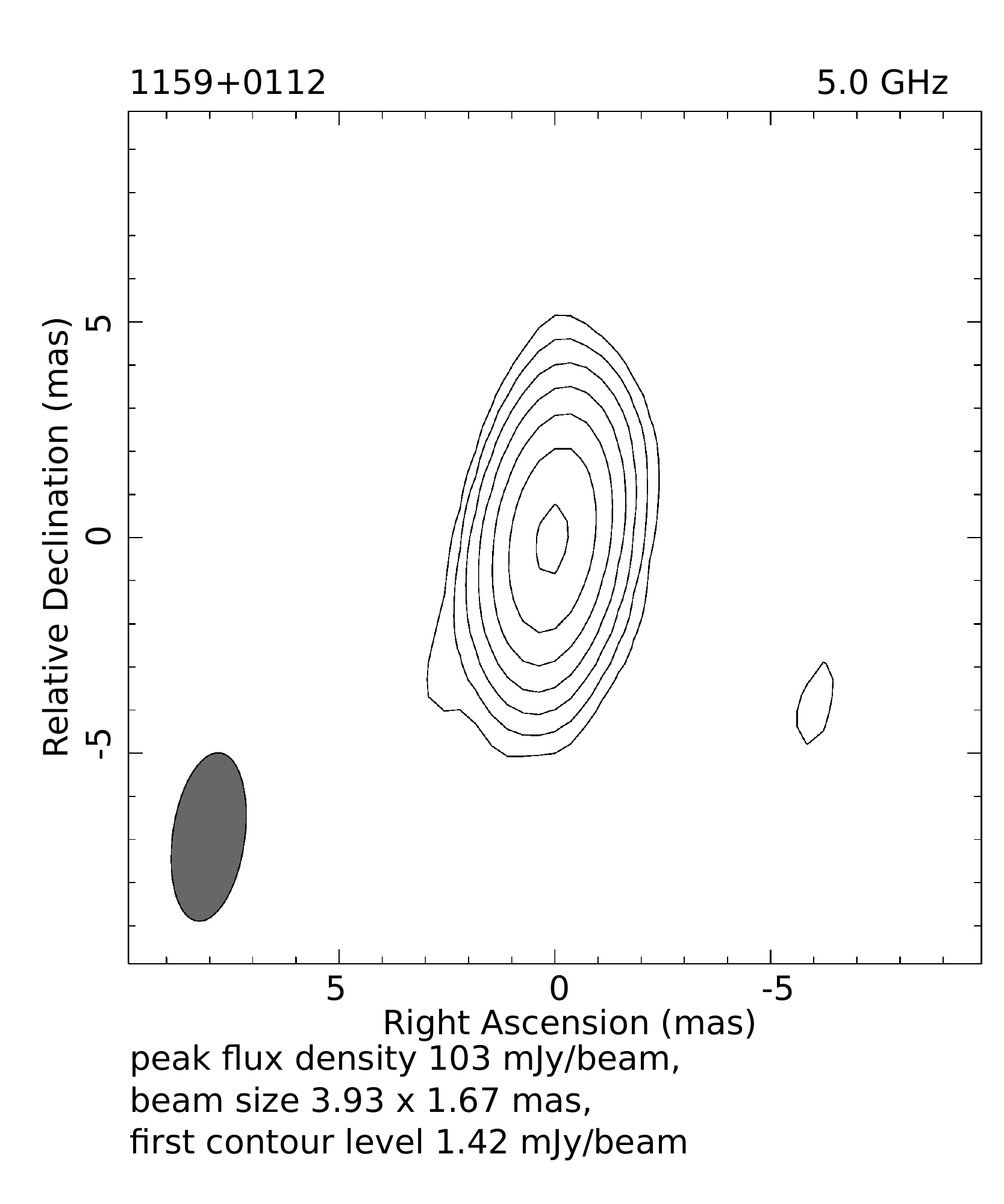}
    \includegraphics[width=0.32\textwidth, height=0.23\textheight]{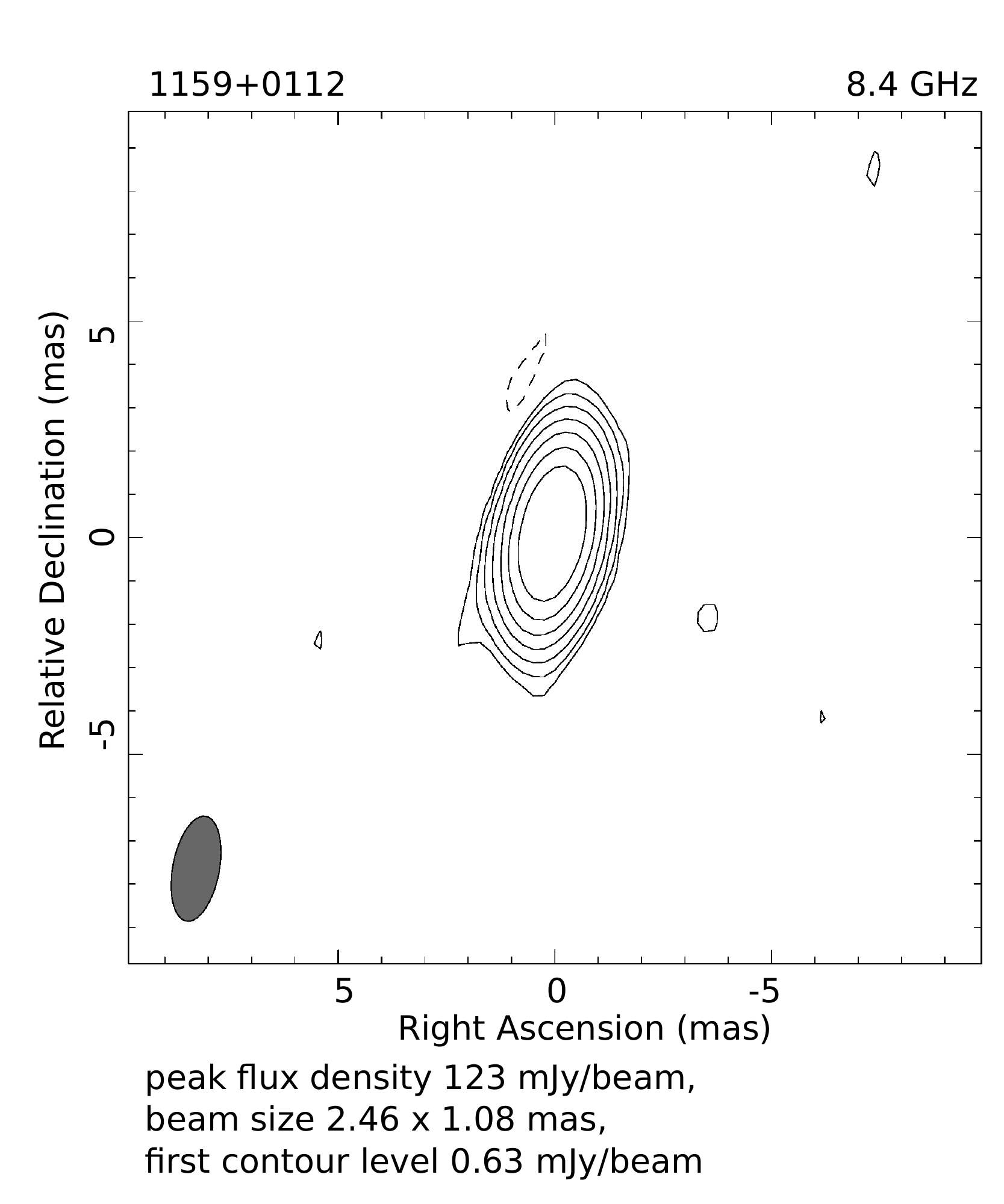}
\caption{Sources observed only with VLBA at 5 and 8.4\,GHz. Contours increase by a factor 2, and the first contour level corresponds to $\approx 3\sigma$.}
\label{images2}
\end{figure*}

\begin{table*}
\caption[]{Flux densities of the principal components of the sources from the 1.7, 5\footnote{Data taken from VIPS are marked with $^{*}$} and 8.5 GHz observations.}
\begin{center}
\label{observations}
\scalebox{0.8}{
\begin{tabular}{@{} c c c c  c c c c c c c c c c c c c c@{}}
\hline
\hline
RA(J2000) & Dec(J2000) &
\multicolumn{3}{c}{Components}&
\multicolumn{1}{c}{${\rm S_{1.7\,GHz}}$}&
\multicolumn{1}{c}{${\rm S_{5\,GHz}}$}&
\multicolumn{1}{c}{${\rm S_{8.4\,GHz}}$}&
 \multicolumn{1}{c}{ $\alpha^{1.7 GHz}_{5 GHz}$}&
 \multicolumn{1}{c}{ $\alpha^{5 GHz}_{8.4 GHz}$}&
\multicolumn{3}{c}{$\theta$}&
\multicolumn{1}{c}{${\rm T_{b}(vlbi)}$}& 
LAS &
LLS&
Type\\
 
 &
 &
\multicolumn{1}{c}{1.7\,GHz}&
\multicolumn{1}{c}{5\,GHz}&
\multicolumn{1}{c}{8.4\,GHz}&
&
&
&
&
&
\multicolumn{1}{c}{{1.7\,GHz}}&
\multicolumn{1}{c}{5\,GHz}&
\multicolumn{1}{c}{8.4\,GHz}&
&
&\\

h~m~s & $\degr$~$\arcmin$~$\arcsec$ & 
&
&
&
\multicolumn{3}{c}{(mJy)}&
&
&
\multicolumn{3}{c}{(mas)}&
\multicolumn{1}{c}{($\rm 10^{9}~K~\delta^{-1}$)}&
(mas)& (pc)& \\

(1)& (2)& \multicolumn{3}{c}{(3)} &(4)&   
(5)&
(6)&
\multicolumn{1}{c}{(7)}&
\multicolumn{1}{c}{(8)}&
\multicolumn{3}{c}{(9)}&
(10)&
(11)&
(12)&
(13)
\\

\hline
\hline
02 17 28.614& -00 52 26.880& -- & C & C&  -- & 29.24  & 24.71 & -- & 0.32 & -- &3.46  &3.30 & 0.2&\multirow{2}{*}{3.8} & \multirow{2}{*}{30.8} & \multirow{2}{*}{CJ} & \\
	&     & -- & W & W &-- & 5.46  & 4.38 & -- & 0.42  & -- & 2.08 & 0.84&\\
\hline	     
07 56	28.26	&37 14	55.800& \multirow{3}{*}{C} &C1&C1 & \multirow{3}{*}{277.40} &  $98.31^{*}$  & 49.00 & -- & 1.34 &\multirow{3}{*}{2.85} &0.53 &0.49 & &\multirow{3}{*}{4.7} &  \multirow{3}{*}{37.9}&\multirow{3}{*}{CJ}\\
	 &    & & C2 & C2 &   & $106.46^{*}$  & 74.50 & -- & 0.69  & &0.19  &0.41& 41.0&\\
	  &   &  &C3&   &   & $10.85^{*}$ &  -- &-- & --  & &1.80  & -- &\\
\hline
08 15 34.184& 33 05 29.280 & \multirow{3}{*}{C1} & C1.1 & C1.1 & 	\multirow{3}{*}{35.62}  & 9.80 & 2.70 & -- & 2.48 &\multirow{3}{*}{10.61} &0.53  &0.34 & &\multirow{7}{*}{726.2}& \multirow{7}{*}{5895.8}& \multirow{7}{*}{O} \\
	&      &                   & C1.2 & C1.2 &  		                & 8.94 & 10.73& --  & -0.35 &                       & 0.48  &0.52 & 3.41\\
	 &     & 		   & C1.3 & C1.3 &                             & 4.04 &  5.28& --  & -0.51 &                        &0.94 &0.68&\\
 	  &    &       C2           &    &  &        5.88            &    --   & --  &   --   &  --& 16.86 & -- & --&\\
	   &   &       N1            &   &  &     55.65                &    --    & --  &   --   &  -- & 28.60&  -- & --&\\
	    &  &       N2            &   &  &      37.7                &    --    & --  &   --   &  -- &31.94 &  -- & --&\\
	    &  &       N3            &   &  &     11.65                &   --     & --  &   --   &  -- &18.58 &   --& --&\\ 
\hline
09 28 24.133& 44 46 04.680 & \multirow{2}{*}{C} &\multirow{2}{*}{C}& C1 & \multirow{2}{*}{231.30} &  \multirow{2}{*}{$245.40^{*}$} & 159.57 &\multirow{2}{*}{-0.05}  & -- &\multirow{2}{*}{1.00} &\multirow{2}{*}{0.39} &0.19 &  & &  \multirow{3}{*}{33.6} &\multirow{3}{*}{CJ} \\
	  &   & &  &C2  &   &   &  133.23 &  & -- & & & 0.24 &168.20& 4.0\\
	  &   & \multirow{1}{*}{E} &\multirow{1}{*}{E} &  E &  \multirow{1}{*}{75.61} & \multirow{1}{*}{$7.6^{*}$} &  3.73 & 2.13  &\multirow{1}{*}{1.37} &\multirow{1}{*}{2.24} & \multirow{1}{*}{1.83}  &1.41 &\\
\hline
10 05 15.961& 48 05 33.210  & \multirow{3}{*}{C} & C1& C1 & \multirow{3}{*}{30.62} & 26.80  & 21.36&  -- & 0.44 & \multirow{3}{*}{3.20}&  0.86 & 0.59 &5.19 & &\multirow{3}{*}{118.3}&\multirow{3}{*}{CJ} \\
	   &  &                    & C2 & C2  &  & 4.00   &  2.67 & -- & 0.78 & & 1.32  &1.15 & &19\\
	   &  &                    & S &   &   & 3.44  & -- &  --  & -- &   & 2.92 & --  &\\
  \hline
10 13 29.931& 49 18 41.110 & \multirow{1}{*}{C}   & C1& C1 &\multirow{1}{*}{55.21}  & $18.87^{*}$   & 32.67 &0.99  & -1.06&\multirow{1}{*}{2.33} &0.23&0.40 &16.40& \multirow{4}{*}{9.0} & \multirow{4}{*}{74.4} & \multirow{4}{*}{CJ}\\
	   &  &   \multirow{3}{*}{N}& N1 & N1  & \multirow{3}{*}{209.77} & $19.27^{*}$  &  12.45 & --  & 0.84 & \multirow{3}{*}{2.53}& 0.73 &1.16&\\
	   &  &                    & N2 &  N2  &   &  $40.14^{*}$  & 26.87 & -- & 0.77  & & 0.77 &1.06&\\
	   &   &                    & N3 &  N3  &   & $34.56^{*}$  & 17.96 &--  &  1.26 & & 1.65 &1.65&\\
  \hline
10 18 27.837& 05 30 29.900& \multirow{3}{*}{}   &   \multirow{2}{*}{C1}        &  C1.1    &                  &\multirow{2}{*}{327.33}  &261.29   &   --   & -- & -- & \multirow{2}{*}{0.53} & 0.34 &167.00 & &\multirow{3}{*}{28.5}&\multirow{3}{*}{CJ}\\
	   & &			    &    			   & C1.2   &                     &                        &  65.86  & -- & --&  --& &0.83& &3.4\\
	   &   &                    &           C2  	           &  C2    &                    & 	31.08	            &  14.33 &  --   & 1.49 & -- &2.46& 3.13& \\
  \hline
10 42 57.598& 07 48 50.600 & \multirow{2}{*}{}   & N1&  &\multirow{2}{*}{--}  &   39.74  & -- & -- & -- & -- &  0.98 & --& \multirow{4}{*}{--}&\multirow{4}{*}{13.7}& \multirow{4}{*}{108.9}&\multirow{4}{*}{O}\\
	&     &                     & N2 &   &                   &   24.43  & --   &  --   & -- & -- & 3.32 & --&\\
	&     &   \multirow{2}{*}{}& S1 &   & \multirow{2}{*}{--} &  109.2   &  -- & --  & -- & --   &1.06& --&\\
	&     &                    & S2 &    &                    &    5.91& -- & -- & -- &  --&  1.99&  --&\\
  
  \hline
10 57 26.608& 03 24 48.470  & \multirow{3}{*}{}  & C1& C1 &\multirow{3}{*}{--}  &   12.50  & 12.74  &--     &   -0.04 & -- & 1.84&1.05 &1.11&&\multirow{3}{*}{68.1}&\multirow{3}{*}{CJ}\\
	&     &                     & C2 &  &                   &  5.27    &  --  &    --  & -- & -- &2.00 & --  & &8.7 \\
	&     &                     & C3 &    &                   &   1.40    &  --  &    --      & -- & -- &2.08 & -- & \\
  \hline
11 03 44.536& 02 32 09.740 & \multirow{3}{*}{}  & C1& \multirow{3}{*}{C1} &\multirow{3}{*}{--}  &  64.24  &  51.26  & --   &   0.43 & -- &1.84 &1.57 & 1.83 &&\multirow{3}{*}{82.2}&\multirow{3}{*}{CJ}  &\\
	&     &                     & C2 &  &                   &  10.22   &  --  &   --   & -- & -- & 1.97 & --  & &10.2\\
	&     &                     & C3 &    &                   &   13.09   &  --  &     --     & -- & -- & 4.51 & -- & \\	 	     
  \hline
  11 59 44.832& 01 12 06.870 & \multirow{1}{*}{C}  & C & C &  -- &   110.44  & 134.39   & --  &  -0.38  &  -- & 0.59 & 0.41& 6.02 & 0.4 & 3.3&S\\

  \hline
12 23 43.165& 50 37 53.490  & \multirow{2}{*}{C1}  & C1.1& C1.1 &\multirow{2}{*}{127.73}  &   $102.52^{*}$ &  122.04 &  --  & -0.33  &\multirow{2}{*}{2.17} & 0.17& 0.23 &260.00&\multirow{4}{*}{71.2}&\multirow{4}{*}{521.7} &\multirow{4}{*}{CJ}\\
	&     &                     & C1.2 & C1.2 &                         &   $29.59^{*}$  &  11.00  &  --   & 1.89 & & 1.17& 2.47 & \\
	&     &  	C2	    &   &   & 		15.25	 &    --         & --  &   --    & -- & 3.50 &  -- & -- && \\
	&     &  E		    &   &  & 	 25.49	     &     --        &  --  &       --      &  --& 3.21&  --  & --& \\
	       \hline
14 05 07.795& 40 56 58.060 & \multirow{1}{*}{C1}  & C1 & C1 &\multirow{1}{*}{316.39}  &   $185.62^{*}$  &  204.86  &   0.49 &  -0.19 &2.27 &0.19 & 0.43 &83.20&\multirow{2}{*}{2.8}& \multirow{2}{*}{23.4}&\multirow{2}{*}{CJ}\\
 	&      &                     & C2 &  & --                   &  $2.64^{*}$   &   --   &--    & --  & --  &  0.15& -- &\\
 	&      &          C3           & &  & 5.87                   &  --   &   --   &--    & --  & 1.93  & -- & --  &\\
	   \hline
14 32 43.322& 41 03 28.040   & \multirow{8}{*}{C} & C1 & C1 & \multirow{8}{*}{162.75}  & 20.00 & 16.5 & -- & 0.37&\multirow{8}{*}{3.13} &0.58 &0.49 & 5.12&\multirow{10}{*}{66.5}& \multirow{10}{*}{557.5} &\multirow{10}{*}{CJ} \\
	&      &                   & C2 & C2 &  		          & 2.44 & -- &   --    & -- & & 0.92 &-- &\\
	&      & 		   & C3 &    &                           &   11.76 &  4.91& --  & 1.68 & &0.47  &0.58 &\\
	&       & 		   & C4 & C4 &                           &  20.14 & 14.38 & --  & 0.65 & & 0.55 &0.64 &\\
	&      &                   & C5 & C5  &                           &  7.13 &  3.07& --    & 1.62 & & 1.07 &0.60 &\\
	&      &                   &  C6  & C6  &                           &   4.58    &  1.94 &--   &  1.65 & &1.40  &1.72 &\\
	&      &                   &  C7  &  &                          &   2.83    & --  & --  &  --& &1.28& -- &\\
	&      &                   &  C8  & C8 &                          &    4.70   &    2.17& --  & 1.49 & &0.89 &1.56 &\\
	&	&\multirow{2}{*}{S} & S1   &  S1 & \multirow{2}{*}{37.44}&   9.77     &  4.41 &--   &  1.53 &\multirow{2}{*}{4.40} & 2.87 &2.19 &\\
	&      &                   &  S2    &  S2   &                       &    4.46  &  1.25 &  --   &  2.45& &4.57&1.39 &\\
\hline 
\hline

\end{tabular}

}
\end{center} 

\begin{minipage}{185mm}
{ 
Description of the columns:
(1) \& (2) source coordinates (J2000) extracted from FIRST, (3) components as indicated on the images, (4) flux density measured with the EVN on 1.7 GHz ,
(5) \& (6) flux density measured with the VLBA on 5.0 GHz and  8.4 GHz, (7) spectral index between
1.7 and 5.0\,GHz calculated using flux densities in columns (4) and (5), (8) spectral index between
5.0 and 8.4\,GHz calculated using flux densities in columns (5) and (6),
(9) deconvolved major axis of the Gaussian fit on 1.7, 5 and 8.4 GHz, 
(10) brightness temperature calculated based on the component's size and flux density measured at 
the highest resolution map available using equation \ref{vlbi},
(11) largest angular size (LAS) measured at resolving frequency,
LAS is defined as a separation between the two outermost Gaussian components, 
(12) largest linear size (LLS) calculated based on the LAS,
(13) radio morphology: CJ - core-jet, O - other, S - single .
}
\end{minipage}
\label{observation}

\end{table*} 


\begin{landscape}
\begin{table}
\caption[]{Candidates for variable sources}
\label{variable}
\begin{tabular}{ccccccccccccccccc}
\hline
RA(J2000) & Dec(J2000)&
\multicolumn{1}{c}{\it z}&
\multicolumn{1}{c}{F$_{peak}$}&
\multicolumn{1}{c}{$\sigma_{F_{peak}}$}&
\multicolumn{1}{c}{$\rm Epoch_{F}$}&
\multicolumn{1}{c}{N$_{int}$}&
\multicolumn{1}{c}{$\sigma_{N_{int}}$}&
\multicolumn{1}{c}{$\rm Epoch_{N}$}&
\multicolumn{1}{c}{AI}&
\multicolumn{1}{c}{BI}&
\multicolumn{1}{c}{log $R_{I}$}&
\multicolumn{1}{c}{${\rm T_{b}(var)}$}&
\multicolumn{1}{c}{$\rm \delta_{1}$}&
\multicolumn{1}{c}{$\theta_{1}$}&
\multicolumn{1}{c}{$\rm \delta_{2}$}&
\multicolumn{1}{c}{$\theta_{2}$}\\

h~m~s & $\degr$~$\arcmin$~$\arcsec$ & 
\multicolumn{1}{c}{}&
\multicolumn{1}{c}{(mJy)} &
\multicolumn{1}{c}{(mJy)}&
\multicolumn{1}{c}{} & 
\multicolumn{1}{c}{(mJy)}&
\multicolumn{1}{c}{(mJy)}&
\multicolumn{1}{c}{} & 
\multicolumn{1}{c}{}&
\multicolumn{1}{c}{} & 
\multicolumn{1}{c}{}&
\multicolumn{1}{c}{($\rm 10^{11}~K~\delta^{-3}$)}&
\multicolumn{1}{c}{}&
\multicolumn{1}{c}{(deg)}&
\multicolumn{1}{c}{}&
\multicolumn{1}{c}{(deg)}\\

(1)& (2)& (3) &(4)&   
\multicolumn{1}{c}{(5)}&
\multicolumn{1}{c}{(6)}& 
\multicolumn{1}{c}{(7)}&
\multicolumn{1}{c}{(8)}&  
\multicolumn{1}{c}{(9)}&
\multicolumn{1}{c}{(10)}&
\multicolumn{1}{c}{(11)}&
\multicolumn{1}{c}{(12)}&
\multicolumn{1}{c}{(13)}&
\multicolumn{1}{c}{(14)}&
\multicolumn{1}{c}{(15)}&
\multicolumn{1}{c}{(16)}&
(17)\\

\hline 
07	53	10.42	&	21	02	44.31	&	2.29	&	16.78	&	0.15	&	1998.695	&	14.4	&	0.6	&	1993.836	&	7479	&	3526	&	1.95	&	5.7	&	$>$1.79	&	$<$34.0	&	$<$1	&	$-$ \\                       
08	11	02.93	&	50	07	24.52	&	1.84	&	23.07	&	0.19	&	1997.262	&	19.5	&	0.7	&	1993.874	&	1573	&	371	&	2.14	&	12.1	&	$>$2.29	&	$<$25.9	&	$>$1.06	&	$<$70.6\\
08	53	42.02	&	06	56	55.18	&	2.39	&	6.89	&	0.14	&	2000.091	&	5.6	&	0.4	&	1993.874	&	1980	&	275	&	1.65	&	2.9	&	$>$1.43	&	$<$44.4	&	$<$1	&	$-$\\
11	44	36.66	&	09	59	04.80	&	3.15	&	12.83	&	0.14	&	2000.045	&	11.2	&	0.5	&	1995.159	&	783	&	105	&	1.48	&	6.9	&	$>$1.91	&	$<$31.6	&	$<$1	&	$-$\\
14	01	26.16	&	52	08	34.63	&	2.97	& 	36.18	&	0.14	&	1997.337	&	30.4	&	1	&	1993.874	&	517	&	70	&	2.02	&	42.8	&	$>$3.50	&	$<$16.6	&	$>$1.62	&	$<$38.1\\
14	41	36.26	&	63	25	18.76	&	1.78	&	7.69	&	0.23	&	2002.594	&	4.4	&	0.4	&	1993.896	&	2838	&	150	&	1.50	&	1.6	&	$>$1.17	&	$<$58.7	&	$<$1	&	$-$\\
14	59	26.33	&	49	31	36.79	&	2.37	&	5.22	&	0.14	&	1997.291	&	3.8	&	0.4	&	1995.194	&	14001	&	9419	&	1.24	&	19.6	&	$>$2.69	&	$<$21.8	&	$>$1.25	&	$<$53.1\\
15	37	03.95	&	53	32	19.93	&	2.40	&	9.28	&	0.14	&	1997.347	&	7.1	&	0.4	&	1993.874	&	2398	&	934	&	1.30	&	11.5	&	$>$2.26	&	$<$26.3	&	$>$1.05	&	$<$72.2\\
21	07	57.68	&	-06	20	10.49	&	0.65	&	20.29	&	0.15	&	1997.145	&	12.4	&	0.6	&	1993.720		&	1309	&	559	&	1.41	&	3.4	&	$>$1.51	&	$<$41.5	&	$<$1	&	$-$\\
\hline	
02	25	56.50	&	-07	43	07.34	&	2.45	&	7.11	&	0.15	&	1997.166	&	5.5	&	0.5	&	1993.720		&	1133	&	0	&	1.91	&	8.9	&	$>$2.07	&	$<$28.9	&	$<$1	&	$-$\\
07	48	23.87	&	33	20	51.38	&	2.99	&	8.14	&	0.15	&	1995.128	&	5.9	&	0.4	&	1993.956	&	292	&	0	&	1.78	&	151.3	&	$>$5.33	&	$<$10.8	&	$>$2.47	&	$<$23.9\\
07	56	28.26	&	37	14	55.80	&	2.51	&	238.60	&	0.17	&	1994.560		&	216.2	&	6.5	&	1993.956	&	841	&	0	&	3.46	&	4243.7	&	$>$16.19	&	$<$3.5	&	$>$7.51	&	$<$7.7\\
08	21	44.01	&	36	44	08.87	&	1.88	&	17.50	&	0.15	&	1994.549	&	15.5	&	0.6	&	1993.956	&	1058	&	0	&	2.12	&	234.8	&	$>$6.17	&	$<$9.3	&	$>$2.86	&	$<$20.5\\
08	28	17.25	&	37	18	53.64	&	1.35	&	21.18	&	0.13	&	1994.561	&	14.8	&	0.6	&	1993.956	&	1356	&	0	&	2.47	&	388.1	&	$>$7.29	&	$<$7.9	&	$>$3.39	&	$<$17.2\\
08	33	50.61	&	38	39	22.71	&	2.01	&	4.24	&	0.14	&	1994.602	&	2.7	&	0.4	&	1993.956	&	628	&	0	&	0.97	&	172.4	&	$>$5.57	&	$<$10.3	&	$>$2.58	&	$<$22.8\\
09	05	52.41	&	02	59	31.62	&	1.82	&	43.54	&	0.14	&	1998.542	&	36.4	&	1.2	&	1993.874	&	256	&	0	&	1.48	&	12.7	&	$>$2.33	&	$<$25.4	&	$>$1.08	&	$<$67.8\\
10	50	44.24	&	06	09	58.24	&	3.27	&	13.49	&	0.14	&	2000.109	&	11.4	&	0.5	&	1995.159	&	319	&	0	&	2.02	&	9.2	&	$>$2.09	&	$<$28.6	&	$<$1	&	$-$\\
11	45	53.69	&	-00	33	04.84	&	2.06	&	3.91	&	0.16	&	1998.623	&	2.1	&	0.4	&	1995.159	&	1576	&	0	&	1.48	&	7.3	&	$>$1.94	&	$<$31.0	&	$<$1	&	$-$\\
12	14	46.08	&	53	20	23.73	&	2.15	&	13.02	&	0.22	&	1997.346	&	8.9	&	0.9	&	1993.874	&	748	&	0	&	1.77	&	17.9	&	$>$2.62	&	$<$22.4	&	$>$1.21	&	$<$55.7\\
12	17	29.30	&	06	07	50.75	&	2.10	&	25.54	&	0.14	&	2000.105	&	13.5	&	0.6	&	1995.159	&	677	&	0	&	2.22	&	24.7	&	$>$2.91	&	$<$20.1	&	$>$1.35	&	$<$47.8\\
14	19	11.60	&	52	05	45.48	&	1.71	&	14.71	&	0.17	&	1997.336	&	12.6	&	0.5	&	1993.874	&	1407	&	0	&	2.37	&	6.1	&	$>$1.83	&	$<$33.1	&	$<$1	&	$-$\\
14	44	34.82	&	00	33	05.49	&	2.04	&	12.76	&	0.15	&	1998.567	&	10.5	&	0.5	&	1995.159	&	1926	&	0	&	1.90	&	9.3	&	$>$2.10	&	$<$28.4	&	$<$1	&	$-$\\
14	58	15.23	&	00	39	08.90	&	2.02	&	20.48	&	0.13	&	1998.565	&	18.2	&	0.7	&	1995.159	&	410	&	0	&	2.06	&	9.2	&	$>$2.10	&	$<$28.4	&	$<$1	&	$-$\\
23	31	32.84	&	01	06	20.94	&	2.64	&	42.40	&	0.13	&	1995.790		&	35.4	&	1.1	&	1993.874	&	713	&	0	&	2.05	&	143.5	&	$>$5.23	&	$<$11.0	&	$>$2.43	&	$<$24.3\\
23	36	34.08	&	-09	43	18.69	&	2.22	&	96.41	&	0.16	&	1997.226	&	82.4	&	2.5	&	1993.720		&	960	&	0	&	2.84	&	63.3	&	$>$3.99	&	$<$14.5	&	$>$1.85	&	$<$32.7\\

\hline
\end{tabular}
\begin{minipage}{215 mm}
{Description of the columns:
(1) \& (2) source coordinates (J2000) extracted from FIRST, (3) redshift as measured from the SDSS, 
(4) FIRST peak flux density, (5) uncertainty of the FIRST peak flux density, (6) FIRST observation time, (7) NVSS integrated flux density, (8) uncertainty of the
NVSS integrated flux density, (9) NVSS observation time, (10) absorption index AI taken from \cite{trump06},(11) balincity index BI taken from \cite{trump06},
(12) radio-loudness, the radio-to-optical (i-band) ratio of the 
quasar core \citep{kimball2011}, which were calculated
from {\it z}, ${\rm S_{1.4\,GHz}}$, $\rm M_{i}$ taken from \citet{schneider2007}, and the assumption of a radio core spectral index of 0 and
an optical spectral index of -0.5, (13) lower limit of the brightness temperature calculated using equation \ref{var_eq}, (14) minimum Doppler factor calculated using the equipartition
brightness temperature value of $\rm 10^{11}\,K$, (15) maximum viewing angles estimated as described in section \ref{var} using the value of $\rm \delta_{1}$, 
(16) minimum Doppler factor calculated using the inverse Compton brightness temperature value of $\rm 10^{12}\,K$, (15) maximum viewing angles estimated as described 
in section \ref{var} using the value of $\rm \delta_{2}$.
}
\end{minipage} 
\end{table} 
\end{landscape}

{\bf 1042+0748}. This source has been observed only at 5\,GHz and it has been complex radio structure. 
Integrated flux density from the 5-GHz VLBA image is comparable with single dish measurement (Table~\ref{basic}) and thus we have detected 
its whole structure. The overall spectrum of 1042+0748 is steep.
The classification of this source is difficult and thus we marked it as 'Other' in Table~\ref{observations}.

{\bf 1057+0324}. This source has three components on the
C-band map which are resolved out in 8.4-GHz observations. 
The flat spectrum component C1 is likely a radio core. The 5-GHz integrated flux density is significantly smaller ($\sim$ 27\%) 
compared to a single dish observations (Table~\ref{basic}).
However, we are unable to unambiguously determine whether it is a result of luck of short spacings or due to intrinsic variability.
The overall spectrum of 1057+0324 is steep.
Based on its structure and spectral index we classified this source as a core-jet.

{\bf 1103+0232}. This source was observed only at 5 and 8.4-GHz with VLBA (Fig.~\ref{images2}).
The brightest component C1 is radio core and steep spectrum components C2 and C3 are probably parts of a radio jet.
Its morphology can be classified as a core-jet.

{\bf 1159+0112}. This source was observed by us only at 5 and 8.4\,GHz with VLBA and remained unresolved (Fig.~\ref{images2}).
Spectral index $\alpha^{5}_{8.5} $ = -0.37 indicates this object is core dominated.
\citet{bruni12} compared their flux densities measurements of 1159+0112 with those previously reported by \citet{monte08}. The higher frequency they contrasts the more significant variability occurred. Nevertheless they could not conclude whether it is intrinsic
or due to different resolution (dissimilar VLA configuration). Based on the complex spectral energy distribution \citet{monte08} concluded
that the source may consists of more extended, diffuse emission and compact part with a peak frequency
of $\nu_{p}$ = 6.3 GHz implying the young age of the compact structure. 
1159+0112 has been also observed by \citet{taka13} on L, C and
X band with VLBA. Their radio map on 1.7\,GHz proves that this source posses extended emission on lower frequencies and our images at 5\,GHz
and 8.4\,GHz are zoom in version of probable radio core. Interestingly, \citet{taka13} detected additional component on C-band in a core proximity
 - likely a radio jet. It is not visible on our map. This may suggest that this part of radio jet was very recently lunched from the core and such
events in the past could be responsible for flux density variability detections. Based on the overall structure of 1159+0112 \citet{taka13} suggested 
a re-activation scenario for this object with the compact core being the new activity phase.

{\bf 1223+5037}. The EVN 1.7\,GHz image shows double structure with a weak eastern component E, which is not present in 5- and 8.4-GHz observations.   
Our VLBA 8.4\,GHz map is consistent with that obtained in the 
VIPS project \citep{helm07} at 5\,GHz and we present all three images in Fig.~\ref{images1}.
The inverted spectrum component C1.1 is probably a radio core, components C1.2 and C2 are parts of the radio jet and E 
might be a part of radio lobe. We lost $\sim$ 26\% of the total flux density at 1.7\,GHz probably connected with the extended emission that can be present between the eastern and central components.
The overall spectrum of 1223+5037 is steep. 
We have classified this source as a core-jet.

{\bf 1405+4056}. Our 1.7- and 8.4-GHz images are consistent with the 5-GHz VIPS image \citep{helm07}. They show double structure with component C1
being a radio core (Fig.~\ref{images1}). The same result was obtained in VLBA observations made by \citet{taka13}. 
Contrasting single dish flux density measurements with those reported in this paper and provided by \citet{taka13} meaningful differences
($\sim$ 30\%) are clearly visible at 1.7 and 5\,GHz.
However, we are unable to unambiguously determine whether it is a result of different {\it uv}-coverage  or due to intrinsic variability. 
1405+4056 is a candidate for GPS object \citep{marecki99}.

{\bf 1432+4103}. This source posses double structure in the 1.7-GHz EVN map which has been resolved into complex structure in VLBA 5- and 8.4-GHz
observations.
Component C1 with the most flat spectrum of all detected parts is probably a radio core, the other components are parts of the one-sided jet. 
We lost ($\sim$ 23\%) of the 1.7-GHz total flux density probably due to poor {\it uv}-coverage.
The overall spectrum of 1432+4103 is steep. We have classified this source as a core-jet.

\section{Results \& discussion}
\label{dis}

\subsection{Parameters from VLBI imaging}
\label{observatioins}

During three observational campaigns 14 out of 16 sources were detected.
All 14 sources but one (1159+0112) were resolved, the majority of which fall under core-jet classification (11 out of 14) thus confirming the existence of non-thermal jets in BALQSOs.
We have radio maps for four sources at three different frequencies. For the rest other than one (1042-0748), we obtained 
images on C and X-band. The measured and derived quantities for individual components of the sources as well as their classification are presented in Table~\ref{observation}.

When creating the samples of BI and AI quasars we did not put any constrains on the shape of the spectrum. It has
already been  shown by \citet{becker00} that there is wide variety of spectral indices among the BALQSOs. However, most of the observed AI sources possess steep spectra and in the 
few cases (0217-0052, 0815+3305, 1005+4805, 1057+0324) we have noticed significant lack of flux density comparing to VLA or single dish observation. 
This is probably connected with more diffuse structures present in these sources which we did not fully detect in our observations due to their weakness. One source, 1159+0112, remained
unresolved even at 8.4\,GHz. It looks however, that due to the luck of short spacings we lost extended structures present in this source which in turn can suggest 
previous phase of the activity \citep{taka13}.
Similar findings concerning BALQSOs have been recently reported in the work of \citet{bruni13}. The missing flux density between VLA and VLBI observations in some cases can be attributed to the low frequency remnant of the previous phase of the radio source activity \citep{kun10b,kun10}.

Finally, two of our quasars (0928+4446, 1018+0530) have been classified as flat-spectrum radio quasars (FSRQ) and one (1405+4056)
is a GPS candidate \citep{marecki99}. The radio morphology of these three sources (0928+4446, 1018+0530, 1405+4056) and their significant flux density variations reported
in the literature imply they are blazar candidates (\citet{taka13}, see also Table~\ref{basic}). 
Comparing radio morphology of our sources with QSO structures \citep{kimball2011} we conclude that they represent
typical quasar geometries.

\subsubsection{VLBI brightness temperature}
\label{vlbi_tb}
Additional parameter which can shed some light on the concept of orientation scenario is the brightness temperature 
which can be derived directly from interferometric observation or from the flux density variability.
By imaging analysis, the brightness temperature in the rest frame is calculated using following equation:

\begin{equation}
T_b (vlbi) = \frac{1+z}{\delta} \frac{c^2 F_\nu}{2k_{\rm B} \pi \nu^2 (\theta/2)^2}
\label{vlbi}
\end{equation}
 
where {\it c} is the speed of light, $k_{\rm B}$ is the Boltzmann constant, $F_\nu$ is the observed flux density at frequency $\nu$, $\delta$ is 
a Doppler factor, and $\theta$ is the angular diameter of a source. 
Using values from Table \ref{observation} we 
estimated $T_{b}$ for components likely representing radio cores in our AI quasars. In the case of three flat spectrum sources (0928+4446, 1018+0530, 1223+5037)
the value of the brightness temperature is the highest among our AI quasars and exceeds $10^{11} $K.

The intrinsic brightness temperature of extragalactic radio sources should be in the range of $\rm 10^{11} - 10^{12}\,K$.
The value of $\rm 10^{12}\,K$ is a theoretical limit which when it is exceeded leads to the well-known inverse Compton catastrophe. 
The $\rm 10^{11}\,K$ is an empirical value derived for a
sample of variable extragalactic radio sources by \citet{lahteenmaki1999}. 
Taking the second limit we estimated the minimum 
Doppler factor that avoids the inverse Compton catastrophe as follows: $\rm \delta_{1}=\frac{T_{b}(vlbi)}{10^{11}\,K}$.
Afterwards we estimated the viewing angle $\theta$ of each 
object, defined as an angle between the jet axis and the observer,
as the maximum of the following function \citep{ghosh07}:

\begin{equation}
cos\theta \leqslant \max\left[ \frac{\delta - \sqrt{1-\beta^2}}{\delta \beta}\right]
\label{theta}
\end{equation}

where $\beta$ is the velocity of the jet and $\delta$ is a Doppler factor.

The obtained values of the viewing angle are $36.5\degr$, $37\degr$ and $22.6\degr$ for 0928+4446, 1018+0530 and 1223+5037, respectively. 
These are, however, the maximum allowed values which means that the viewing angles of the above quasars are $<37\degr$. This result is in agreement
with previously reported large flux density variations found in these objects. This in turn may imply that the
sources can be larger and older than estimated by us based on the projected angular size.

\begin{figure}
\centering
   \includegraphics[width=\columnwidth]{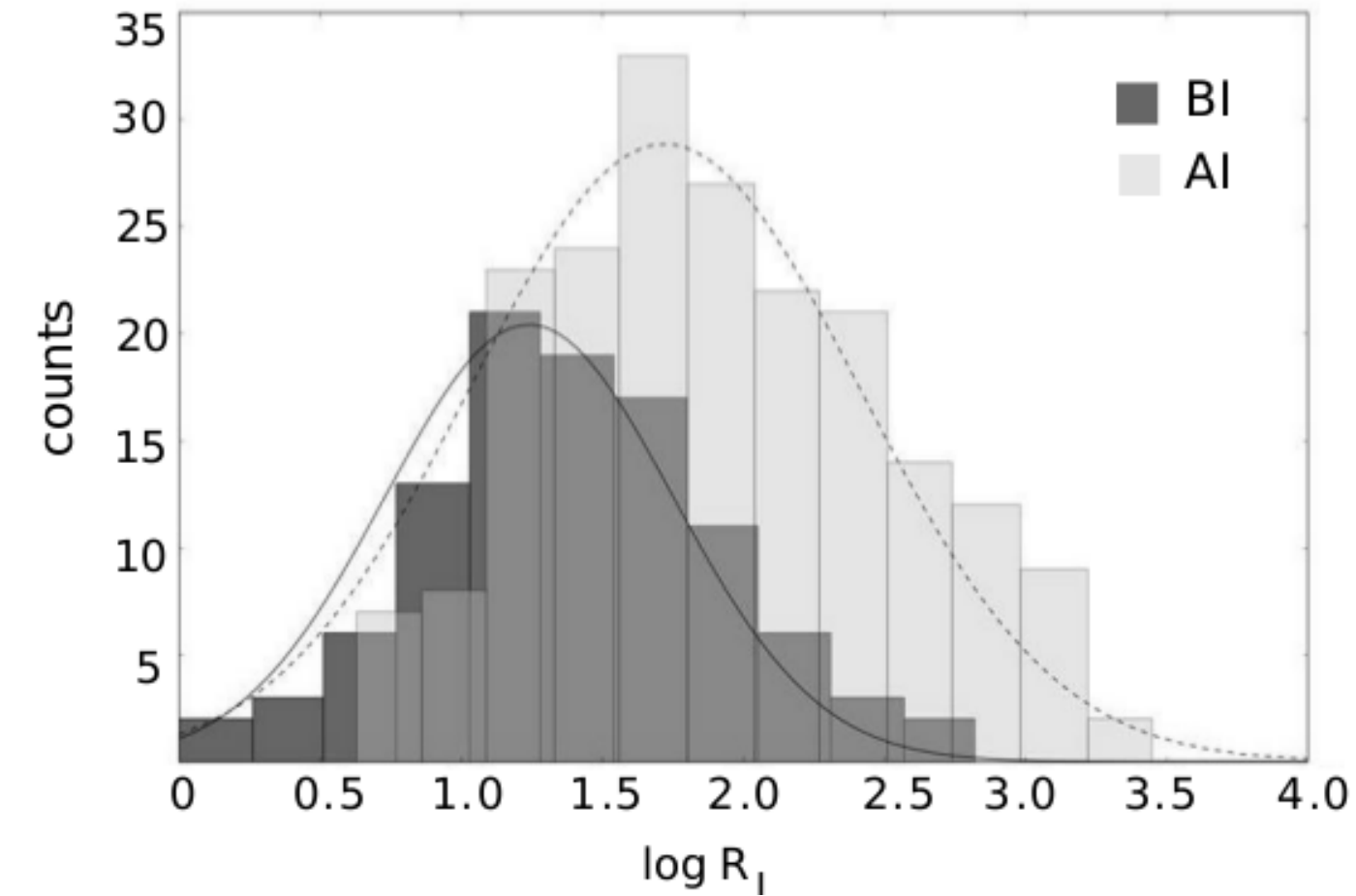}
\caption{Radio-loudness, log\,$\rm R_{I}$, distribution for AI and BI quasars from parent sample. 
Gaussian function was fitted to both distributions (see sec.~\ref{radioL}).}
\label{counts2}
\end{figure}

\subsection{Statistical analysis of parent sample}

In the next step, using the parent sample of BALQSOs (309 sources) selected from \citet{trump06} 
we performed statistical studies concerning orientation of these objects.
We also compared the properties of the two subgroups of sources, namely the AI quasars (204 objects) and BI quasars (105 objects).

\subsubsection{Radio-loudness distribution}
\label{radioL}

The radio-to-optical ratio of the quasar
core is thought to be a strong statistical indicator of orientation \citep{wills95}.
An analysis of its distribution among BALQSOs can be another way to deal with the enigma of their nature. 
We adopted radio-loudness definition from \citet{kimball2011}: $\rm R_{I} = (M_{radio}-M_{i})/-2.5$,
where $\rm M_{radio}$ is a K-corrected radio absolute magnitude and $\rm M_{i}$ is a Galactic reddening corrected and K-corrected i-band absolute magnitude.
If indeed radio-loudness parameter is indicative of line-of-sight orientation, then its large numbers,
possibly $\rm log\,R_{I}>2.5$, could mean close to the radio jet axis orientation. Quasars
with $\rm log\,R_{I}<1.5$ are thought to be viewed at small angles relative to the plane of the disk \citep{kimball2011}.
Since all our sources are unresolved by FIRST we used their integrated 1.4-GHz flux densities as a core flux and assume the radio core 
spectral index and optical spectral index to be 0 and -0.5 respectively. 
The histogram in Fig.~\ref{counts2} 
shows the number of BALQSOs versus the radio-loudness parameter $\rm R_{I}$ for both, BI and AI, samples.
The result of Kolmogorov-Smirnov (K-S) test (D=0.34) implies that both distributions are dissimilar 
at the 0.05 confidence level.
Histogram clearly hints that while $\rm R_{I}$ rises AI sample outnumbers BI. While BI distribution peaks between
1.0 and 1.5 (fitting Gauss results in $\rm (log\,R_{I})_{max}$ = 1.24 $\pm$ 0.30) 
AI distribution is shifted and apex between 1.5 and 2.0 (fitting Gauss results in  $\rm (log\,R_{I})_{max}$ = 1.72 $\pm$ 0.06). Thus, on average BI quasars are viewed closer to the disk plane than AI sources.

The majority of AI objects observed by us in the VLBI technique have $\rm log\,R_{I}>2.5$ so they belongs to the tail of log\,$\rm R_{I}$ distribution
for AI quasars and constitute the most luminous subgroup of AI population.

\subsubsection{Variability brightness temperature}

\label{var}

Significant flux density variations can signal Doppler boosting, which in turn indicates that a jet points close to the line of sight toward the observer.
Therefore the flux density variability is an another, independent way of orientation determination by estimating variability brightness temperature:
\begin{equation}
T_b (var) \geq \frac{1}{\delta^3 (1+z)} \frac{F_\nu D_{L}^2}{2k_{\rm B} \pi \nu^2 (\Delta t)^2}
\label{var_eq}
\end{equation}

where $D_{\rm L}$ is the luminosity distance and $\Delta t$ is the variability time scale in the observer's
frame.

Following \citet{zhou06} we performed radio variability analysis for the whole parent sample (309 sources).
We compared flux density of quasars present in both, FIRST and the NRAO VLA Sky Survey (NVSS;\citet{condon1998}) catalogues, to find significant changes.     
To classify the radio-loud BALQSO as variable source we computed the variability ratio (VR) and the significance of the radio flux density
variability ($\sigma_{var}$) for each quasar as proposed by \citet{zhou06}.
BALQSOs with $\rm VR>1$ and $\sigma_{var}>3$ where selected as candidate for variable sources.
While VR $>$ 1 may hint intrinsic flux density changes, value of significance proposed by \citet{zhou06} 
$\sigma_{var}$ = 3, which suggests variability is instead taken a priori.
Nevertheless, to provide complementary data for sake of informative 
discussions we applied this threshold. Total of 26 BALQSOs, 9 BI and 17 AI objects, have been selected for further studies.

\begin{figure}
\centering
\includegraphics[width=\columnwidth]{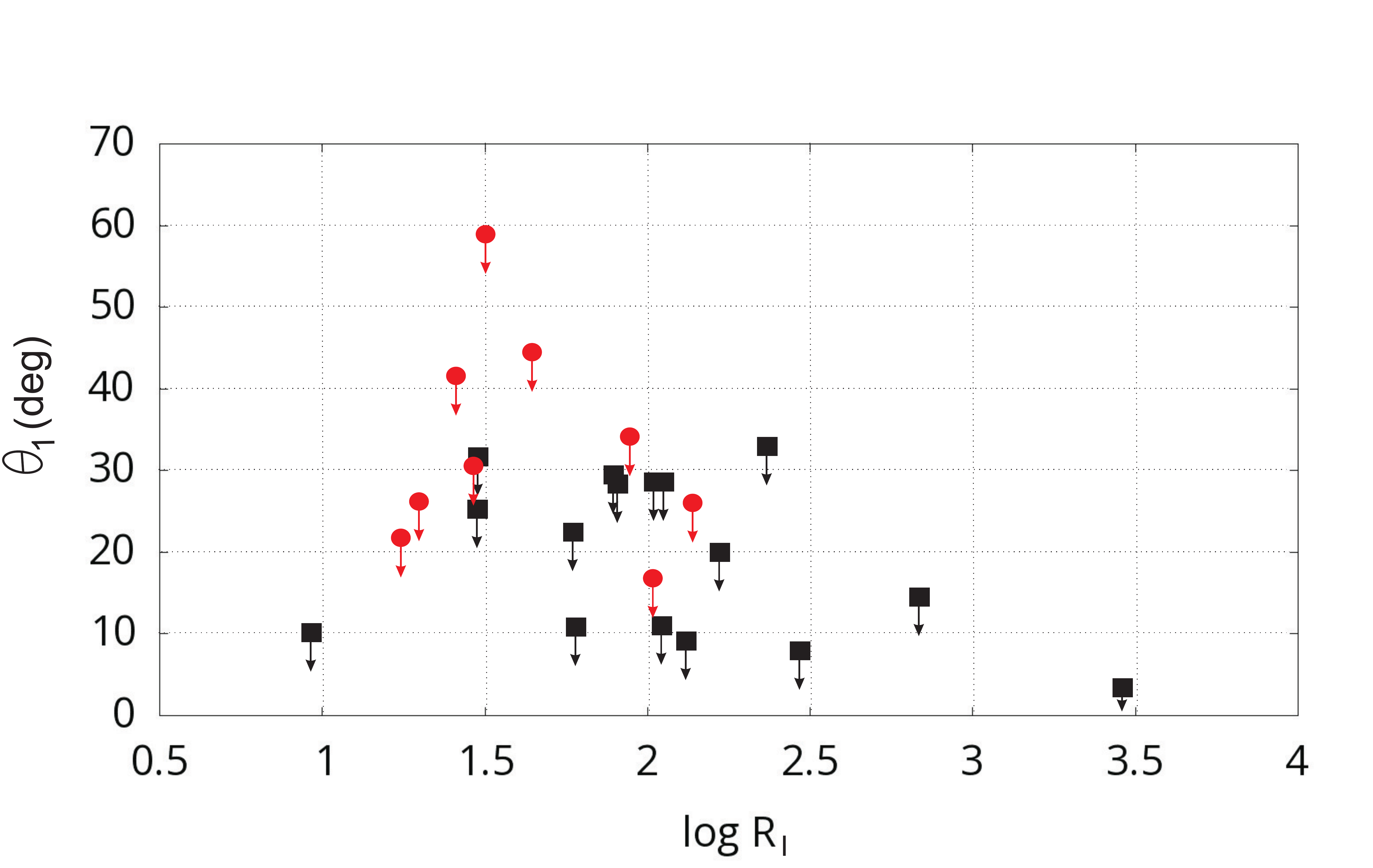}
\caption{Upper limits of the viewing angles ($\theta_{1}$) vs. radio-loudness parameter $\rm R_{I}$ for sources from Table~\ref{variable}. BI and AI quasars are indicated as circles and squares, respectively.}
\label{range}
\end{figure}

In the next step using equation \ref{var_eq} we calculated variability brightness temperature for all quasars and estimated the viewing angle $\theta$ of each 
object using formula~\ref{theta} as discussed in section~\ref{vlbi_tb}.
This time we took both, $\rm 10^{11}\,K$ and $\rm 10^{12}\,K$ limits, and we estimated the minimum 
Doppler factor as follows: $\rm \delta_{1}=\Big(\frac{T_{b}(var)}{10^{11}\,K}\Big)^{1/3}$ and 
$\rm \delta_{2}=\Big(\frac{T_{b}(var)}{10^{12}\,K}\Big)^{1/3}$. 
The maximum value of function \ref{theta} for both, $\rm \delta_{1}$ and $\rm \delta_{2}$, allowed us to 
determine the range where the maximum viewing angle of the quasar should be found.

The obtained range of viewing angles is wide, reaching very large values in a few cases (Table~\ref{variable}).
One of the AI quasars from our VLBI sample, namely 0756+3714, also show flux density variability and is listed in Table~\ref{variable}.
We classified this source as a core-jet, although an alternative interpretation as double-lobed young radio source has been proposed by \citet{bruni13}. The value
of the VLBI brightness temperature of 0756+3714 is not high, but it is close to the equipartition limit of $\rm 5\times10^{10}\,K$ proposed by \citet{read94}. It is however,
much lower than the variability brightness temperature we obtained for this object. We suggest that the possible flux density variations (sec.~\ref{not}) and very high value 
of the radio-loudness parameter allow for the core-jet classification of 0756+3714. 
VLBI observations with higher angular resolution may help to resolve puzzle of radio morphology of this source.

\begin{figure}
\centering
\includegraphics[width=\columnwidth]{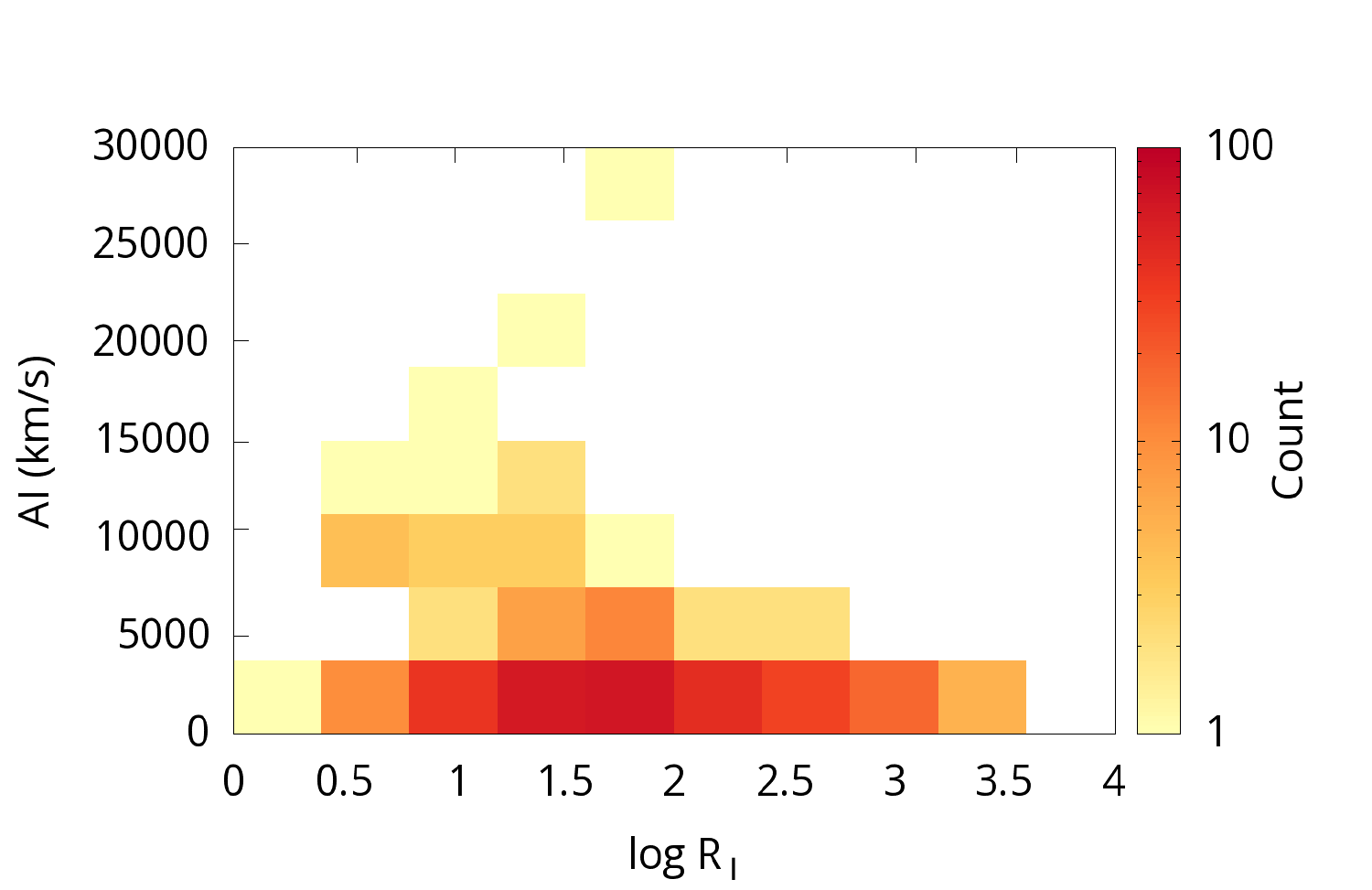}\\
\includegraphics[width=\columnwidth]{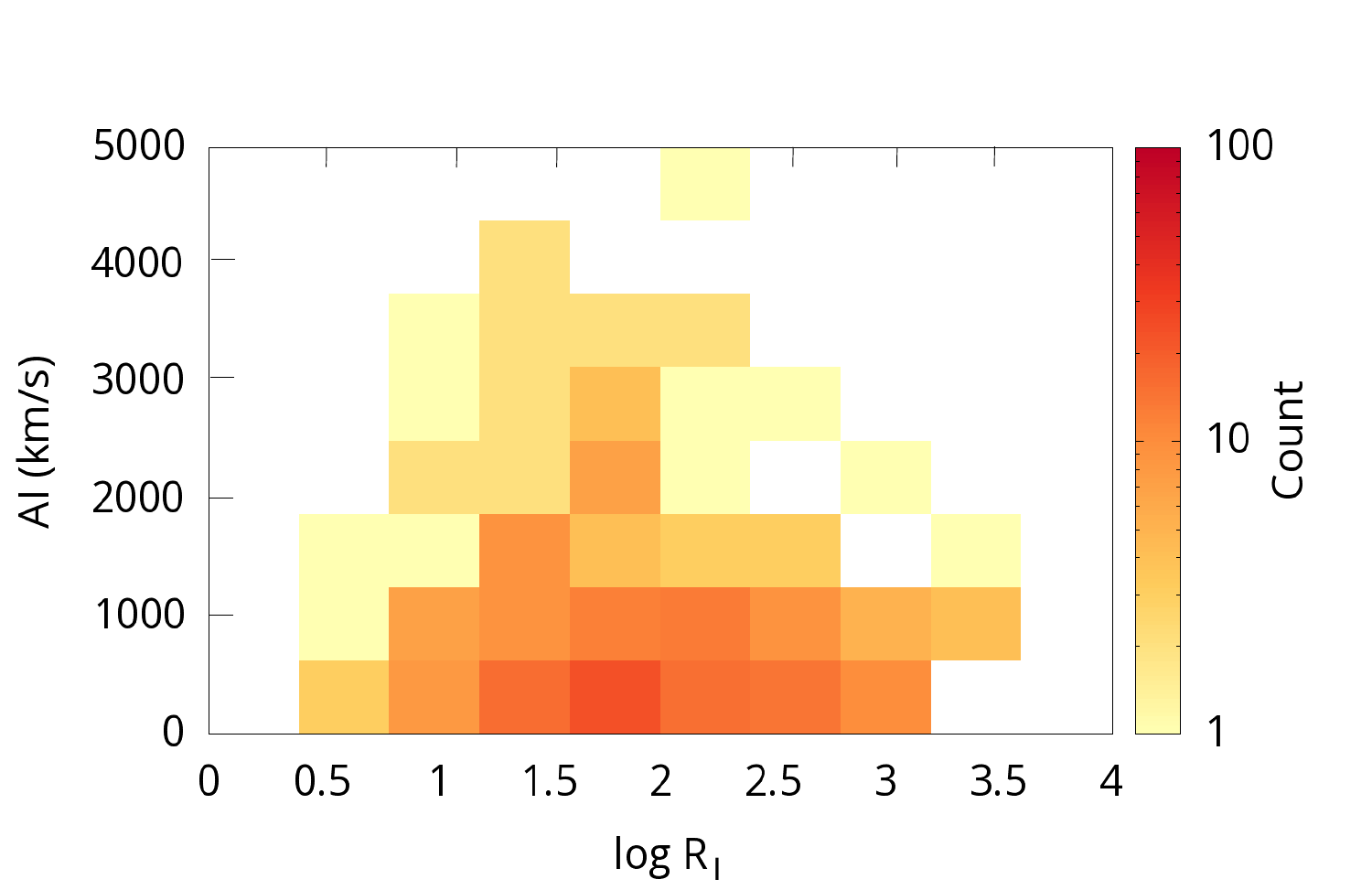}\\
\includegraphics[width=\columnwidth]{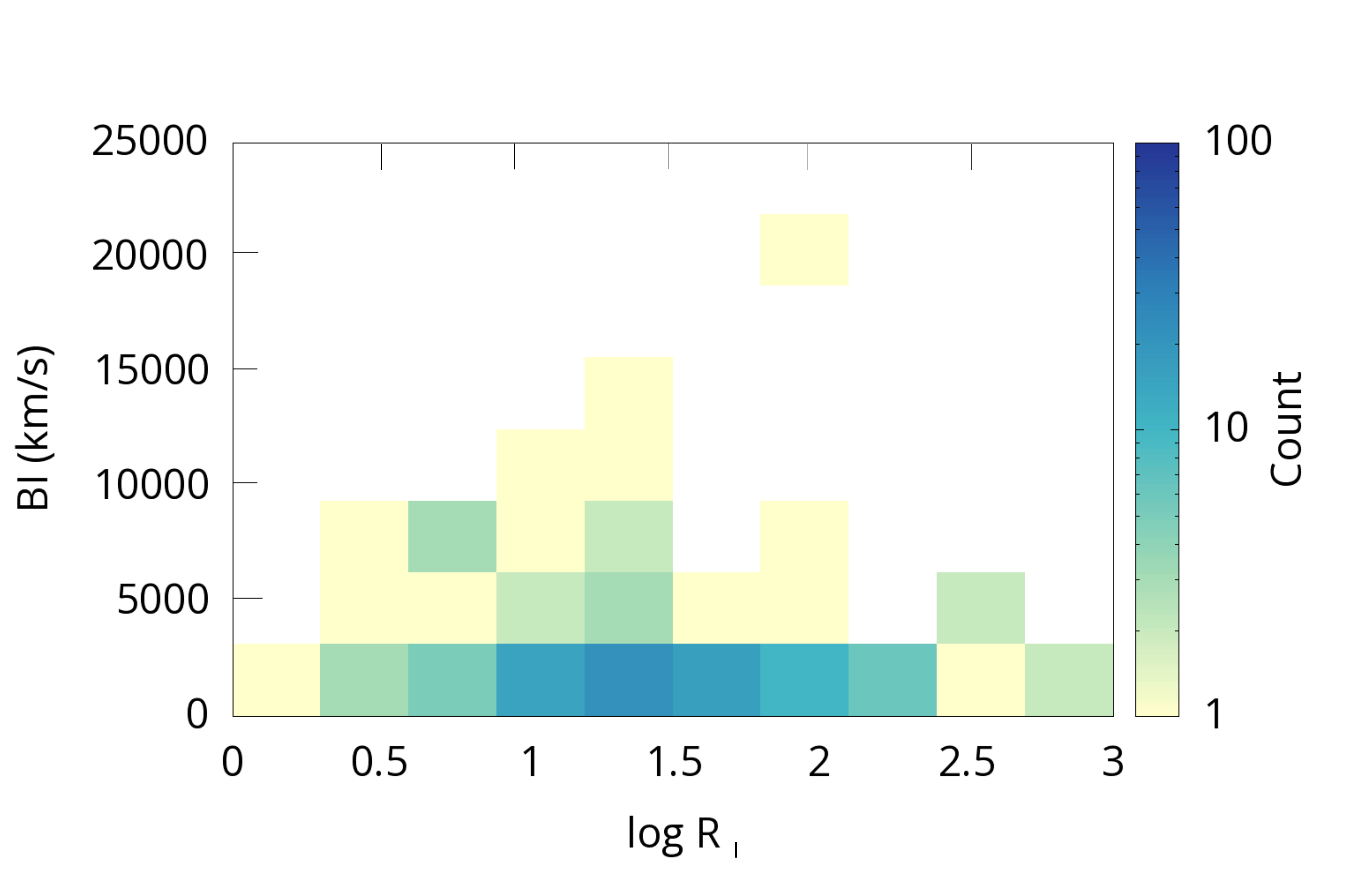}
\caption{
Radio-loudness plotted versus the value of: absorption index (AI) for the whole parent sample (top) and AI quasars only (middle), and balnicity index (BI) for BI quasars only (bottom).}
\label{counts3}
\end{figure}

\subsubsection{Orientation of BALQSOs}

 Estimations of the upper limit of wieving angles $\theta$ show that $\theta$ can reach value as large as $\sim 70\degr$ 
 for AI and BI quasars with $\rm log\,R_{I} < 1.5$ (Table~\ref{variable}). This range changes as a function of radio-loudness parameter, 
 log\,$\rm R_{I}$, what is well visible in Figure~\ref{range}.
 For the middle group, $\rm 1.5<log\,R_{I}<2.5$, where the peak of the AI quasars distribution is present, the viewing angles are 
 less than $\rm 45 - 56\degr$. We are aware that the viewing angle analysis based only on the upper limits might be burdened with significant error.
 However, the obtained trend seems to be in agreement with recent report about radio-loudness parameter as an orientation indicator \citep{kimball2011}.

We contrasted then the radio-loudness values with the absorption index (AI) for the whole parent sample (Fig.~\ref{counts3}, top panel). 
It is visible that stronger absorption is associated with lower values of radio-loudness parameter, log\,$\rm R_{I}<1.5$. 
The same trend is present for BI quasars only (Fig.~\ref{counts3}, bottom panel). This relationship however, is not so obvious 
for radio stronger AI quasars (Fig.~\ref{counts3}, middle panel). Nevertheless it can be noticed that the fraction of quasars with AI value grater than 2000 starts to drop for log\,$\rm R_{I}<2$. 

Additionally VLBI observations of three AI quasars allowed us to verify this conclusion by independent viewing angles determination. 
The estimated values of the viewing angles of 
0928+4446, 1018+0530 and 1223+5037 amounts to $\theta<37\degr$ for $\rm log\,R_{I}>2.4$. And the values of absorption index for these
three objects are the ones of the lowest in our sample of AI quasars
indicating relatively weak absorption (Table~\ref{basic}).

Based on our studies presented here we conclude that there exist a preferable orientation, 
possibly $\theta\ga37\degr$, in which absorbing screen reaches maximum covering factor.

\subsection{AI versus BI quasars}

In the QSO unification scheme of \citet{elvis00}, both broad and narrow absorption lines are assumed to be formed in the same disc wind and orientation
is the parameter which
determines our ability to detect them as AI or BI sources. On the other hand, radio, optical and X-ray studies of BI and AI quasars revealed many differences between them indicating that they 
constitute two independent classes \citep{knigge2008,strebl10} or are connected by the 'evolution of the flow' scenario, namely the 
weaker and much narrower absorption lines may represent the late evolutionary stages of classical BALs \citep{ganguly07, strebl10}.
The strong influence of evolution on broad absorption lines phenomenon is supported also by the radio observations of BALQSOs \citep{becker00,gregg06,hewett03}. The fact that most of radio-loud BALQSOs are compact, so young sources, introduced the scenario in which absorption lines are present in the early evolution phase of quasars.

The results presented in this paper confirm the fact that AI and BI quasars are statistically independent. Both groups of sources differ in the 
value of the absorption, its distribution versus the radio-loudness parameter and the radio-loudness distribution itself. 
Stronger absorption is associated with smaller values of $\rm log\,R_{I}<1.5$ which in turn can indicate small inclination angles with respect
to the disk plane. Orientation is then indeed an important parameter determining the possibility of absorption line detection.
However, there is another important conclusion that can be drawn from our studies of the relationship between the radio-loudness parameter and absorption. 
We have to emphasize that there is no correlation between the radio-loudness parameter and the AI/BI index since a large
span of AI/BI values occur in each bin of log\,$\rm R_{I}$. Therefore there has to be an additional factor which plays important role in BALQSO phenomenon.
We think that the fact that strong absorption is connected with weak radio emission is not without significance here. We suggest that the radio characteristics of BALQSOs can be well explained by the already established evolutionary scheme of radio-loud AGNs and by the scenario of intermittent 
activity proposed for the weak compact ones \citep{kun10, kun10b}. Such intermittent behaviour can be directly connected 
with the properties of the black hole and accretion process but not with the radio evolution itself.
For instance recent UV observations studies of radio-loud and radio-quiet BALQSOs show that they are statistically identical \citep{roch14}.
Both classes of objects differ only in radio emission which, as we discussed, is weak for some reason.

\section{Conclusions}

BALQSO phenomenon is usually explained by either an orientation scenario in which structures responsible for 
BALs are visible under specific inclinations or by an evolutionary scenario which connects BAL with young stage of QSO evolution. Despite long
and thorough discussion over the last 20 years, the ambiguity in the origin of broad absorption
lines in AGN spectra remains. To address this, we have presented 1.7-, 5- and 8.4-GHz
interferometric observations of 14 AI quasars (BI = 0) and compared properties of AI and BI
sources from a newly selected sample. Our main conclusions are as follows:

\begin{itemize}
\item {All AI quasars but one have been resolved in VLBI observations showing typical for QSOs, core-jet morphology. 
Their high radio luminosities and compact sizes indicate they belong to the population of young AGNs. 
Simultaneously the bright sources are a minority among absorption quasars and thus these AI objects belong to the tail of radio power distribution for AI quasars.}

\item {The statistical analysis of AI and BI sources shows that for the radio-loudness parameter, log\,$\rm R_{I}$, the distribution of AI and BI quasars 
differ significantly, peaking at 1.2 and 1.7 for BI and AI, respectively. Notice that the strong absorption, which is exclusively visible in BI quasars, is connected with  lower values of log\,$\rm R_{I}$ and weak radio emission.}

\item {The radio-loudness parameter is thought to be a good indicator of source orientation and therefore its low values, log\,$\rm R_{I}\lesssim1.5$ may imply large viewing angles. 
Since, the AI quasars have on average larger values of log\,$\rm R_{I}$, the orientation can mean that we see them less absorbed.}

\item {Orientation is an important but not the only one parameter that determines our ability to detect absorption lines. We suggest also that the radio evolution itself is not directly connected with BAL phenomenon
but is rather superimposed on the radio-loud BALQSO population.}

\end{itemize}

\section{Acknowledgments}
This work was supported by the National Scientific Centre under grant DEC-2011/01/D/ST9/00378.\\
The research leading to these results has received funding from the European Commission Seventh Framework Programme (FP/2007-2013) under grant agreement No 283393 (RadioNet3).\\
The European VLBI Network is a joint facility of European, Chinese, 
South African and other radio astronomy institutes funded by their national research councils.\\
The National Radio Astronomy Observatory is a facility of the National Science Foundation 
operated under cooperative agreement by Associated Universities, Inc.\\
The research described in this paper makes use of Filtergraph, an online data 
visualization tool developed at Vanderbilt University through the Vanderbilt 
Initiative in Data-intensive Astrophysics (VIDA).\\

\label{lastpage}

\end{document}